\documentclass[10pt]{article}

\usepackage{graphicx}
\usepackage{epsfig,cite}
\usepackage{amssymb}
\usepackage{amsmath}
\usepackage{dsfont}
\usepackage{multirow}
\usepackage{color}
\usepackage{subfigure,amstext,alltt,setspace}
\usepackage{amsbsy}
\usepackage{comment}
\usepackage{fullpage}
\usepackage{array}
\usepackage{booktabs,multirow,tabularx}
\usepackage{hyperref}

\newcommand{\amc}{{\sc MadGraph5\_aMC@NLO}}

\newcommand{\mc}{{\sc MC@NLO}}

\newcommand{\fr}{{\sc FeynRules}}
\newcommand{\fa}{{\sc FeynArts}}
\newcommand{\ufo}{{\sc UFO}}
\newcommand{\aloha}{{\sc ALOHA}}
\newcommand{\nlo}{{\sc NLOCT}}
\newcommand{\gev}{\,\textrm{GeV}}
\newcommand{\tev}{\,\textrm{TeV}}
\newcommand{\br}{\textrm{BR}}
\newcommand{\perc}{\%}
\newcommand{\hbt}{\ensuremath{H^-\bar{b}t}}
\newcommand{\mH}{\ensuremath{m_{H^-}}}

\usepackage{url}

\begin{document}

\begin{flushright}
IPPP/15/41 --- DCPT/15/82 --- MCNET-15-17 --- Cavendish-HEP-15/04  --- ZU-TH 18/15 \\
\end{flushright}

\vspace*{1cm}

\begin{center}
 {\Large \bf{Heavy charged Higgs boson production at the LHC}}
\end{center}

\vspace*{1cm}

\begin{center}
 C\'{e}line Degrande$^{[a]}$, 
 Maria Ubiali$^{[b]}$, Marius Wiesemann$^{[c]}$ and Marco Zaro$^{[d,e]}$
\end{center}

\vspace*{1cm}

\noindent
{\small
$^{[a]}$ Institute for Particle Physics Phenomenology, Department of Physics,\\ 
Durham University, Durham DH1 3LE, United Kingdom\\
$^{[b]}$ Cavendish Laboratory, HEP group,\\
University of Cambridge, J.J. Thomson Avenue, Cambridge CB3 0HE, United Kingdom\\
$^{[c]}$ Physik-Institut, Universit\"at Z\"urich,\\ 
Winterthurerstr. 190, 8057 Zurich, Switzerland\\
$^{[d]}$ Sorbonne Universit\'es, UPMC Univ. Paris 06,\\ 
UMR 7589, LPTHE, F-75005, Paris, France\\
$^{[e]}$ CNRS, UMR 7589, LPTHE, F-75005, Paris, France}
%%celine.degrande@durham.ac.uk
%%ubiali@hep.phy.cam.ac.uk
%%marco.zaro@lpthe.jussieu.fr

\vspace*{2cm}

\begin{center}
{\bf Abstract}
\end{center}

\noindent
In this paper we study the production of a heavy charged Higgs boson in association with heavy quarks at the LHC, 
in a type-II two-Higgs-doublet model. We present for the first time fully-differential results obtained in 
the four-flavour scheme at NLO accuracy, both at fixed order and 
including the matching with parton showers.
Relevant differential distributions are studied for two values 
of the charged boson mass and a thorough comparison is performed between 
predictions obtained in the four- and five-flavour 
schemes. 
We show that the agreement between the two schemes is improved by NLO(+PS) corrections for observables inclusive in 
the degrees of freedom of 
bottom quarks. We argue that the four-flavour scheme leads to more reliable predictions, thanks to its accurate
description of the bottom-quark kinematics and its small dependence on the Monte Carlos, which in turn is rather 
large in the five-flavour scheme. A 
detailed set of recommendations for
the simulation of this process in experimental analyses at the LHC is provided. 

\pagebreak

\tableofcontents
%%%%%%%%%%%%%%%%%%%%%%%%%%%%%%%%%%%%%%%%%%%%%%%%%%%%%%%%%%%%%%%%%%%%%5
\section{Introduction}
\label{sec:intro}
%%%%%%%%%%%%%%%%%%%%%%%%%%%%%%%%%%%%%%%%%%%%%%%%%%%%%%%%%%%%%%%%%%%%%5
Charged Higgs bosons appear in several extensions of the scalar sector of the Standard Model (SM). 
In particular, as the SM does not include any elementary charged scalar particle, the observation of a 
charged Higgs boson would necessarily point to the presence of new physics.

In this paper we focus on one of the simplest extensions of the SM featuring a charged scalar, namely
the two-Higgs-doublet model (2HDM). Within this class of models, the existence of five physical Higgs bosons, 
including two (mass-degenerate) charged particles $H^{\pm}$, is foreseen. 
Imposing flavour conservation, there are four different ways 
to couple the SM fermions to the two Higgs doublets. Each of these four ways 
of assigning the couplings gives rise to a different phenomenology for 
the charged Higgs boson. Here we consider the so-called type-II 2HDM, 
in which one doublet couples to up-type quarks 
and the other to the down-type quarks and the charged leptons.
The Minimal Supersymmetric Standard Model (MSSM), up to SUSY corrections, belongs to the
type-II 2HDM category.

Light charged Higgs scenarios are classified by Higgs boson masses 
smaller than the mass of the top quark (typically $m_{H^\pm}\lesssim 160\gev$),
where the top decay to a charged scalar and a bottom quark is allowed. 
One refers to heavy charged Higgs bosons, on the other hand, 
for masses larger than the top-quark mass (typically $m_{H^\pm}\gtrsim 180\gev$).
In this case, the dominant $H^{\pm}$ production channel at the Large Hadron Collider (LHC) is the associated
production with a top quark/antiquark and a (possibly low transverse momentum) bottom antiquark/quark.
In the intermediate-mass range ($160\lesssim m_{H^\pm} \lesssim 180\gev$) width effects become
important and the full amplitude for $p p\to H^{\pm}W^{\mp} b \bar{b}$ 
(including non-resonant contributions) 
must be taken into account for a proper description. 
The intermediate-mass range has
not been studied so far at the LHC Run I.

Searches at LEP have set a limit $m_{H^\pm} > 80\gev$ on the mass of a charged Higgs 
boson for a type-II 2HDM~\cite{hplep}. 
The Tevatron limits~\cite{hpcdf,hpd0} have been superseded by results of the LHC experiments. 
Recent ATLAS results~\cite{Aad:2014kga} for a type-II 2HDM
based on 19.5 fb$^{-1}$ of $pp$ collisions at 8 TeV exclude 
$\br(t \rightarrow bH^{+})\cdot \br(H^+ \rightarrow \tau^{+}\nu_{\tau})$ larger than $(0.23-1.3)\%$ 
in the low-mass region ($80\gev< m_{H^\pm}<160\gev$). For the first time  
limits on $tH^+$ production times $\br(H^+ \rightarrow \tau^{+}\nu_{\tau})$ 
are provided for $180\gev< m_{H^\pm}< 1000\gev$. Rates above 0.76 pb at low masses 
and 4.5 fb at large masses are excluded.
A preliminary note by CMS~\cite{CMS-PAS-HIG-14-020}, based on the analysis of 19.7 fb$^{-1}$ of $pp$ collisions at 8 TeV
in the same search channel, sets an upper
limit on $\br(t \rightarrow bH^{+})\cdot \br(H^+ \rightarrow \tau^{+}\nu_{\tau})$ between
$(0.16-1.2)\%$ in the low-mass region ($80\gev< m_{H^\pm}<160\gev$). This limit supersedes the one
published in Ref.~\cite{hpcms}. In the high-mass region (180\,GeV $< m_{H^\pm}<$ 600\,GeV), 
limits on $tH^+$ production times $\br(H^+ \rightarrow \tau^{+}\nu_{\tau})$ are
set between 0.38 pb and 26 fb. Finally, CMS has also published a preliminary note~\cite{CMS-PAS-HIG-13-026}
on a direct search for a heavy charged Higgs which decays in both  the 
$H^+\to t\bar{b}$ and the $H^+\to \tau^+\nu_\tau$ channels.

In this work, we consider heavy charged Higgs boson production at hadron colliders and leave the 
intermediate-mass range to future studies~\cite{preparation}. 
In particular, we focus only on the production of a 
negatively charged scalar since the results are identical for a positively charged scalar.
As for any process involving bottom quarks at the matrix-element level, two viable schemes 
exist to compute the production cross section of a heavy charged Higgs boson. These are
usually dubbed as four- and five-flavour schemes. 
In the four-flavour scheme (4FS) the bottom quark mass is considered as a hard scale of 
the process. Therefore, bottom quarks do not contribute to the proton 
wavefunction and can only be generated as massive final states
at the level of the short-distance cross section, entailing that $b$-tagged observables receive contributions 
starting at leading order (LO). In practice, the theory which 
is used in such a calculation is an effective theory with four light quarks, where the massive bottom quark 
is decoupled and enters neither the renormalisation group equation for the running 
of the strong coupling constant nor the 
evolution of the parton distribution functions (PDFs). The LO partonic processes in the case at hand are
\begin{align}
gg\rightarrow \hbt\quad\text{and} \quad q\bar q\rightarrow \hbt\,.
\end{align}
Next-to-leading order (NLO) calculations for the total cross sections in this scheme have been presented in Refs.~\cite{Peng:2006wv,Dittmaier:2009np}.

Conversely, in  five-flavour scheme (5FS), the bottom quark mass is considered 
to be much smaller than the hard scales involved in the process. The simplest definition 
of the 5FS---that suits particularly well perturbative computations---is to
strictly set $m_b=0$ in the short-distance cross section. Consequently, 
bottom quarks are treated on the same footing as all other massless partons. 
The only difference is the presence of a threshold in the bottom-quark PDF and the initial condition of the bottom
quark evolution being of perturbative nature.
The use of $b$-PDFs comes along with the approximation that, 
at leading order, the massless $b$ quark has a small transverse momentum.
In this scheme, the leading logarithms associated to the initial state collinear splitting
are resummed to all orders in perturbation theory by the Dokshitzer-Gribov-Lipatov-Altarelli-Parisi (DGLAP)
 evolution of the bottom densities. The LO partonic process is given by 
\begin{align}
gb\rightarrow H^-t\,.
\end{align}
Next-to-leading order predictions for heavy charged Higgs boson production in the 5FS, 
possibly including the matching to parton-shower Monte Carlos, were studied in 
Refs.~\cite{Zhu:2001nt,Gao:2002is,Plehn:2002vy,Berger:2003sm,Kidonakis:2005hc,Weydert:2009vr,Klasen:2012wq}.
Electroweak corrections~\cite{Beccaria:2009my,Nhung:2012er} and soft gluon resummation 
effects~\cite{Kidonakis:2004ib,Kidonakis:2005hc,Kidonakis:2010ux} have also been included in recent works.

\begin{figure}[t]
\centering
\includegraphics[width=0.3\textwidth]{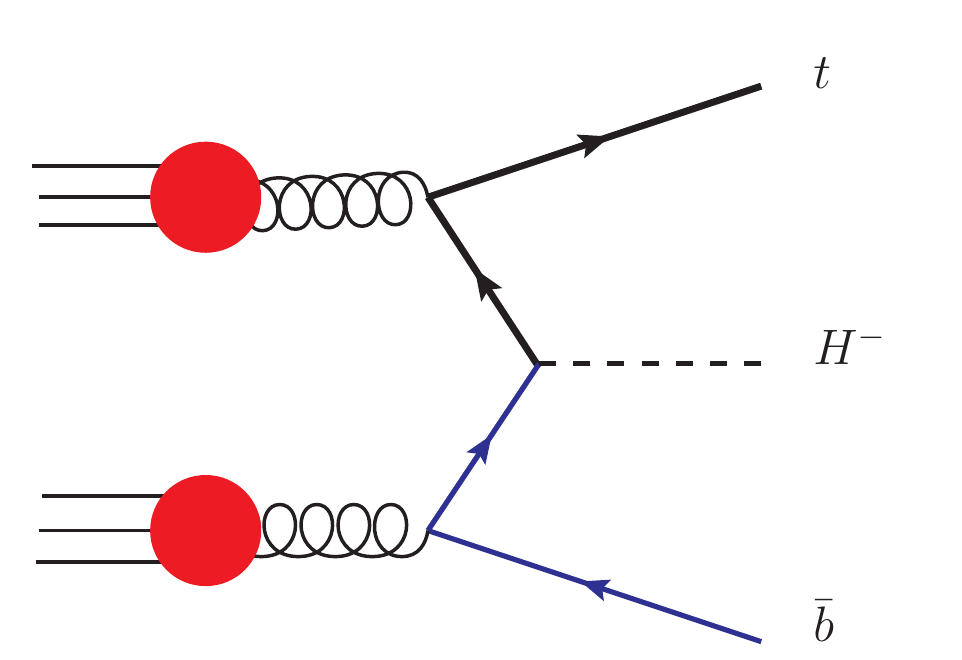}
\includegraphics[width=0.33\textwidth]{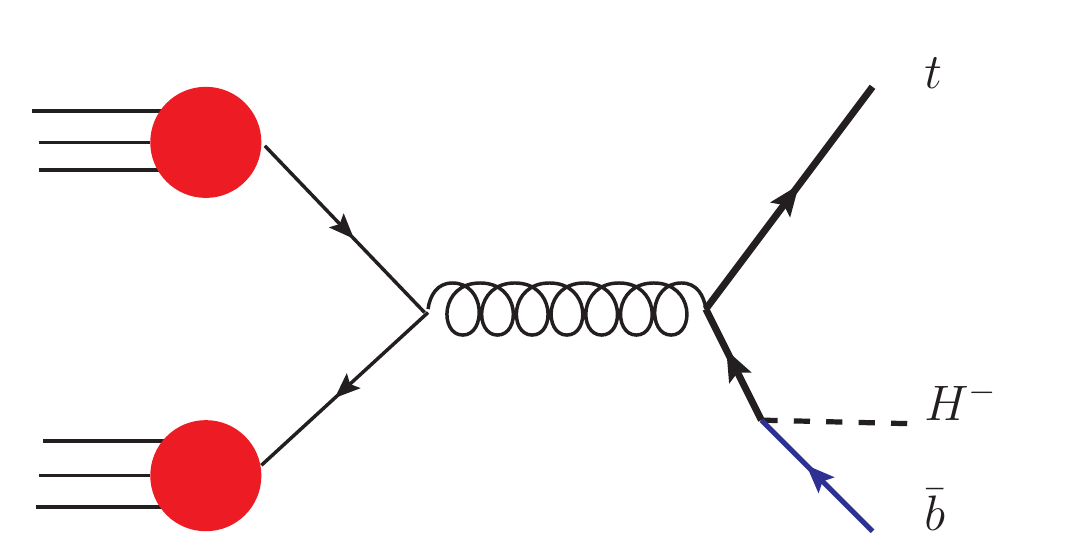}
\includegraphics[width=0.3\textwidth]{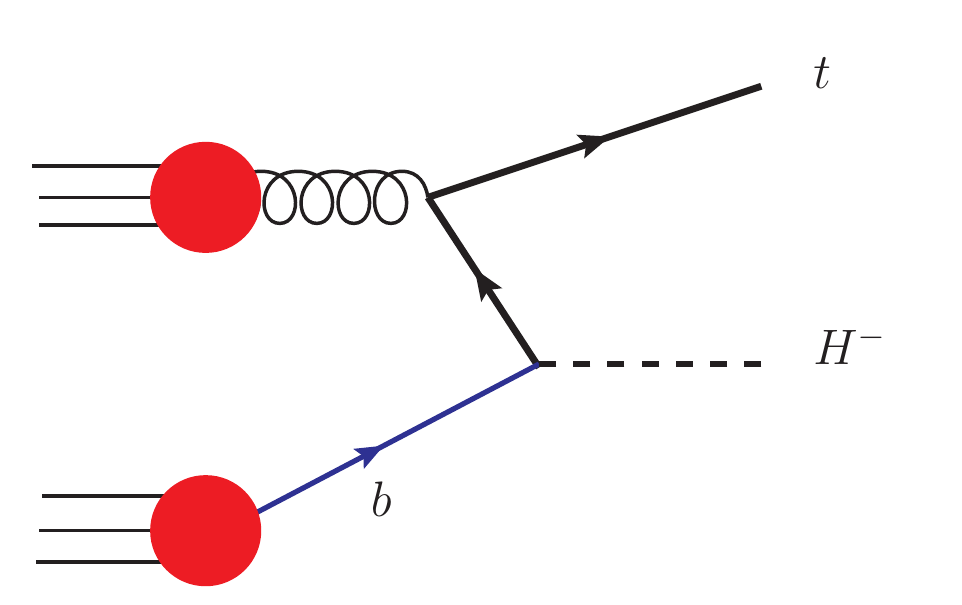}
\caption{ Leading-order diagrams for heavy charged 
Higgs and top associated production, in the 4FS (left and centre) and 5FS (right).}
\label{fig:feynmanLO} 
\end{figure}

The leading-order diagrams in the 4FS and 5FS are displayed in Fig.~\ref{fig:feynmanLO}.
The comparison between the two schemes at the level of total cross section has been performed by 
several groups, see e.g. Ref.~\cite{Dittmaier:2009np} and references therein. 
In a more recent study \cite{Flechl:2014wfa} a thorough combination of all sources of theoretical 
uncertainties is performed, state-of-the-art PDF sets are used, the new scale-setting 
procedure proposed in Refs.~\cite{Maltoni:2012pa,Ubiali:2014cva} is adopted and a 
Santander-matched prediction~\cite{SANTANDER} of the four- and five-flavour 
scheme computations is provided. 
Furthermore, a comparison of five- and six-flavour schemes, 
assessing the effect of the inclusion of top PDFs, has been 
recently performed at the level of total cross sections~\cite{Dawson:2014pea}.
In summary, these studies establish a generally good agreement of total cross section 
predictions in different flavour schemes, once a judicious scale choice is made.

Studies at differential level of heavy charged Higgs production in four- and five-flavour 
schemes are very limited, if not absent. 
NLO plus Parton Shower (PS) accurate predictions are available only in the 5FS 
for both \mc~\cite{Weydert:2009vr} and POWHEG~\cite{Klasen:2012wq}, and their comparison to 
4FS results at the level of differential distributions
has never been performed to date.
Such a differential study is certainly more involved than the 
inclusive one, mainly for two reasons. 
First of all, 
there is a very large number of possibly relevant observables at the differential level. 
Second, differences between the two schemes are generally larger than in the inclusive case. 
As a consequence, an even more careful 
assessment of the related uncertainties is necessary for distributions.
Indeed, 
observables inclusive 
in the degrees of freedom of the bottom quarks, such as total cross sections, 
turn out to be quite similar in four- and 
five-flavour schemes. On the one hand 
power-suppressed terms are small for this class of observables. 
On the other hand 
collinear logarithms are phase-space suppressed~\cite{Maltoni:2012pa} and therefore moderate, 
unless a very heavy charged Higgs is produced.

As far as exclusive observables are concerned, it is vital to investigate 
4FS and 5FS predictions and assess their differences on a case-by-case basis
in an unbiased manner, which is one of the main goals of this paper.
The generic rationale is the following: if power-suppressed terms are 
relevant, the 4FS provides a more reliable prediction, while if collinear 
logarithms are large, the reorganisation of the perturbative series in the 5FS 
improves the stability of the perturbative expansion.

The comparison of the two schemes is further 
complicated by the fact that observables related to light and $b$-flavoured 
jets are associated with different perturbative orders: 
while in charged Higgs production one final-state $b$ quark 
(not accounting for top decays) is already present at LO in the 4FS, 
a final-state $b$ quark enters the 5FS computation only at NLO.
Furthermore, tagging $b$ quarks in a 5FS fixed-order computation necessarily 
leads to unphysical results, since they can be considered 
only as part of jets to retain infra-red cancellations. This issue clearly does 
not affect the 4FS, in which observables related to the $b$ quarks are regulated 
by the $b$ mass. Nevertheless, when considering $b$ quarks at very large 
transverse momenta, $b$-jet observables have to be preferred anyway, otherwise
large logarithms $\log(p_T/m_b)$ would spoil the perturbative convergence. 

Matching to parton showers improves most notably the 5FS predictions, since 
the parton backward evolution of the initial-state bottom quarks turns them 
into massive final states and further generates $b$-flavoured hadrons which 
renders $b$-tagging realistic. One should bear in mind, however, that
the details and the actual implementation of such 
backward evolution are highly non-trivial---being based on DGLAP evolution
equations, which are only LL accurate---and therefore turn out to be widely 
Monte Carlo dependent (see Sect.~3.3 of Ref.~\cite{Frixione:2010ra}, 
and Sect.~\ref{sec:4vs5FS} below).
Moreover, the necessity of reshuffling the massless into massive bottom quarks 
may have significant effects on the kinematics of final-state $B$ hadrons. 
In the 4FS, the shower primarily improves the description of Sudakov-suppressed 
small-$p_T$ radiation, by resumming leading collinear logarithms
to all orders. In both schemes the PS introduces 
additional power-suppressed terms in the 
soft region.

In this paper, we present for the first time NLO+PS accurate 4FS predictions and thoroughly
compare them to the 5FS ones. 
Our primary aim is to acquire a detailed understanding and assess which scheme 
is better suited for the simulation of the charged Higgs production signal,
whose search represents a central part of the physics program in (and beyond) Run II 
at the LHC. 
As will be shown, our analysis confirms and extends the conclusions that have
been drawn in previous studies of differential distributions in the four- and five-flavour schemes, 
in the context of Higgs production
in association with bottom quarks~\cite{Wiesemann:2014ioa} and
Higgs and single-top associated production~\cite{Demartin:2015uha}. 
Similar analyses were performed for single top production~\cite{Campbell:2009ss,Frederix:2012dh}.
Moreover, given that the mass of the charged Higgs boson is still allowed to take almost arbitrarily large 
values, this process is perfectly suited for a case-study of different flavour schemes 
in QCD in a broad range of scales.

The paper is structured as follows: in Sect.~\ref{sec:calculation} an outline of the calculation is given
and details of the implementation of the 4FS calculation at NLO+PS accuracy are provided.
Results are displayed in Sect.~\ref{sec:results}, in which the features of the 4FS calculation are described,
including the size of the interference term that appears in this scheme, and the 4FS and 5FS distributions
are compared. We conclude in Sect.~\ref{sec:conclusion} with our final recommendations for experimental analyses. 

%%%%%%%%%%%%%%%%%%%%%%%%%%%%%%%%%%%%%%%%%%%%%%%%%%%%%%%%%%%%%%%%%%%%%5
\section{Outline of the calculation}
\label{sec:calculation}
%%%%%%%%%%%%%%%%%%%%%%%%%%%%%%%%%%%%%%%%%%%%%%%%%%%%%%%%%%%%%%%%%%%%%5
Our computation takes advantage of a
full chain of automatic tools developed to study the phenomenology of new physics models at NLO QCD accuracy
in the \amc~\cite{Alwall:2014hca} framework. 

%%%%%%%%%%%%%%%%%%%%%%%%%%%%%%%%%%%%%%%%%%%%%%%%%%%%%%%%%%%%%%%%%%%%%5
\subsection{Framework}

Besides the usual tree-level Feynman rules, some extra ingredients have to be 
provided to the matrix-element generator in order to obtain the code for the simulation 
of a new physics process at NLO. These extra ingredients are the Ultra-Violet (UV) renormalisation 
counter-terms and a sub-class of the rational terms 
that enter the reduction of the virtual matrix elements, the so-called $R_2$ terms \cite{Ossola:2008xq}. 
UV and $R_2$ counter-terms can be computed starting from the model Lagrangian 
for any renormalisable theory 
via the \nlo\ package~\cite{Degrande:2014vpa}, based on~\fr~\cite{Alloul:2013bka} and 
\fa~\cite{Hahn:2000kx}. The tree-level and NLO Feynman rules are exported in the Universal 
Feynrules Output (\ufo)~\cite{Degrande:2011ua} format, as a 
{\sc Python} module which can be loaded by matrix-element generators, 
such as \amc. Finally, the \ufo\ information is translated into helicity 
routines~\cite{Murayama:1992gi} by \aloha~\cite{deAquino:2011ub}. 

We remind the reader that \amc\ is a meta-code that automatically generates the code for 
simulating any process at NLO(+PS) accuracy. \amc\ adopts the FKS 
method~\cite{Frixione:1995ms, Frixione:1997np} for the subtraction of the singularities of 
the real-emission matrix elements, as automated in the {\sc MadFKS} 
module~\cite{Frederix:2009yq}; virtual matrix elements are computed via the 
{\sc MadLoop} module~\cite{Hirschi:2011pa}, which relies on the OPP~\cite{Ossola:2006us} 
and Tensor Integral Reduction 
(TIR)~\cite{Passarino:1978jh, Davydychev:1991va} methods, 
as implemented in {\sc CutTools}~\cite{Ossola:2007ax} 
and {\sc IREGI}~\cite{iregi}, supplemented by an in-house implementation of the {\sc OpenLoops}\ procedure~\cite{Cascioli:2011va}; finally, the generation of hard events and their matching to parton-shower 
simulations is 
performed {\it \`a la} \mc~\cite{Frixione:2002ik}. 
The matching to {\sc Herwig6}~\cite{Corcella:2000bw}, 
{\sc Pythia6}~\cite{Sjostrand:2006za} (ordered in virtuality or in transverse momentum, 
the latter only for processes with no light partons in the final state), 
{\sc Herwig++}~\cite{Bahr:2008pv} and {\sc Pythia8}~\cite{Sjostrand:2007gs} is available.
As a consequence, the only inputs needed are the implementation of the model in \fr\ 
and the inclusion of the running of the bottom quark mass, which follows from our 
renormalisation-scheme choice. More details are given in the next section. 
Both the 4FS and the 5FS computations have been performed in the framework specified above, 
with the additional advantage
of ensuring the consistency of all settings and input parameters. 

To guarantee the validity of our analysis, we have performed a detailed comparison of the inclusive cross sections 
obtained with \amc\ against previous results published in literature. 
The calculation in the 4FS has been compared 
to the private code of Ref.~\cite{Dittmaier:2009np} at the level of total rates, 
and a good agreement within the numerical 
uncertainty of the reference code has been found. 
Furthermore, with the bottom quark Yukawa renormalised on-shell 
and with its mass set the value of the top pole mass, 
we reproduce the $t\bar{t}H$ total cross section in the SM 
at the few per-mille level.

%%%%%%%%%%%%%%%%%%%%%%%%%%%%%%%%%%%%%%%%%%%%%%%%%%%%%%%%%%%%%%%%%%%%%5
\subsection{Implementation}

We have used the implementation of the generic 2HDM in \fr\ detailed in Ref.~\cite{Degrande:2014vpa}. 
This model has been converted into a type-II 2HDM by adding $\beta$ 
as an external parameter and by 
restricting the Yukawa couplings accordingly. If top and bottom quarks are assumed to be 
the only massive fermions, the only non-zero entries of the Yukawa coupling matrices to the doublet 
without vacuum expectation value 
in the Higgs basis for the type-II 2HDM are given by
\begin{equation}
G^u_{3,3} = -\sqrt 2  \,\frac{m_t^y}{v} \,\cot\beta\quad \text{and} \quad G^d_{3,3} 
= \sqrt 2 \,\frac{m_b^y}{v} \,\tan\beta,
\end{equation} 
where  $m_{t/b}^y$ are the Yukawa masses of the top and bottom quark. The parameter 
$\tan\beta = v_2/v_1$ is the ratio of the vacuum expectation values $v_1$ and $v_2$ of 
the two Higgs doublets, such that $v^2\equiv v_1^2+v_2^2 = (\sqrt{2}G_F)^{-1}$ is the SM 
Higgs vacuum expectation value, where $G_F$ is the Fermi constant. 
With those restrictions, the $H^-t \bar b$ vertex is given  by 
\begin{equation}
V_{t\bar{b}H^-}= -i \left(y_t P_R\,\frac1{\tan\beta} + y_b P_L\tan\beta\right),
\label{eq:tbhcoupling}
\end{equation}
where $P_{R/L} = (1\pm\gamma_5)/2$ are the chirality projectors and 
$y_{t/b} \equiv \sqrt2\frac{m_{t/b}^y}{v}$ are the corresponding SM Yukawa couplings. 
We strictly separate the Yukawa masses that are used in the computation 
of the couplings between the fermions 
and the scalars from the kinematic masses that are used everywhere else
and set to the on-shell mass.\footnote{They 
appear explicitly separated also in the YUKAWA and MASS blocks of the SLHA cards~\cite{Alwall:2007mw}.}
This distinction allows us to keep a non-vanishing bottom Yukawa  
in the five-flavour scheme as the leading term in the small $m_b$ 
expansion \cite{Plehn:2002vy,Berger:2003sm}. Furthermore, it allows us to 
choose different renormalisation schemes for the bottom quark mass in the matrix
element and in the Yukawa coupling.

The model $R_2$  and UV vertices required for NLO computations in \amc\ 
have been computed using \nlo~\cite{Degrande:2014vpa}. The masses and the
 wave functions are renormalised in the on-shell scheme to avoid the computation 
of loops on external legs. The strong coupling constant is renormalised 
in the $\overline{\text{MS}}$ scheme with the contribution of massive quarks 
subtracted from the gluon self-energy at zero-momentum transfer. 
Therefore, only the massless modes affect the running of $\alpha_s$. 
The renormalisation of the masses in principle fixes 
the renormalisation of the top and bottom Yukawa since
\begin{equation}
\delta y_{t/b} = \sqrt2\frac{\delta m_{t/b}}{v}\label{eq:deltayOS},
\end{equation} 
with 
\begin{equation}
\delta m_{t/b} = -\frac{g_s^2}{12\pi^2}m_{t/b} \left(\frac{3}{\bar\epsilon} +4 - 6 \log\frac{m_{t/b}}{\mu_R}\right)\,\label{eq:deltamOS}
\end{equation} 
in the on-shell scheme. This is the default renormalisation used in \nlo\ 
and it would ensure that nothing but the strong coupling constant 
depends on the renormalisation scale. 
The top mass and Yukawa are always renormalised in this way throughout 
this paper.
Therefore its Yukawa mass is set equal to the  pole mass. 
On the contrary, the bottom quark Yukawa has been renormalised in the 
$\overline{\text{MS}}$ scheme, i.\,e.
\begin{equation}
\delta y_{b} = -\frac{\sqrt2}{v}\frac{g_s^2  m_{b}^y}{4\pi^2 \bar\epsilon} \,. \label{eq:deltayMSbar}
\end{equation} 
This scheme choice has the advantage of resumming potentially large logarithms
 $\log (\mu_R/m_b)$ (with  $\mu_R\sim m_{H^\pm}$) to all orders. 
 The bottom Yukawa mass is set to the value of the running $\overline{\text{MS}}$ mass at the renormalisation scale.
Besides the modifications at the level of the \ufo\ model,
also the code written by \amc\ had to be changed in order to account for the 
additional scale dependence introduced by the $b$-quark Yukawa, in particular for
what concerns the on-the-fly evaluation of scale uncertainties obtained via reweighting~\cite{Frederix:2011ss}. This 
has been done in an analogous way as for bottom-associated Higgs production~\cite{Wiesemann:2014ioa}, 
by splitting 
the cross section in parts that factorise different 
powers of $y_b$, i.\,e. $y_b^2$, $y_b\,y_t$ and $y_t^2$. 

%%%%%%%%%%%%%%%%%%%%%%%%%%%%%%%%%%%%%%%%%%%%%%%%%%%%%%%%%%%%%%%%%%%%%5
\section{Results}
\label{sec:results}
%%%%%%%%%%%%%%%%%%%%%%%%%%%%%%%%%%%%%%%%%%%%%%%%%%%%%%%%%%%%%%%%%%%%%5

In this section, we present four-flavour scheme predictions of charged Higgs boson 
production at NLO matched to parton showers.
This calculation has never been performed before in the literature. 
Several differential distributions that are reconstructed from the final state particles in $tH^-\bar{b}$ production
are studied.
We investigate the role of the shower scale in this process, discuss the
impact of the $y_by_t$ interference term and compare our reference predictions at (N)LO+PS to the f(N)LO results.
For matched predictions, both {\sc Herwig++} and {\sc Pythia8} are employed.
We conclude this chapter with a comprehensive comparison of 4FS and 5FS distributions, in which 
the effects of higher order corrections, the impact of the choice of the shower scale and the dependence of each scheme on the different Monte 
Carlos are analysed.

%%%%%%%%%%%%%%%%%%%%%%%%%%%%%%%%%%%%%%%%%%%%%%%%%%%%%%%%%%%%%%%%%%%%%5
\subsection{Settings}
\label{sec:settings}
We present results for charged Higgs boson production at the LHC Run II 
($\sqrt S_{\rm had}=13 \tev$) by considering two scenarios: a lighter ($m_{H^\pm}=200 \gev$) 
and a heavier ($m_{H^\pm}=600 \gev$) charged Higgs boson. For simplicity, we set $\tan\beta=8$
throughout this paper. At this value, $y_b^2$ and $y_t^2$ terms are of similar size and the 
relative contribution of the $y_by_t$ term to the total cross section is close to its maximum. 
Results for any other value of $\tan\beta$ can be obtained by a trivial overall rescaling of the 
individual contributions according to their Yukawa couplings 
($y_b$ by $\tan\beta$, $y_t$ by $1/\tan\beta$). 
Therefore, we preserve the generality of our results by studying the $y_b^2$, 
$y_t^2$ and $y_by_t$ contributions separately.

We show results obtained with the NNPDF2.3 set~\cite{Ball:2012cx} 
at NLO and the NNPDF3.0~\cite{Ball:2014uwa} set at LO. To obtain consistent 
predictions, parton distribution functions (PDFs) computed in the proper 
flavour number scheme are used: we interface our NLO (LO) calculation with the NNPDF2.3 
(NNPDF3.0) with $n_f=4$ and $n_f=5$ active flavours for the 4FS and 5FS respectively.
The mismatch between the PDF sets used in the LO and NLO computations is
due to the absence of a public set of non-QED LO PDFs in the NNPDF2.3 family. This does not affect the accuracy of our results, given that the LO PDF sets exhibit a theoretical
uncertainty which is larger than the difference between the two NNPDF families. 
The strong coupling constant is consistent with
$\alpha_s(M_Z)=0.118$ for the 5FS NLO parton densities and $\alpha_s(M_Z)=0.1226$ for the 
4FS NLO ones.\footnote{This is the value of $\alpha_s(M_Z)$ associated with the
{\tt NNPDF23\_nlo\_as\_0118\_nf4} set: the 4FS sets are constructed by evolving backwards the 5FS PDFs
and the strong coupling constant from the $Z$ mass to the threshold associated to the bottom PDF. 
They are then evolved upwards from the bottom threshold to higher scales by setting $n_f=4$.}
 
The heavy quark pole masses are set to
\begin{equation}
    m_b^{\rm pole} = 4.75 \gev\; \textrm{(relevant only to the 4FS)},\qquad m_t^{\rm pole} = 172.5\gev.
\end{equation}
At one loop, the value of the bottom pole mass translates into a $\overline{\text{MS}}$ mass
\begin{equation}
    \bar{m}_b(\bar{m}_b) = 4.3377\gev.
\end{equation}
Finally, our central renormalisation and factorisation scales $\mu_R, \mu_F$ are set to
\begin{equation}
    \mu_{R,F}=H_T/3\equiv\frac{1}{3}\sum_i \sqrt{m(i)^2 + p_T(i)^2},
    \label{eq:muFmuR}
\end{equation}
where the index $i$ runs over all final state particles (the top quark, the charged Higgs boson and 
possibly the extra $b$ quark and/or light parton) of the hard process. 
For vanishing transverse momenta of the
external particles, our scale choice 
corresponds to the factorisation scale set in the 4FS calculation of 
Refs.~\cite{Flechl:2014wfa,Dittmaier:2009np}. In the following, scale uncertainties 
are obtained by varying $\mu_F$ and $\mu_R$ independently by a factor of two around 
their central values, given in Eq.~\eqref{eq:muFmuR}.
We have checked that, particularly for our reference 
4FS NLO+PS prediction, the dependence of the distributions on the shower scale 
$\mu_{\text{sh}}$, when varied by a factor in the range $[1/\sqrt 2,\sqrt 2]$, 
is rather mild and significantly smaller 
than uncertainties associated with the renormalisation and factorisation scales; we therefore refrain from including uncertainties associated with $\mu_{\text{sh}}$ in what follows. 
Furthermore, we will not discuss any PDF systematics.\footnote{Note
that scale variations due to $\mu_F$ and $\mu_R$ as well as PDF uncertainties 
are computed at no extra CPU cost using the reweighting procedure of Ref.~\cite{Frederix:2011ss}.}

Jets are reconstructed via the anti-$k_T$ 
algorithm~\cite{Cacciari:2008gp}, as implemented in {\sc FastJet}~\cite{Cacciari:2011ma, Cacciari:2005hq}, with a distance parameter $\Delta R =0.4$ and subject to 
the conditions 
\begin{equation}
    p_T(j)\ge25\, \textrm{GeV}, \qquad |\eta(j)|\le 2.5.
\end{equation} 
For fixed-order computations jets are clustered from partonic final states, while in simulations matched 
to parton showers jets are made up of hadrons; $b$ jets are defined to contain 
at least one $b$ quark (at fixed order) or $B$ hadron (in matched simulations).

In our simulations we keep the charged Higgs boson stable, while we decay the top quark 
leptonically (although the leptons from the decay will not affect any observable 
we consider) in order to keep as much control as possible on the origin of QCD radiation. 
The task to decay the top quark is performed by the parton shower for (N)LO+PS runs,
while at fixed order we simulate the decay $t\to bW$ in an isotropic way (in the $t$ rest frame) at the analysis level.\footnote{
    Such an approach neglects spin-correlation in the decay of the top quark. However, 
    within the \amc\ framework, spin correlation can be included 
    in (N)LO+PS runs by decaying the top quark with {\sc MadSpin}~\cite{Artoisenet:2012st}.} 
No simulation of the underlying event is performed by the parton shower. 

Let us conclude this section by addressing one further point, which is 
crucial when processes with final-state $b$ quarks are matched to
parton showers: the choice of the shower starting scale $\mu_{\text{sh}}$. 
Such processes are known to prefer much lower values of the renormalisation and factorisation 
scales than the one naively identified as the hard scale of the process ($\hat s$). 
In fact, the shower starting scale and the factorisation scale emerge both from the same concept, 
namely the 
separation of soft and hard physics. Furthermore, it has been argued in Ref.~\cite{Wiesemann:2014ioa} 
for the associated production of a neutral Higgs boson with bottom quarks that the shower starting scale 
(limiting the hardest emission that the shower can generate) should be set at similar values, i.\,e. 
well below $\hat s$. 
Following the arguments made in Ref.~\cite{Wiesemann:2014ioa}, we 
check their validity in the case of charged Higgs boson production.
We shall stress at this point that the following discussion applies both 
to our reference scenarios with $m_{H^-}=200\gev$ and 
$m_{H^-}=600\gev$, although we refrain from showing 
explicit results for the latter.

%%%%%%% MATCHING PLOT FOR 4FS FRAC 4 VS FRAC 1
\begin{figure}[t!]
\centering
\includegraphics[width=0.48\textwidth, clip=true, trim=0.5cm 4cm 0.7cm 1cm]{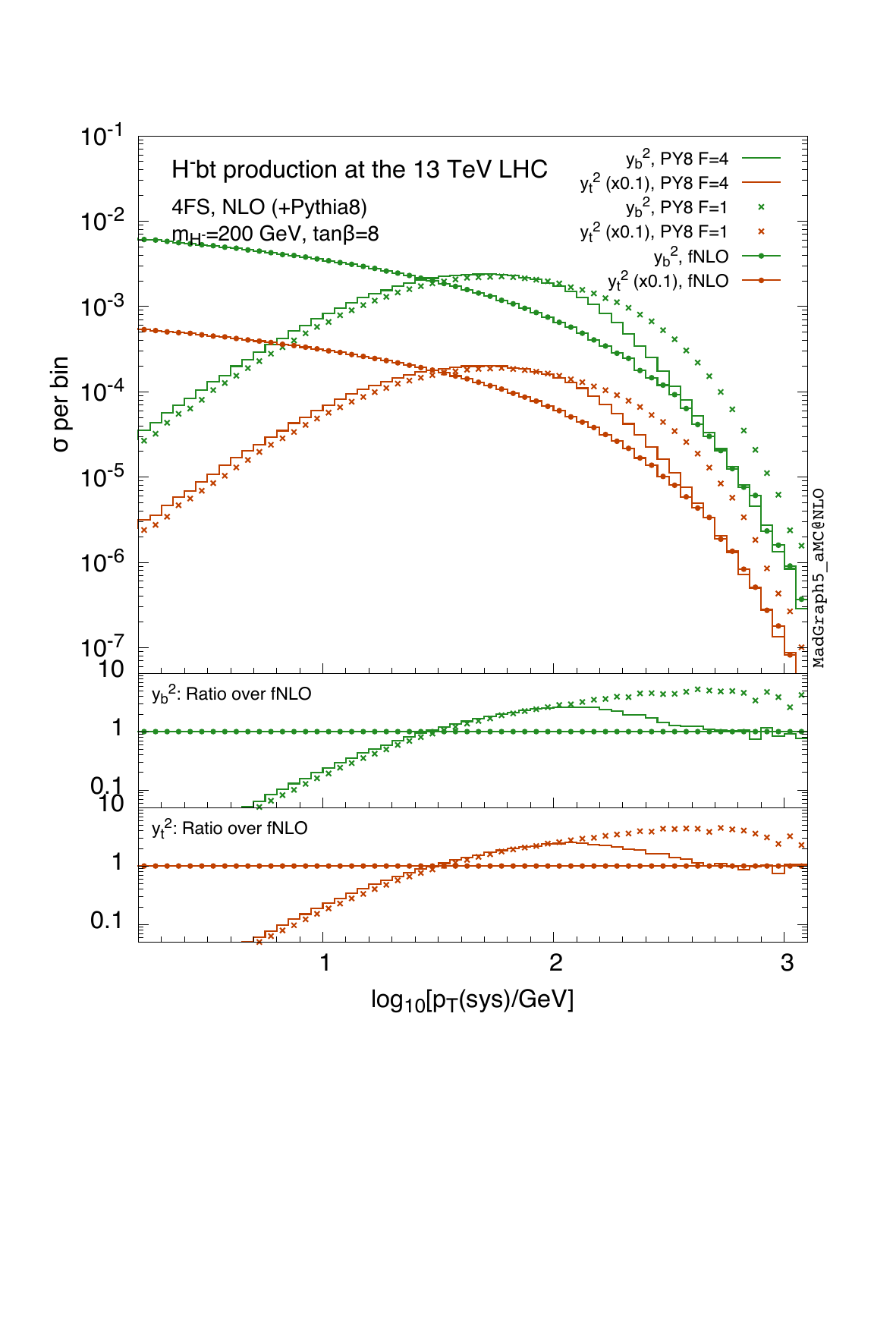}
\includegraphics[width=0.48\textwidth, clip=true, trim=0.5cm 4cm 0.7cm 1cm]{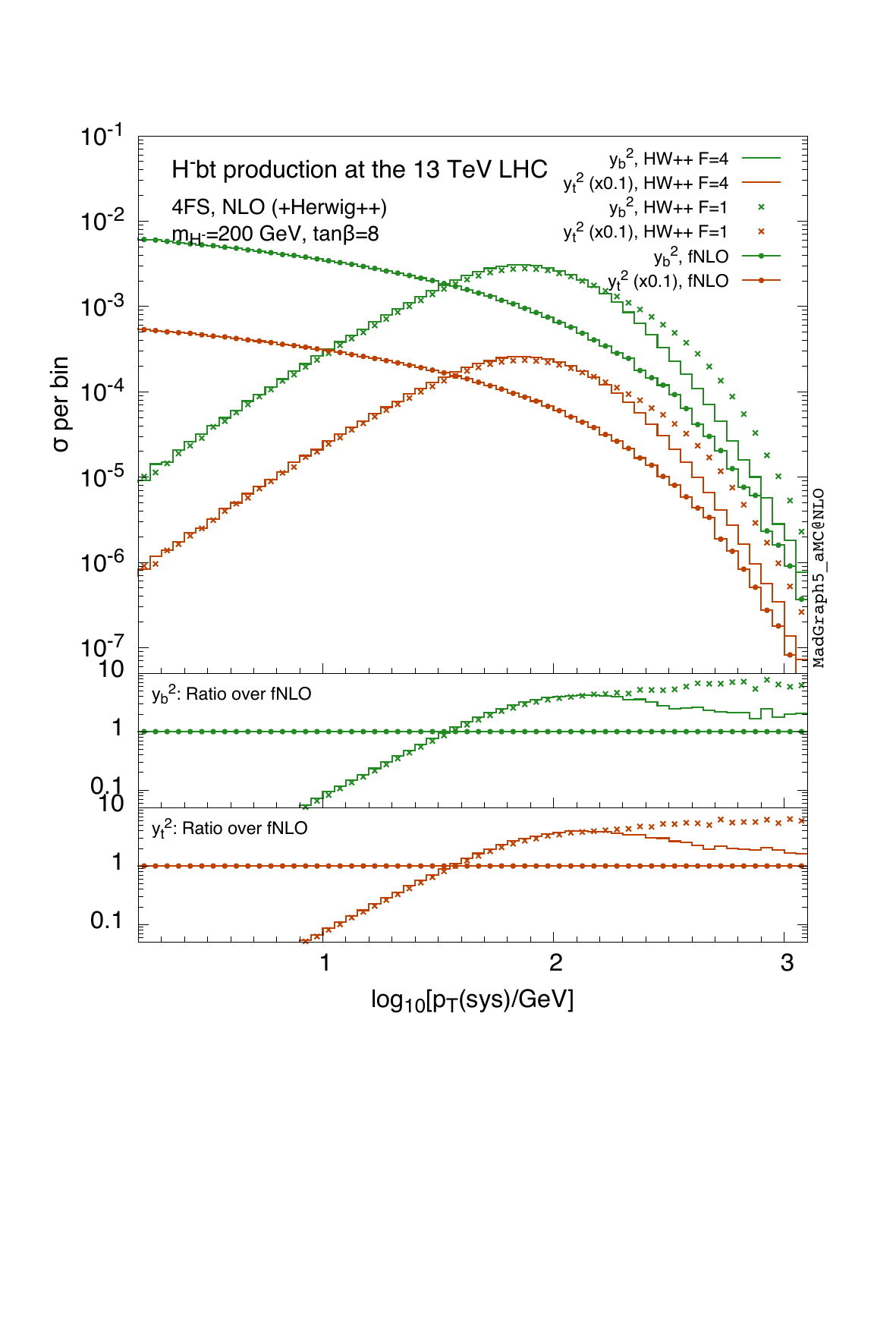}
\caption{\label{fig:logptsys} Transverse momentum of the \hbt{} system for $m_{H^-}=200\gev{}$ in the 4FS at fNLO
(green dotted-solid for the $y_b^2$ term, orange dotted-solid for the $y_t^2$ term), and at NLO+PS with $F$= 1 
(green dots for the $y_b^2$ term, orange dots for the $y_t^2$ term) and $F$= 4 (green solid for the $y_b^2$ term, 
orange solid for the $y_t^2$ term). We show predictions matched with {\sc Pythia8}\ (left) and {\sc Herwig++}\ (right). The insets show the ratio of the curves in the main frame 
over the fNLO prediction, for both the $y_b^2$ and the $y_t^2$ terms.}
\end{figure}

\amc{} assigns a dynamical shower scale chosen from a distribution in the range\footnote{See Ref.~\cite{Alwall:2014hca} for further details.}
\begin{equation}
    \frac{0.1}{F} \hat s \le \mu_{\text{sh}} \le  \frac{1}{F}\hat s,
\end{equation}
where $F$ is a parameter that drives the bounds of the distribution, and whose default value is $F=1$. 
With such a default setting the effective value of $\mu_{\text{sh}}$, namely the maximum of the 
$\mu_{\text{sh}}$ distribution (which for simplicity we will refer to as just $\mu_{\text{sh}}$ 
in the following), is indeed much larger than 
$\mu_{F,R}$. Furthermore, considering the transverse momentum distribution of the Born-level ``system" 
($p_T(\textrm{sys})$),\footnote{Note that the Born-level system is unambiguously defined only 
in a fixed-order calculation, being in our case the charged Higgs accompanied by the final state 
top and bottom quark. At NLO+PS we define it to include the hardest $B$ hadron 
(instead of the bottom quark), 
which does not originate from the top decay; in this case, MC-truth is used.} 
which is maximally sensitive to the interplay between the fixed-order prediction and the shower, 
the NLO+PS distribution (in particular in the 4FS) does not match the fixed-order NLO (fNLO) one at large $p_T$ 
for $F=1$. This can be deduced from Fig.~\ref{fig:logptsys}, when comparing the crosses (NLO+PS for $F=1$)
to the solid curves overlayed with points (fNLO). 
On the contrary, we observe a clearly improved high-$p_T$
matching of the NLO+PS results to the fixed-order ones 
by choosing a reduced shower scale corresponding to $F=4$ (solid curves).\footnote{Our focus here is on the 
4FS prediction. However, similar conclusions, if less stringent, can be drawn 
from the corresponding plots in the 5FS.}
Indeed, such a choice brings $\mu_{\text{sh}}$ much closer to the value of the renormalisation and factorisation scales. We have also checked that the agreement among {\sc Pythia8} and {\sc Herwig++}
improves (although often only marginally) when differential observables in the 4FS are computed with $F=4$.

In conclusion, although for this process we do not reproduce all results of Ref.~\cite{Wiesemann:2014ioa} with the same significance, we still find sufficient evidence that $F=4$ is favourable in many respects 
and make it our default choice. 
In Sect.~\ref{sec:4vs5FS}, we shall further study the impact of this choice when comparing the 
4FS and 5FS results: by setting $F=4$ an improved agreement 
between the two schemes at the level of shapes is observed.

%%%%%%%%%%%%%%%%%%%%%%%%%%%%%%%%%%%%%%%%%%%%%%%%%%%%%%%%%%%%%%%%%%%%%5
\subsection{Four-flavour scheme results}
We now turn to our phenomenological results for charged Higgs boson production. 
Let us first consider state-of-the-art 4FS predictions, which, as will be shown, constitute 
the most reliable differential results for observables exclusive in the degrees 
of freedom of final-state bottom quarks. We split this section 
into two parts: in Sect.~\ref{sec:4FSresults} we limit our study to the dominant $y_b^2$ and $y_t^2$ 
contributions, while the $y_b\,y_t$ contribution is considered in Sect.~\ref{sec:ybyt}.

%%%%%%%%%%%%%%%%%%%%%%%%%%%%%%%%%%%%%%%%%%%%%%%%%%%%%%%%%%%%%%%%%%%%%5
\subsubsection{$y_b^2$ and $y_t^2$ contributions at NLO+PS}
\label{sec:4FSresults}
%%%%%%%%%%%%%%%%%%%%%%%%%%%%%%%%%%%%%%%%%%%%%%%%%%%%%%%%%%%%%%%%%%%
\begin{table}[h]%[ph]
\begin{center}
\begin{tabular}{cl|cccc}
\toprule
\multicolumn{2}{c|}{\multirow{2}{*}{ $\sigma(\mH=200\gev)$ [fb]}} & 
\multicolumn{2}{c}{NLO} & \multicolumn{2}{c}{LO} \\
& & $y_b^2$ & $y_t^2$ & $y_b^2$ & $y_t^2$ \\
\midrule
\multicolumn{2}{c|}{Inclusive} & 
$50.40 ^{+17.8\perc} _{-18.6\perc}$ &
$42.43 ^{+12.4\perc} _{-13.1\perc}$ & 
$42.12 ^{+52.2\perc} _{-31.9\perc}$ &
$28.68 ^{+36.3\perc} _{-24.7\perc}$ \\[7pt]
\multirow{3}{*}{$\ge 1j_b$} &
F.O.\ &
$45.47 ^{+17.5\perc} _{-18.4\perc}$ &
$38.31 ^{+12.2\perc} _{-13.0\perc}$ &
$38.26 ^{+51.9\perc} _{-31.8\perc}$ &
$26.09 ^{+36.1\perc} _{-24.6\perc}$ \\
 & 
Pythia8\ &
$43.44 ^{+17.4\perc} _{-18.4\perc}$ &
$36.67 ^{+12.0\perc} _{-13.0\perc}$ &
$36.81 ^{+52.0\perc} _{-31.8\perc}$ &
$25.09 ^{+36.1\perc} _{-24.7\perc}$ \\
 & 
Herwig++\ &
$42.64$ &
$36.04$ &
$36.08$ &
$24.61$ \\[7pt]
\multirow{3}{*}{$\ge 2j_b$} &
F.O.\ &
$11.55 ^{+10.9\perc} _{-15.4\perc}$ &
$ 9.76 ^{+ 6.5\perc} _{-10.0\perc}$ &
$11.22 ^{+50.4\perc} _{-31.2\perc}$ &
$ 7.79 ^{+35.0\perc} _{-24.1\perc}$ \\
 & 
Pythia8\ &
$12.55 ^{+15.3\perc} _{-17.4\perc}$ &
$10.67 ^{+10.4\perc} _{-12.1\perc}$ &
$11.73 ^{+51.2\perc} _{-31.5\perc}$ &
$ 8.12 ^{+35.6\perc} _{-24.4\perc}$ \\
 &
Herwig++\ &
$11.03$ &
$ 9.33$ &
$10.09$ &
$ 7.00$ \\
\bottomrule
\end{tabular}\\[10pt]

\begin{tabular}{cl|cccc}
\toprule
\multicolumn{2}{c|}{\multirow{2}{*}{ $\sigma(\mH=600\gev)$ [fb]}} & 
\multicolumn{2}{c}{NLO} & \multicolumn{2}{c}{LO} \\
& & $y_b^2$ & $y_t^2$ & $y_b^2$ & $y_t^2$ \\
\midrule
\multicolumn{2}{c|}{Inclusive} & 
$2.400 ^{+20.3\perc} _{-20.1\perc}$ &
$2.117 ^{+13.1\perc} _{-14.2\perc}$ & 
$1.794 ^{+54.9\perc} _{-33.0\perc}$ &
$1.339 ^{+40.1\perc} _{-26.5\perc}$ \\[7pt]
\multirow{3}{*}{$\ge 1j_b$} &
F.O.\ &
$2.187 ^{+19.9\perc} _{-19.9\perc}$ &
$1.925 ^{+12.6\perc} _{-14.0\perc}$ &
$1.649 ^{+54.7\perc} _{-32.9\perc}$ &
$1.232 ^{+39.9\perc} _{-26.5\perc}$ \\
 & 
{\sc Pythia8}\ &
$2.115 ^{+19.9\perc} _{-19.9\perc}$ &
$1.865 ^{+12.5\perc} _{-14.0\perc}$ &
$1.601 ^{+54.8\perc} _{-32.9\perc}$ &
$1.197 ^{+40.0\perc} _{-26.5\perc}$ \\
 & 
{\sc Herwig++}\ &
$2.077$ &
$1.836$ &
$1.570$ &
$1.175$ \\[7pt]
\multirow{3}{*}{$\ge 2j_b$} &
F.O.\ &
$0.630 ^{+12.6\perc} _{-17.0\perc}$ &
$0.548 ^{+ 5.9\perc} _{-10.8\perc}$ &
$0.548 ^{+53.8\perc} _{-32.6\perc}$ &
$0.413 ^{+39.2\perc} _{-26.2\perc}$ \\
 & 
{\sc Pythia8}\ &
$0.697 ^{+16.7\perc} _{-18.6\perc}$ &
$0.611 ^{+ 9.6\perc} _{-12.6\perc}$ &
$0.588 ^{+54.3\perc} _{-32.8\perc}$ &
$0.443 ^{+39.6\perc} _{-26.3\perc}$ \\
 &
{\sc Herwig++}\ &
$0.602$ &
$0.532$ &
$0.498$ &
$0.376$ \\
\bottomrule
\end{tabular}
\end{center}
\caption{\label{tab:rates}
4FS predictions for total rates (in fb) for $\tan\beta=8$.}
\end{table}

We begin our analysis by studying total rates for the production of charged Higgs 
bosons with a mass of $200\gev$ and $600\gev$ in Table~\ref{tab:rates}.
We consider three possibilities: the fully inclusive case, the case in which we
require at least one $b$ jet, and the one in which two or more $b$ jets are tagged. 
All results are given at both LO and NLO accuracy. The cross sections 
in which one or two $b$ jets
are required depend on the approximation and Monte Carlo under 
consideration. We thus report separately results obtained at
fixed order, with {\sc Pythia8} and with {\sc Herwig++}.
Any quoted uncertainty is due to scale variation, 
evaluated as detailed in Sect.~\ref{sec:settings}; they are indicated only 
at fixed order and for results matched with {\sc Pythia8}, 
since they show little dependence on the specific
Monte Carlo. Results for $y_b^2$ and $y_t^2$ terms are presented 
separately. Let us summarize the conclusions to be drawn from Table~\ref{tab:rates} as follows:
\begin{itemize}
\item The scale uncertainty of NLO predictions is substantially smaller than that of the LO ones; 
at NLO the scale uncertainty is larger for the $y_b^2$ than for the $y_t^2$ contribution 
($\sim 15$-$20\%$ and $\sim 10$-$15\%$, respectively), due to the different renormalisation 
schemes used for the bottom and top Yukawa couplings.
\item Because of our default choice of $\tan\beta=8$, $y_b^2$ and $y_t^2$ predictions 
are of similar size at NLO (only $\sim 15\%$ different); the difference
is larger at LO ($\sim 30$\%). 
As a consequence, the $K$-factors are generally different
between the $y_b^2$ and $y_t^2$ terms; for $\mH=200\gev$, the inclusive $y_b^2$ $K$-factor is close to 1.2, 
while for the $y_t^2$ term the NLO 
corrections have a larger impact, with $K\approx 1.5$;
for $\mH=600\gev$, NLO corrections are larger in both cases: 
$K\approx 1.3$ and $K\approx 1.6$ for the $y_b^2$ and $y_t^2$ terms respectively. 
The smaller $K$-factor for the $y_b^2$ term is a consequence of the fact that,
by renormalising the bottom Yukawa in the $\overline{\text{MS}}$ scheme, the LO 
predictions already include a class of higher order corrections.
\item {\sc Pythia8} and {\sc Herwig++} predictions for cross sections with $b$-jet requirements are consistent 
with each other and well within quoted uncertainties; this holds true both at the LO and at the NLO.  
Even the agreement with the fixed-order results is rather good overall and largely within uncertainties. 
Higher $b$-jet multiplicities tend to deteriorate the agreement, although only to a small extent.
\item The effect of the cuts on the number of $b$ jets is quite moderate: given that there 
    is generally a hard $b$ jet from the top quark decay, the inclusive rate is reduced by only $\sim 10\%$,
when at least one $b$ jet is required. On the other hand, requiring a second $b$ jet reduces the cross section 
by a factor of 4-5. Scale uncertainties slightly decrease at NLO for cross sections within cuts.
\end{itemize}

\noindent 
We now turn to differential observables in the 4FS.
We only discuss results at fixed order and matched with {\sc Pythia8}. 
Any differences to the matched {\sc Herwig++} predictions
will be explicitly mentioned below.  
Besides, the Monte Carlo dependence will be investigated in more 
detail when comparing distributions in the two flavour schemes in Sect.~\ref{sec:4vs5FS}.

\begin{figure}[p!]
\centering
\includegraphics[width=0.48\textwidth, clip=true, trim=0.5cm 2.5cm 0.7cm 1cm]{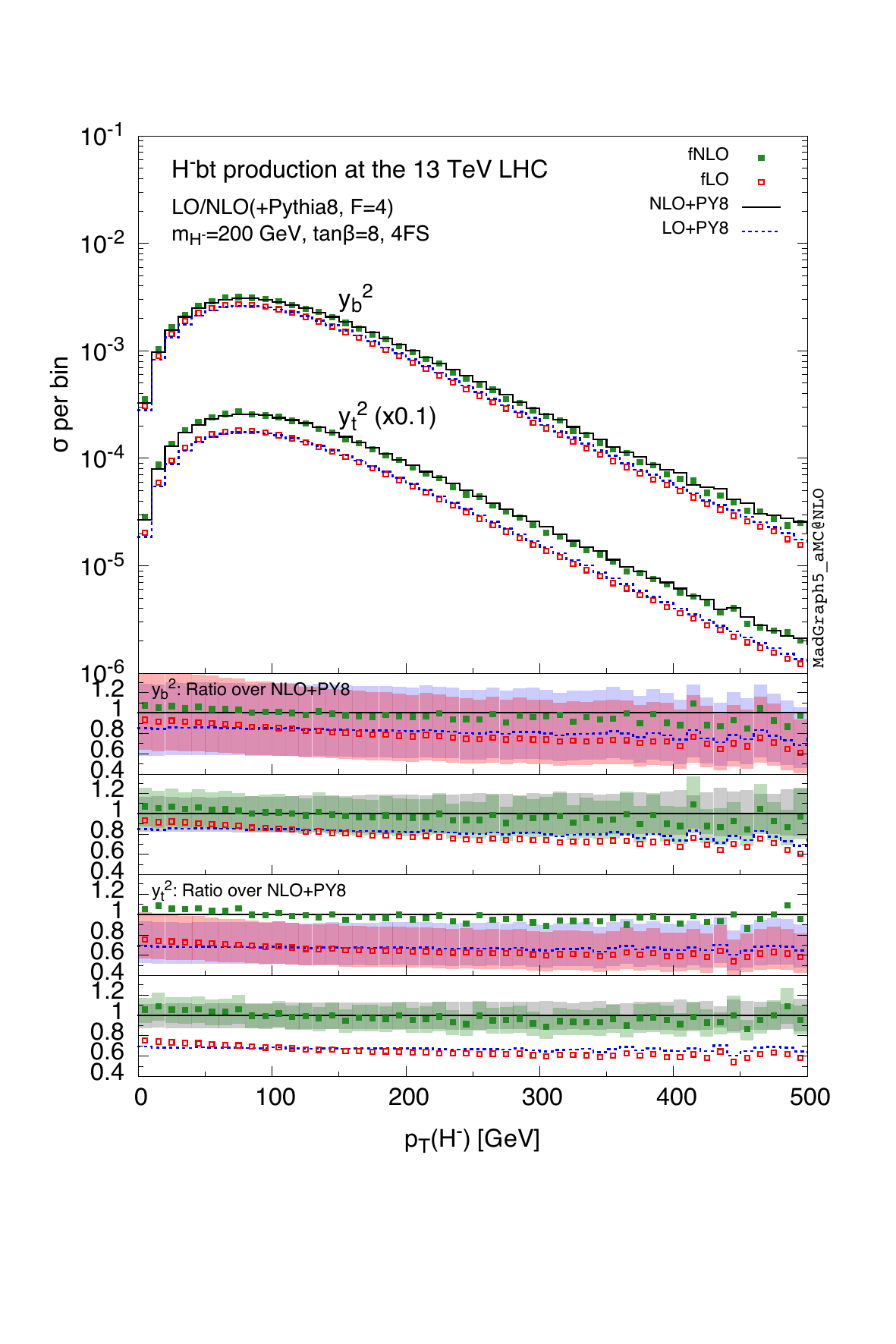}
\includegraphics[width=0.48\textwidth, clip=true, trim=0.5cm 2.5cm 0.7cm 1cm]{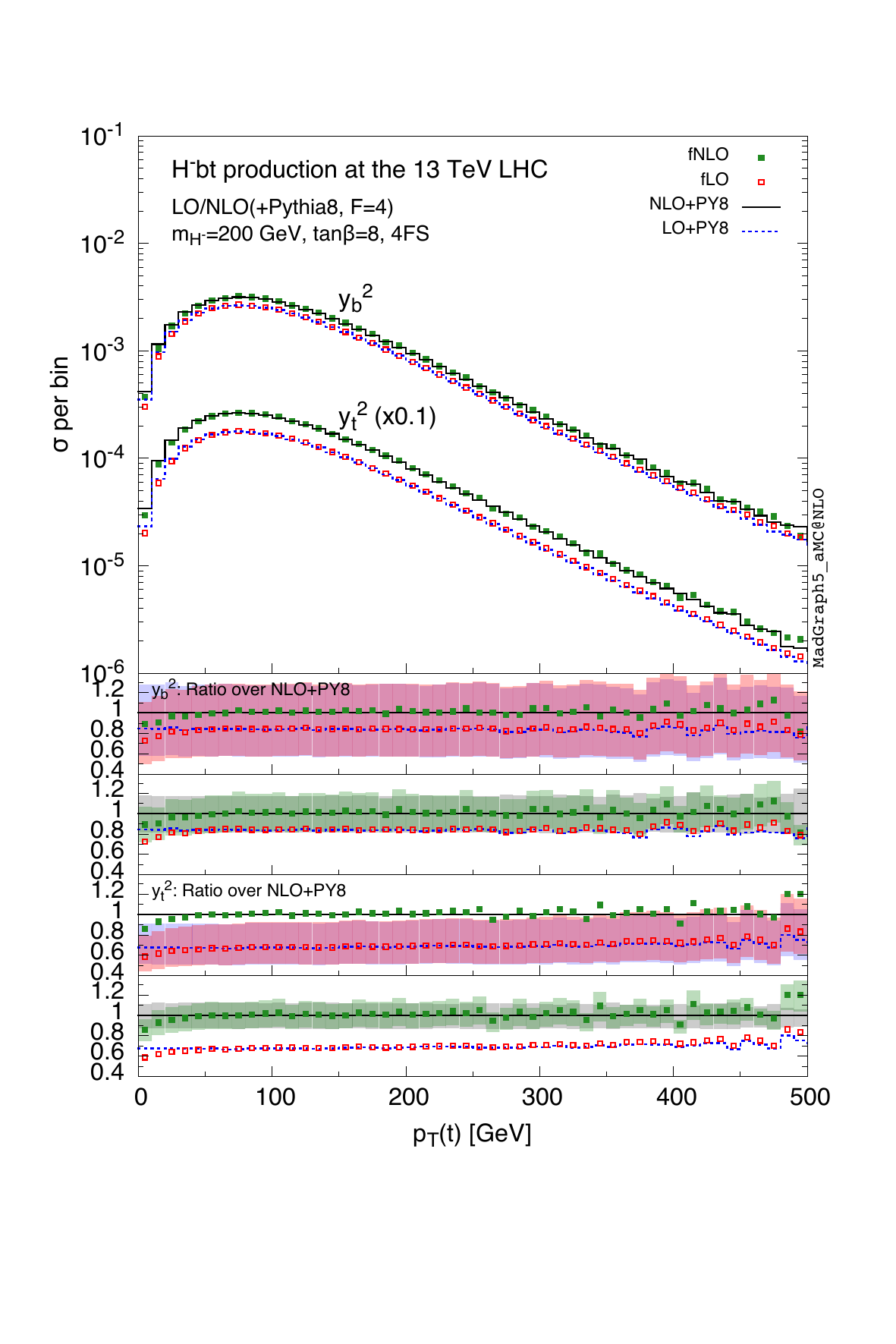}\\
\includegraphics[width=0.48\textwidth, clip=true, trim=0.5cm 2.5cm 0.7cm 1cm]{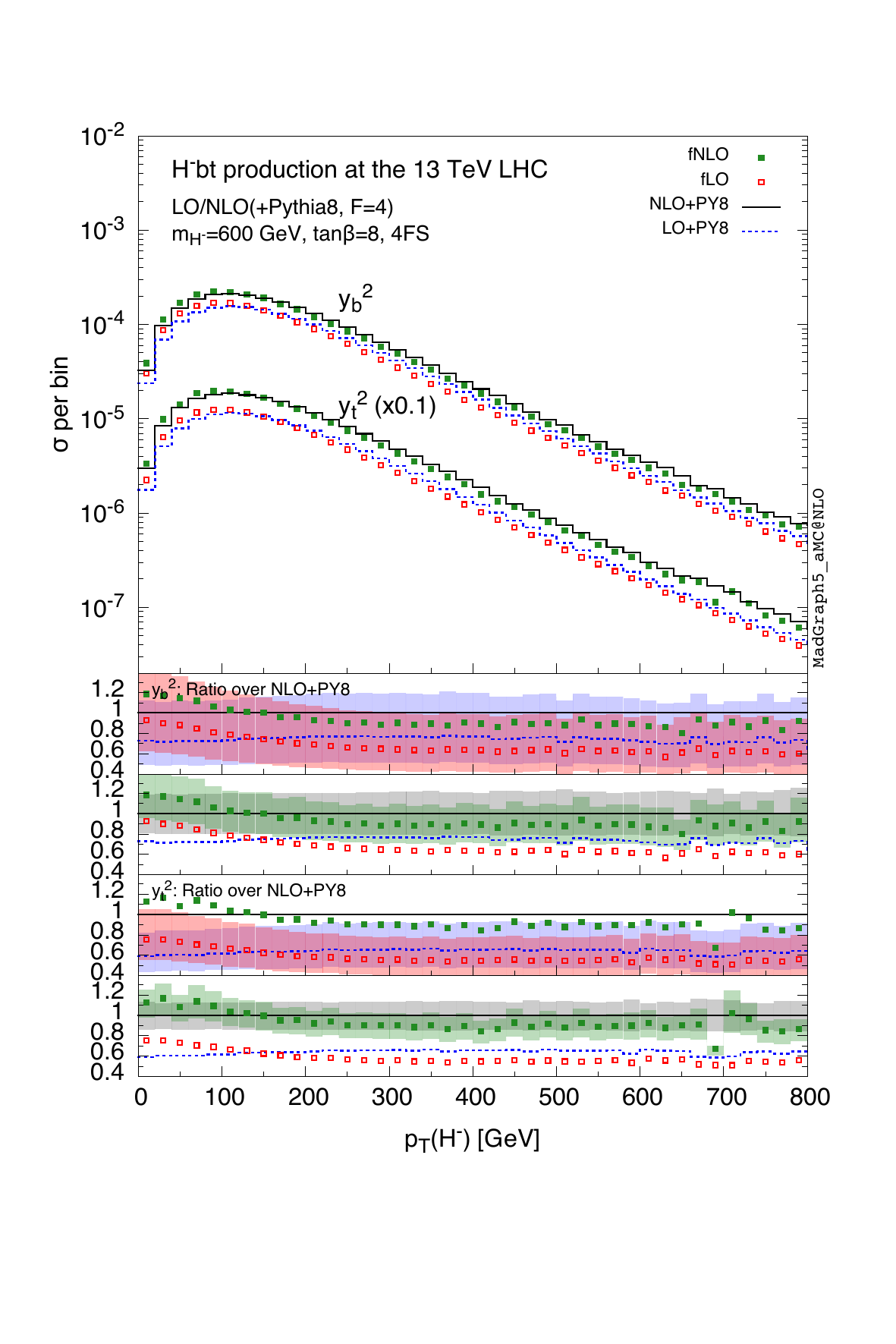}
\includegraphics[width=0.48\textwidth, clip=true, trim=0.5cm 2.5cm 0.7cm 1cm]{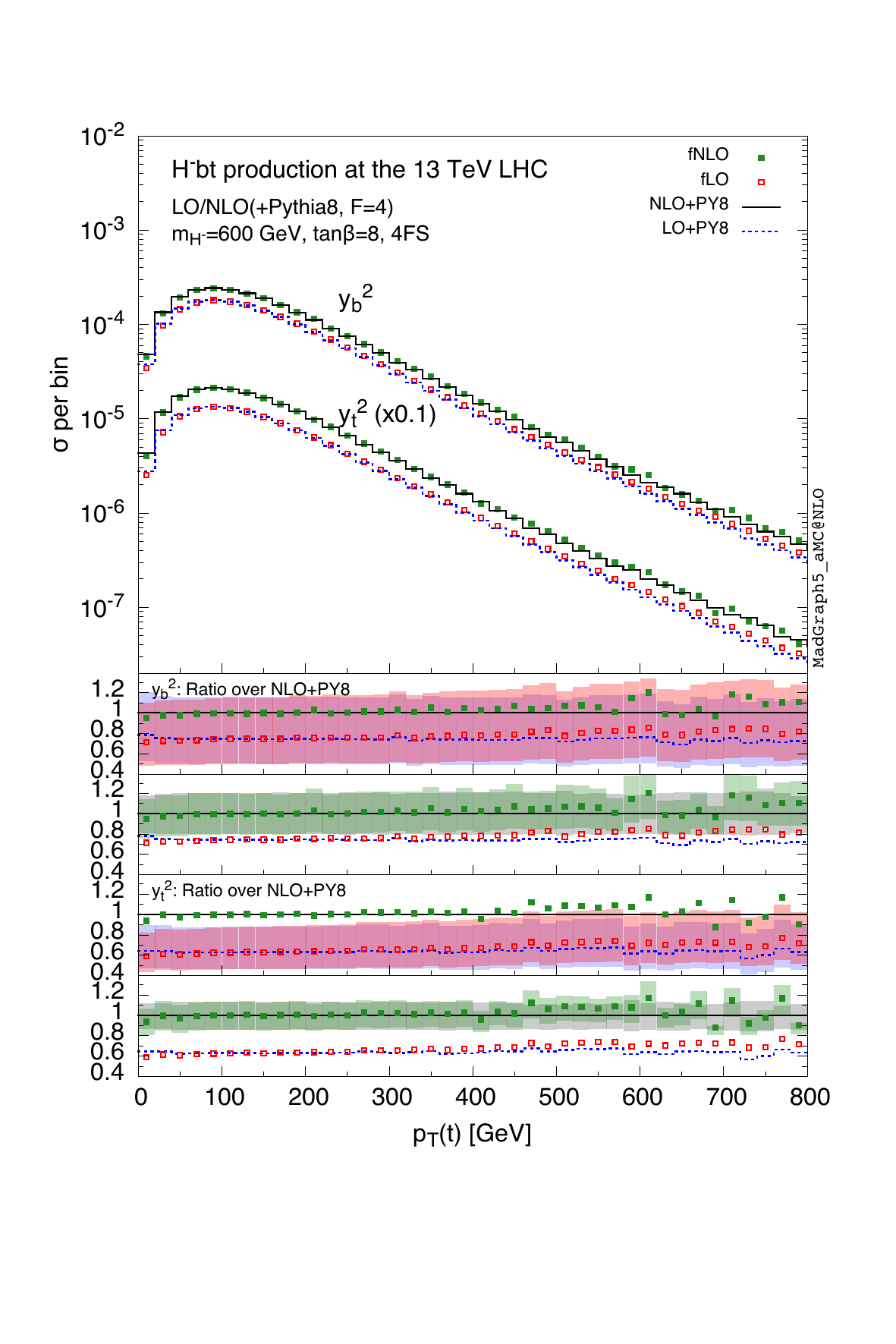}
\caption{\label{fig:4fpth} Transverse momentum distributions of the charged Higgs (left panels) and the top quark (right panels) for
$m_{H^-}=200\gev$ (upper panels) and $m_{H^-}=600\gev$ (lower panels); 
predictions at fixed order (LO red empty boxes, NLO green full boxes) as well as matched to {\sc Pythia8}
(LO blue dashes, NLO black solid)  are shown in the main frame, for the $y_b^2$ and $y_t^2$ terms separately. 
The first and second (third and fourth)
insets show the ratios of the curves in the main frame over the matched NLO prediction, 
for the $y_b^2$ ($y_t^2$) term respectively, together 
with the LO and NLO uncertainty bands.}
\end{figure}

The figures throughout this section are organised according to the following pattern: 
the main frame displays the relevant predictions in absolute size as cross section per 
bin\footnote{The sum of all bins is equal to the total rate, possibly within cuts.} 
for the $y_b^2$ and $y_t^2$ terms. For the sake of readability, they have been 
separated by reducing the $y_t^2$ curve by a factor 
of ten. The fixed-order results are shown using 
boxes, while simulations matched to parton showers are displayed as lines: filled green boxes for 
fNLO, open red boxes for fLO, a black solid line for NLO+PS and a blue dotted 
curve for LO+PS. In the first two insets we display the bin-by-bin ratio of all 
$y_b^2$ histograms in the main frame normalised to the corresponding NLO+PS result. 
The difference between the two insets are the uncertainty bands, which in the first one
are displayed for the LO predictions, while in the second for the NLO 
ones. The third and fourth insets are identical to first two, but for the 
$y_t^2$ contribution.

Let us start in Fig.~\ref{fig:4fpth} with the transverse momentum spectrum of a $200\gev$ 
Higgs (top left panel) and the associated top quark (top right panel). The two plots 
are in fact quite similar and can be discussed simultaneously: the agreement between 
the fixed-order and the PS-matched simulations is close-to-excellent in all cases. Matching to the PS 
has only minor effects on the shape of the distributions, notable only at small transverse momenta.
In other words, the resummation effects of the shower are extremely small, as can 
be expected from observables that are practically not affected by any large 
Sudakov logarithm. The NLO corrections mostly affect the normalisation and reduce 
theoretical uncertainties, reflecting
the numbers quoted in Tab.~\ref{tab:rates}.
We stress here two general features regarding the relation 
between the $y_b^2$ and $y_t^2$ terms which shall hold true for all subsequent 
observables under consideration: 
first, the $\overline{\text{MS}}$ renormalisation makes the 
uncertainty associated with the $y_b^2$ curves  larger, due to  
the variation of $\mu_R$ in $y_b\sim\bar{m}_b(\mu_R)$; second, 
the ratio of the $y_b^2$ and $y_t^2$ contributions at NLO
is generally flat. More details will be given in Sect.~\ref{sec:4vs5FS}.

For a heavier charged Higgs boson with mass $\mH=600\gev$, the lower panels of Fig.~\ref{fig:4fpth}
display a similar behaviour as in the $\mH=200\gev$ case. 
Larger effects due to the PS are visible in the Higgs $p_T$
spectrum at low transverse momenta.

\begin{figure}[t]
\centering
\includegraphics[width=0.47\textwidth, clip=true, trim=0.5cm 2.5cm 0.7cm 1cm]{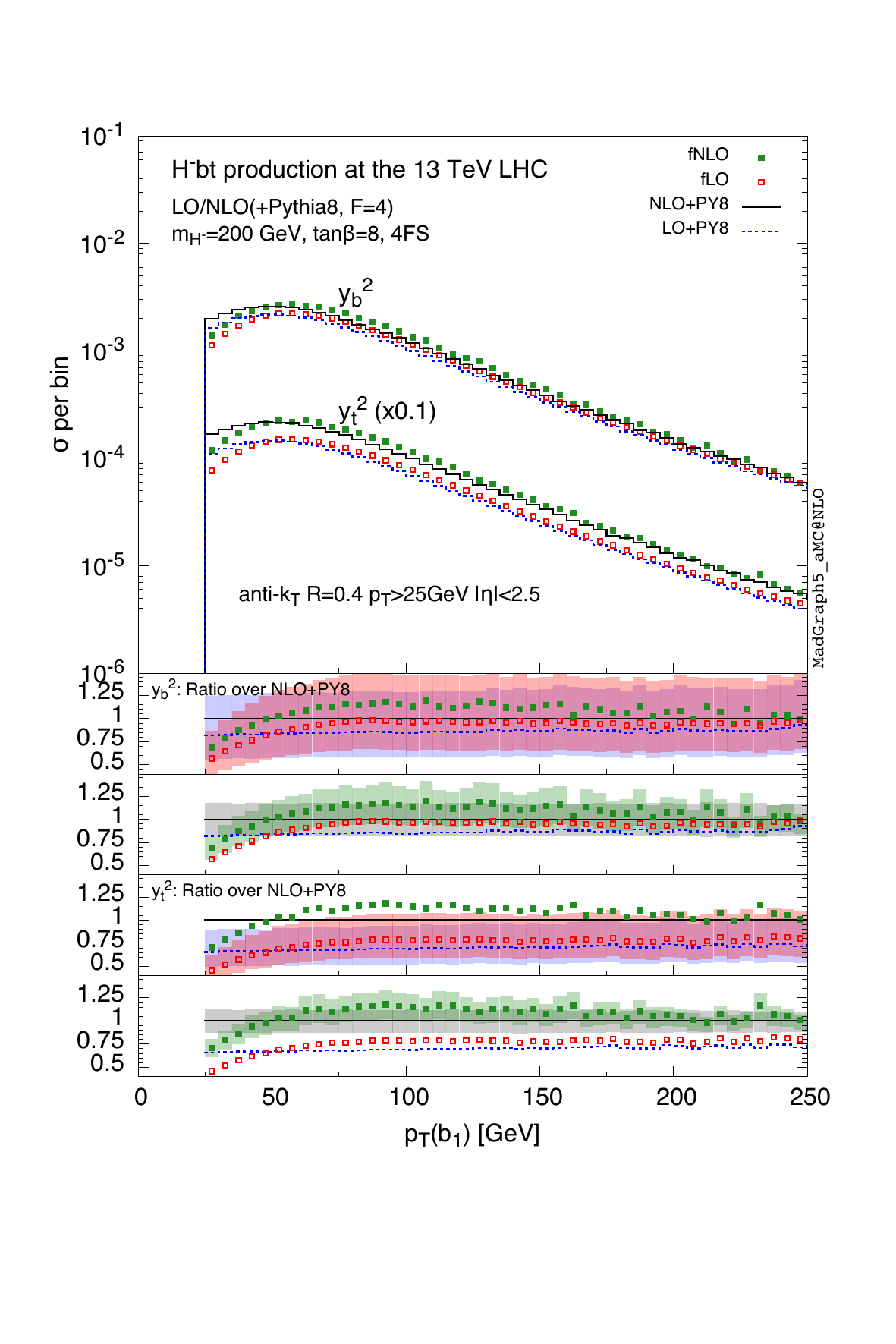}
\includegraphics[width=0.47\textwidth, clip=true, trim=0.5cm 2.5cm 0.7cm 1cm]{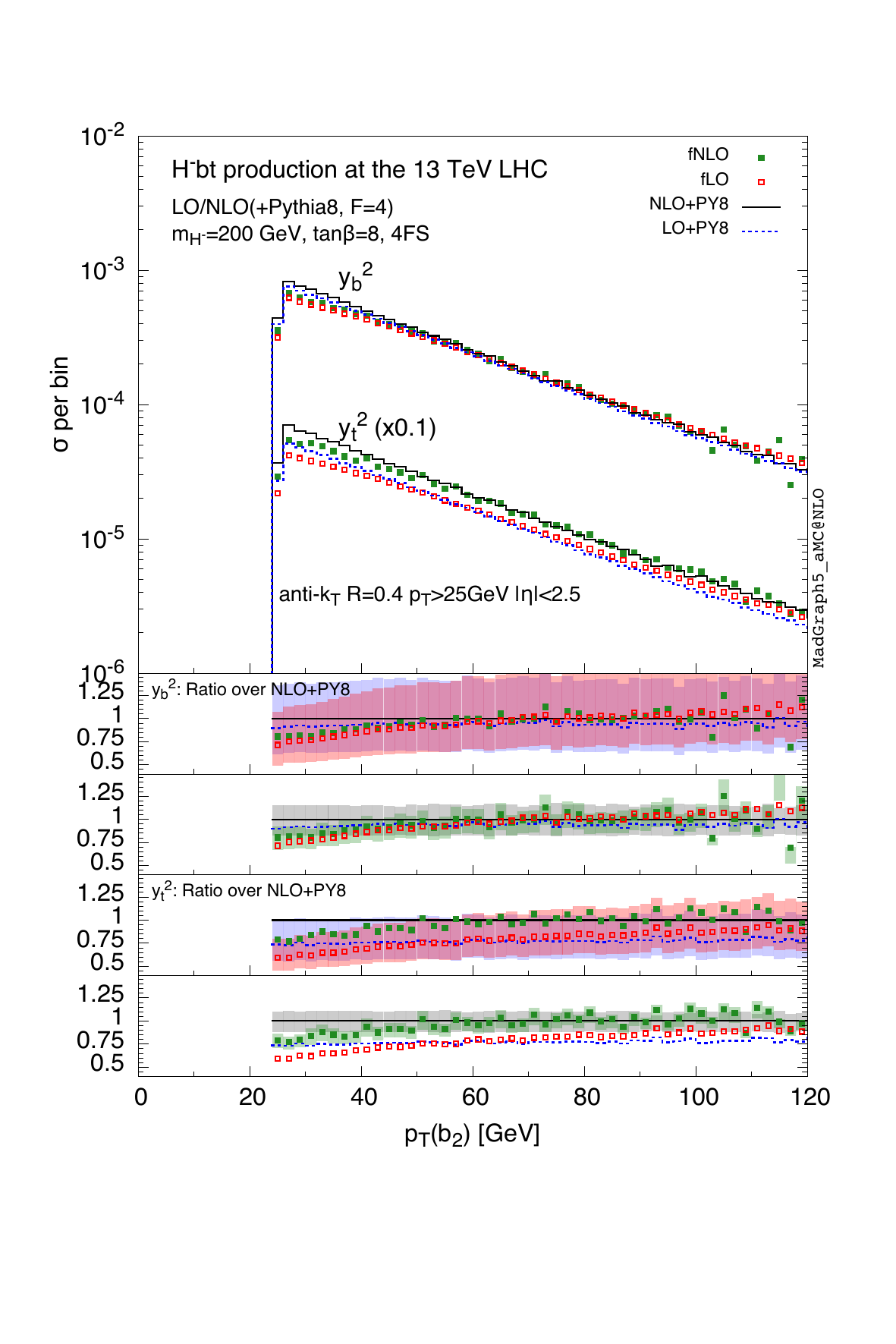}
\caption{\label{fig:4fptbjh1} Same as the upper panel of Fig.~\ref{fig:4fpth}, 
but for the transverse momentum of hardest (left panel) and second-hardest (right panel) $b$ jet.}
\end{figure}

We continue our presentation of the 4FS results with the transverse momentum 
spectra of the hardest ($p_T(b_1)$) and second-hardest ($p_T(b_2)$) $b$ jet in 
Fig.~\ref{fig:4fptbjh1}; see Sect.~\ref{sec:settings} for our jet (and $b$ jet) 
definition. In this case, we limit our discussion to $\mH=200\gev$ 
results, since, apart from a naturally smaller cross section, the relative behaviour 
of the different curves is extremely similar for a heavier charged scalar.
Although the $p_T(b_1)$ and $p_T(b_2)$ distributions in the main frame 
develop a quite different behaviour in terms of their 
shapes, the ratios in the insets exhibit comparable features: 
in all cases, the tail of the spectra 
is driven by the order relevant to the simulation and NLO corrections slightly soften the spectra. 
In other words, the fixed-order 
results agree rather well with the corresponding {\sc Pythia8} ones in the tail. On the other
hand, close to threshold ($p_T=25$\,GeV), where resummation effects are 
enhanced, non-showered and showered results exhibit sizeable differences, in particular for the hardest $b$ jet. 
\begin{figure}[t]
\centering
\includegraphics[width=0.47\textwidth, clip=true, trim=0.5cm 2.5cm 0.7cm 1cm]{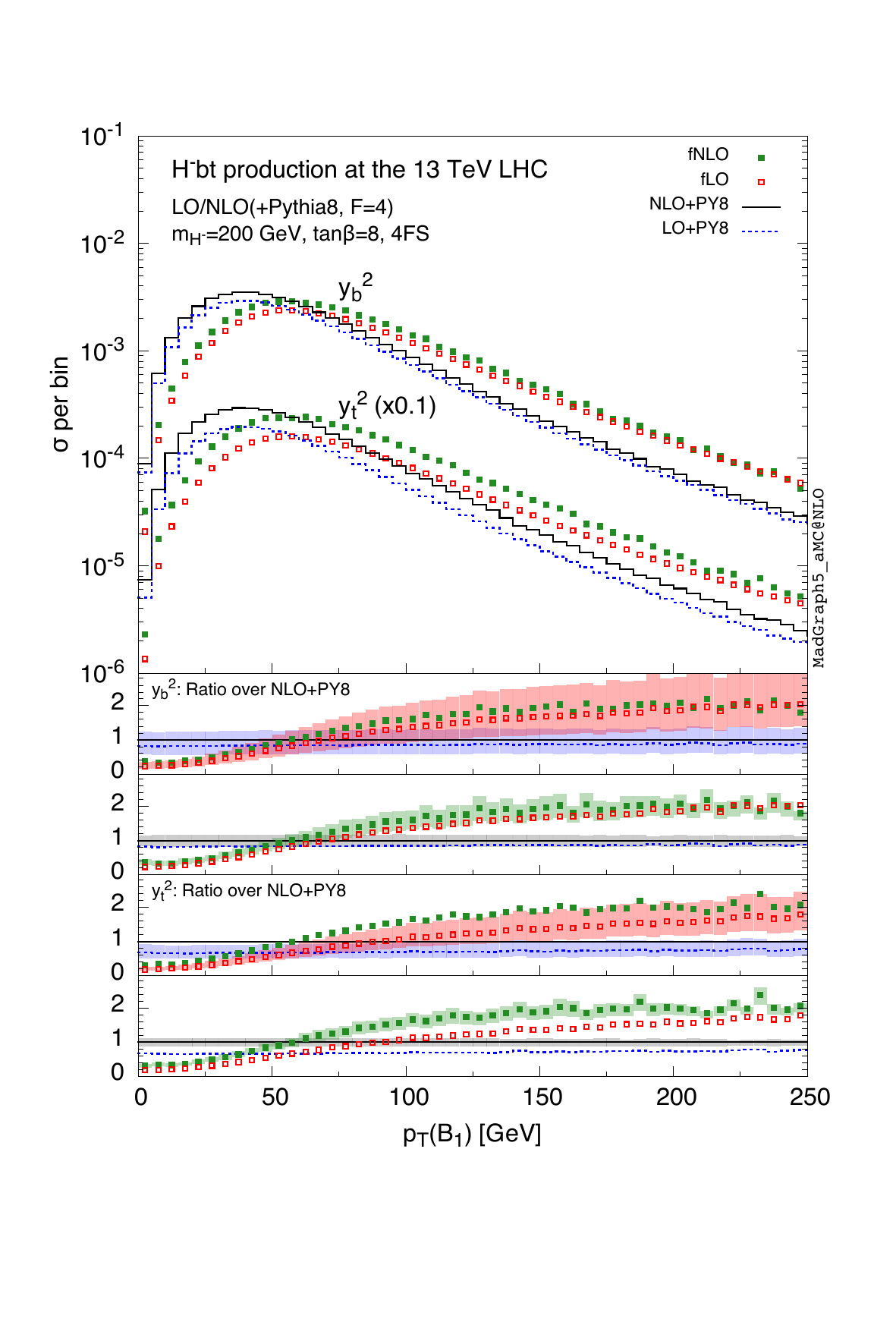}
\includegraphics[width=0.47\textwidth, clip=true, trim=0.5cm 2.5cm 0.7cm 1cm]{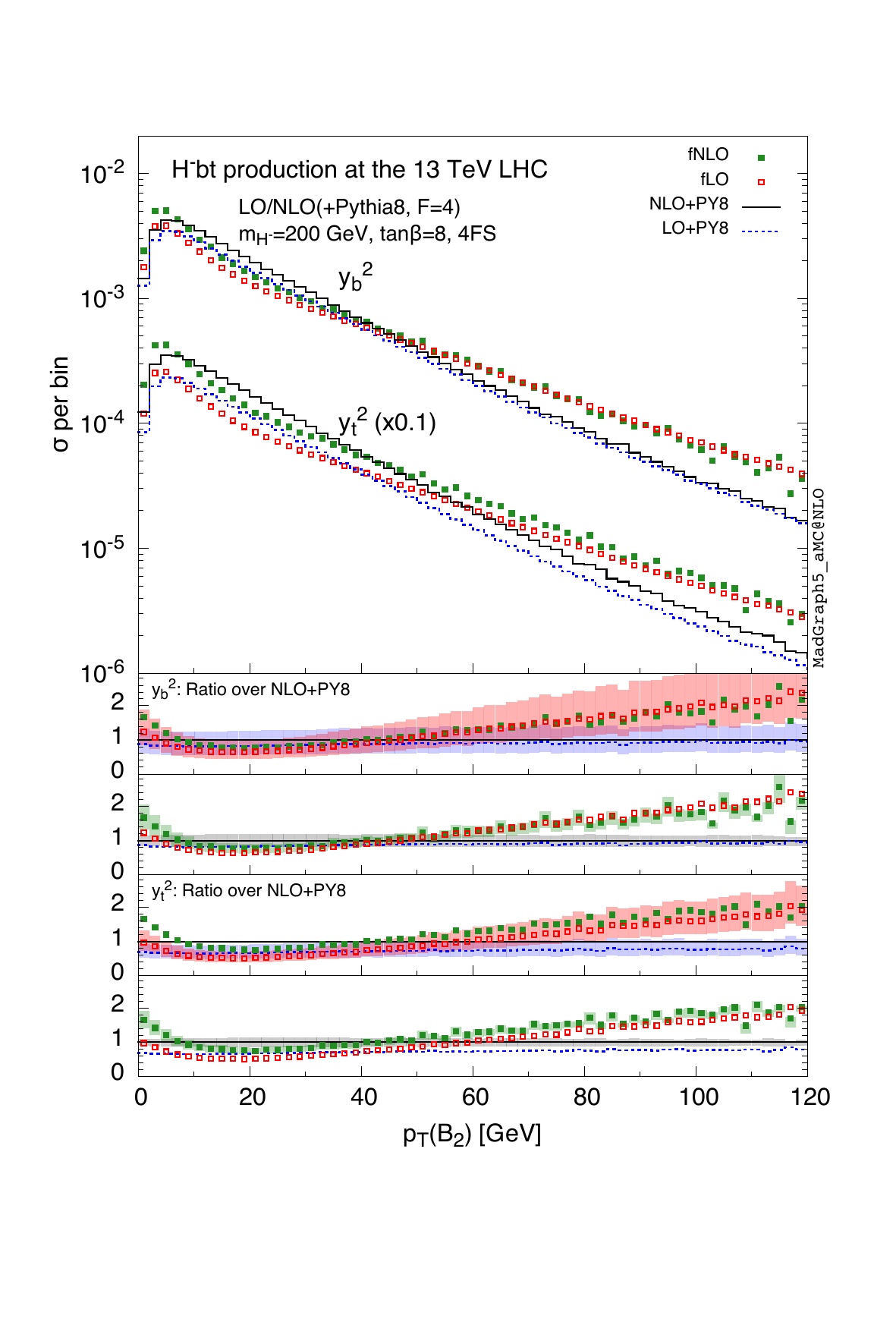}
\caption{\label{fig:4fptbjh2} Same as the upper panel of Fig.~\ref{fig:4fpth},
but for the transverse momentum of hardest (left panel) and second-hardest (right panel) $B$ hadrons.}
\end{figure}

Turning now to somewhat related observables in Fig.~\ref{fig:4fptbjh2}---the transverse momentum 
distributions of the hardest and second-hardest $B$ hadron---, one 
may expect rather similar features to the $b$-jet transverse momentum spectra. 
On the contrary, their pattern is actually very different; the salient feature 
being that showered results are vastly softer than the fixed-order ones and a 
substantial shape distortion due to the matching with parton showers is observed. 
In fact, even the peak of the $p_T(B_1)$ distribution is moved by $\sim 25\gev$ towards the left by the shower.
However, one should bear in mind 
that we compare bottom quarks at parton level for the f(N)LO predictions with $B$ hadrons at 
(N)LO+PS. The observed differences
unavoidably lead to 
the conclusion that fragmentation effects become significant for such exclusive observables. 
Otherwise, the pattern of the {\sc Pythia8} results is very much reminiscent 
of $b$-jet spectra, displaying a slightly harder LO+PS shape than at NLO+PS.
Generally speaking, the relative 
behaviour 
of the $y_b^2$ and $y_t^2$ curves is pretty much alike, including the peculiar 
increase of the f(N)LO cross section towards vanishing $p_T(B_2)$.
Again, we refrain from showing explicit results for a $\mH=600\gev$ charged 
Higgs boson, since the pattern of the various curves turns out to be very similar 
to the $\mH=200\gev$ case; the only difference to be pointed out is a slightly 
reduced gap between showered and fixed-order results for $\mH=600\gev$.

We investigated a vast number of differential observables, the majority of which 
follows the same pattern as illustrated in Figs.~\ref{fig:4fpth} and~\ref{fig:4fptbjh1}: 
the NLO corrections are rather flat and lie within the LO uncertainty bands,  
shower effects are moderate and become more substantial the more exclusive the observable is with respect to the
bottom-quark degrees of freedom.
\begin{figure}[t]
\centering
\includegraphics[width=0.47\textwidth, clip=true, trim=0.5cm 2.5cm 0.7cm 1cm]{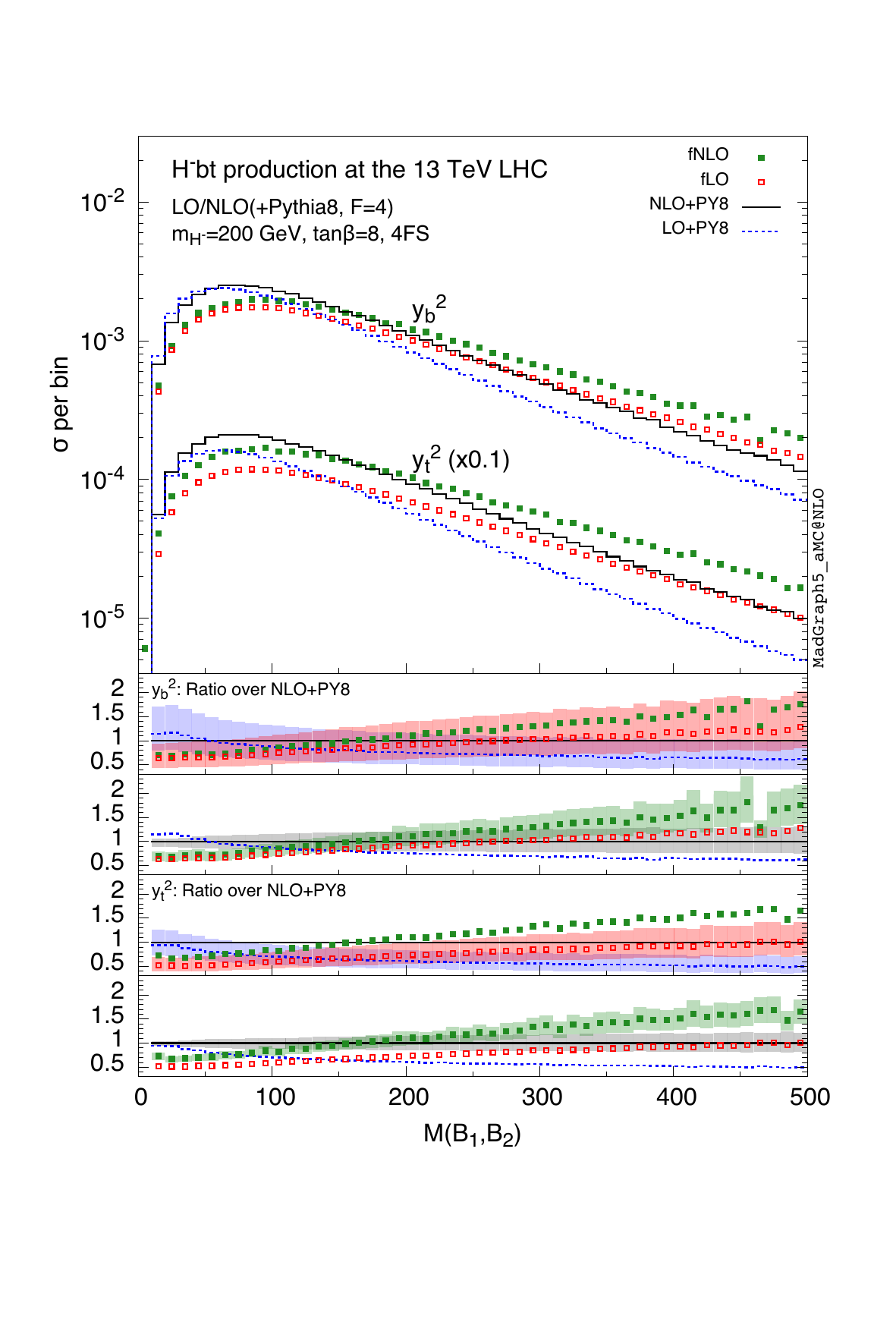}
\includegraphics[width=0.47\textwidth, clip=true, trim=0.5cm 2.5cm 0.7cm 1cm]{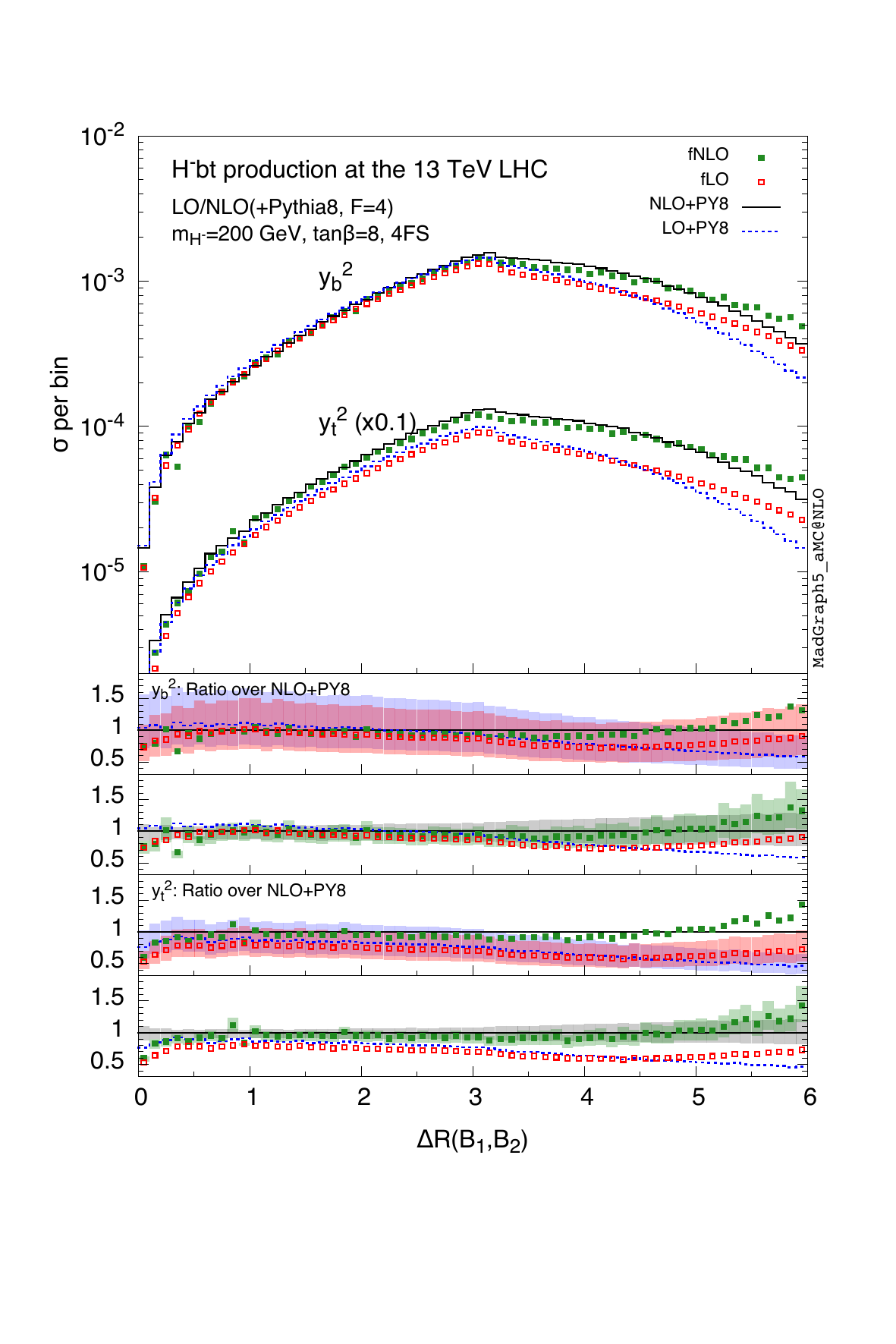}
\caption{\label{fig:4fdrbjh} Same as the upper panel of Fig.~\ref{fig:4fpth}, but for the invariant mass of the two hardest $B$ hadrons (left panel) and their distance in the 
$\eta-\phi$ plane (right panel).}
\end{figure}
Based on our findings, we do not expect larger effects for $b$-jet observables than for distributions relevant to $B$ hadrons. Therefore it is worth to consider the invariant mass $M(B_1,B_2)$ of 
and the distance $\Delta R(B_1,B_2)$ in the $\eta-\phi$ plane 
between the two hardest $B$ hadrons, which are displayed in the left and right panels of Fig.~\ref{fig:4fdrbjh}
respectively. The reader should keep in mind that in the 
fixed-order cases the two hardest $b$ quarks are used instead. 
Since there are no salient differences for larger Higgs masses
we only discuss the $\mH=200\gev$ results.

Although the invariant mass is quite a different observable than the transverse 
momentum of the hardest $B$ hadron, our findings are actually rather similar, but less pronounced:
the shower substantially affects the distributions, by causing a significant softening
whose size exceeds the uncertainty bands of the fixed-order calculations.
This effect reflects the loss of energy of the $b$ quarks due to their fragmentation into $B$ hadrons. These 
observations are to a good extend independent 
of the considered contribution ($y_b^2$ or $y_t^2$). 

In contrast, the effect of the shower in the $\Delta R(B_1,B_2)$ distribution is smaller and becomes 
relevant 
only at large separations ($\Delta R \ge 4.5$). We point out that 
such large separations are at or beyond the coverage edge of the $b$ tracking system of LHC experiments, 
and that if two $b$ jets are explicitly required these differences are reduced.
Effects due to the parton shower are also visible at small separations, where secondary $g\to b \bar b$ splittings 
can be important. In this case, the two $B$ hadrons are likely to be clustered in the same $b$ jet.
The salient differences between $y_b^2$ and $y_t^2$ 
contributions are related to the behaviour of the respective LO curves, which again depends 
on their different relative normalisations; their shapes, on the contrary, are quite similar. 
\begin{figure}[p!]
\centering
\includegraphics[width=0.47\textwidth, clip=true, trim=0.5cm 2.5cm 0.7cm 1cm]{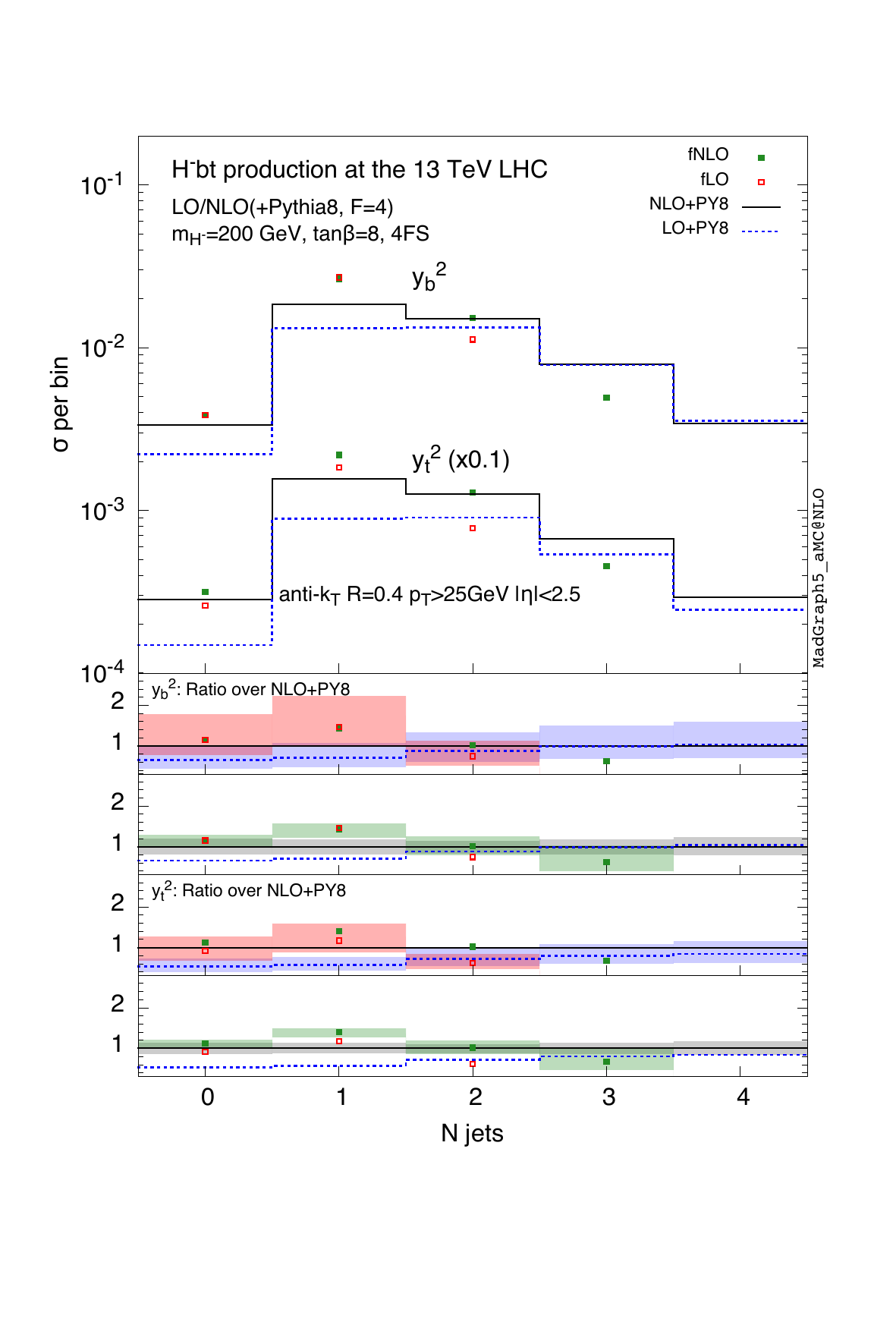}
\includegraphics[width=0.47\textwidth, clip=true, trim=0.5cm 2.5cm 0.7cm 1cm]{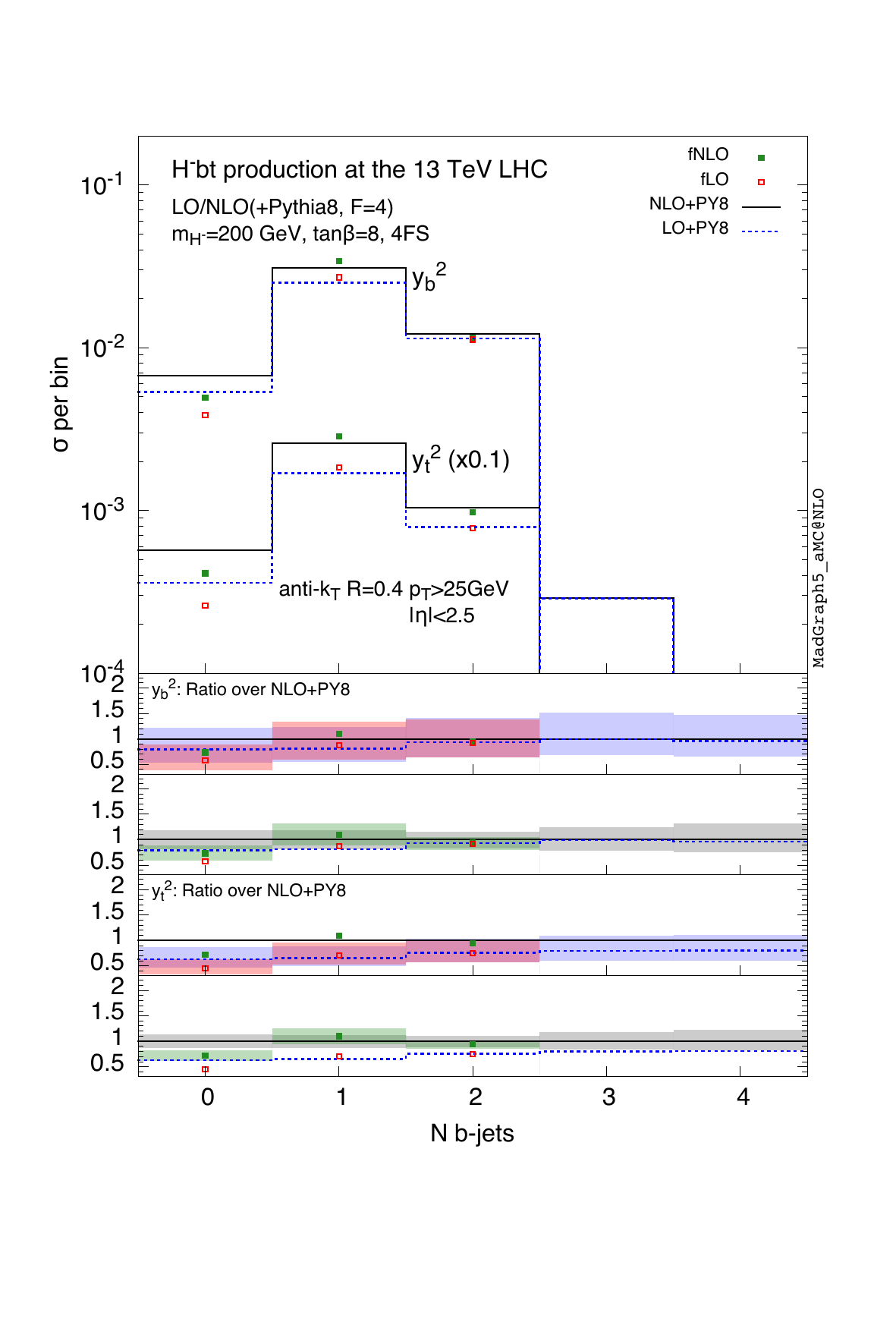}\\
\includegraphics[width=0.47\textwidth, clip=true, trim=0.5cm 2.5cm 0.7cm 1cm]{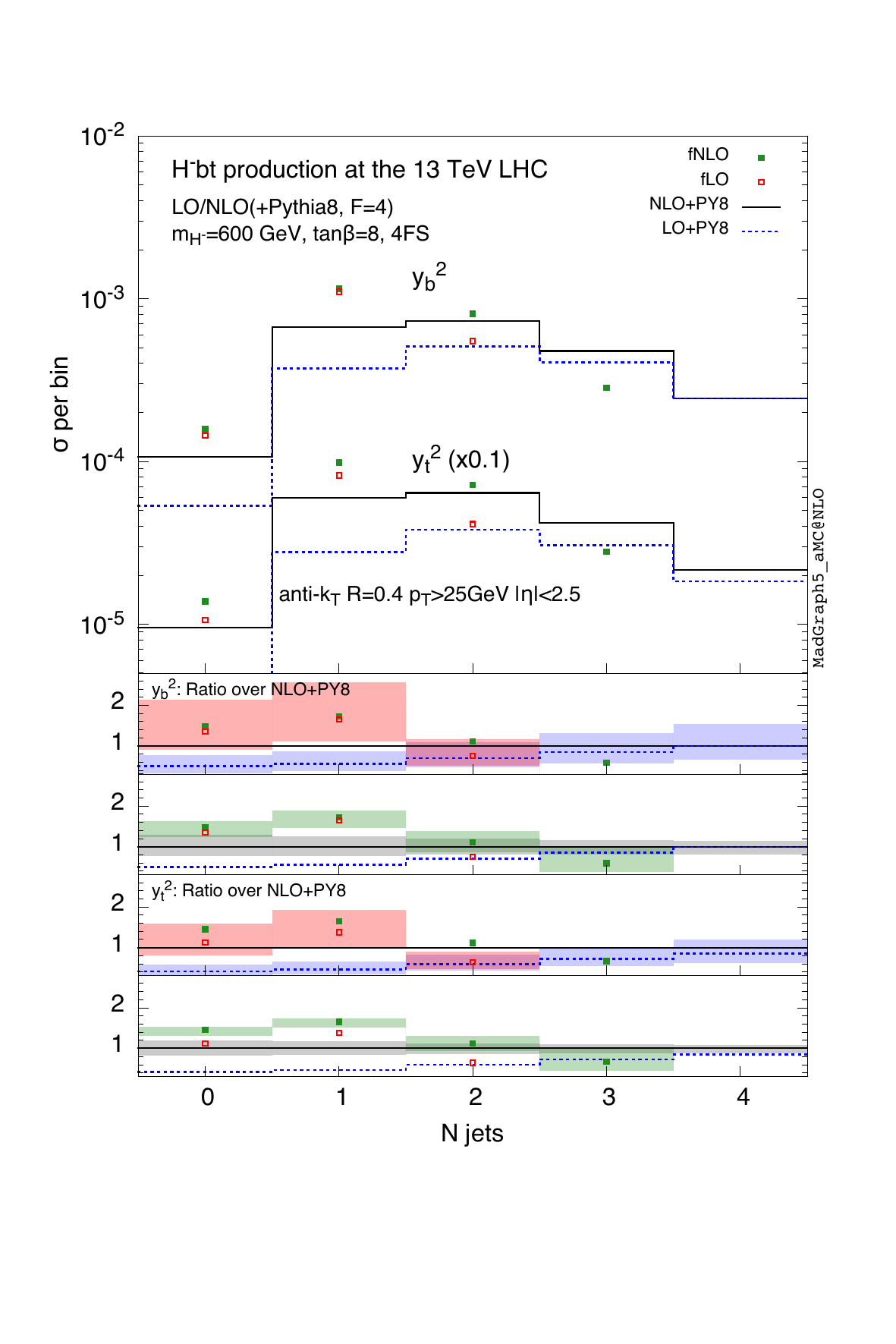}
\includegraphics[width=0.47\textwidth, clip=true, trim=0.5cm 2.5cm 0.7cm 1cm]{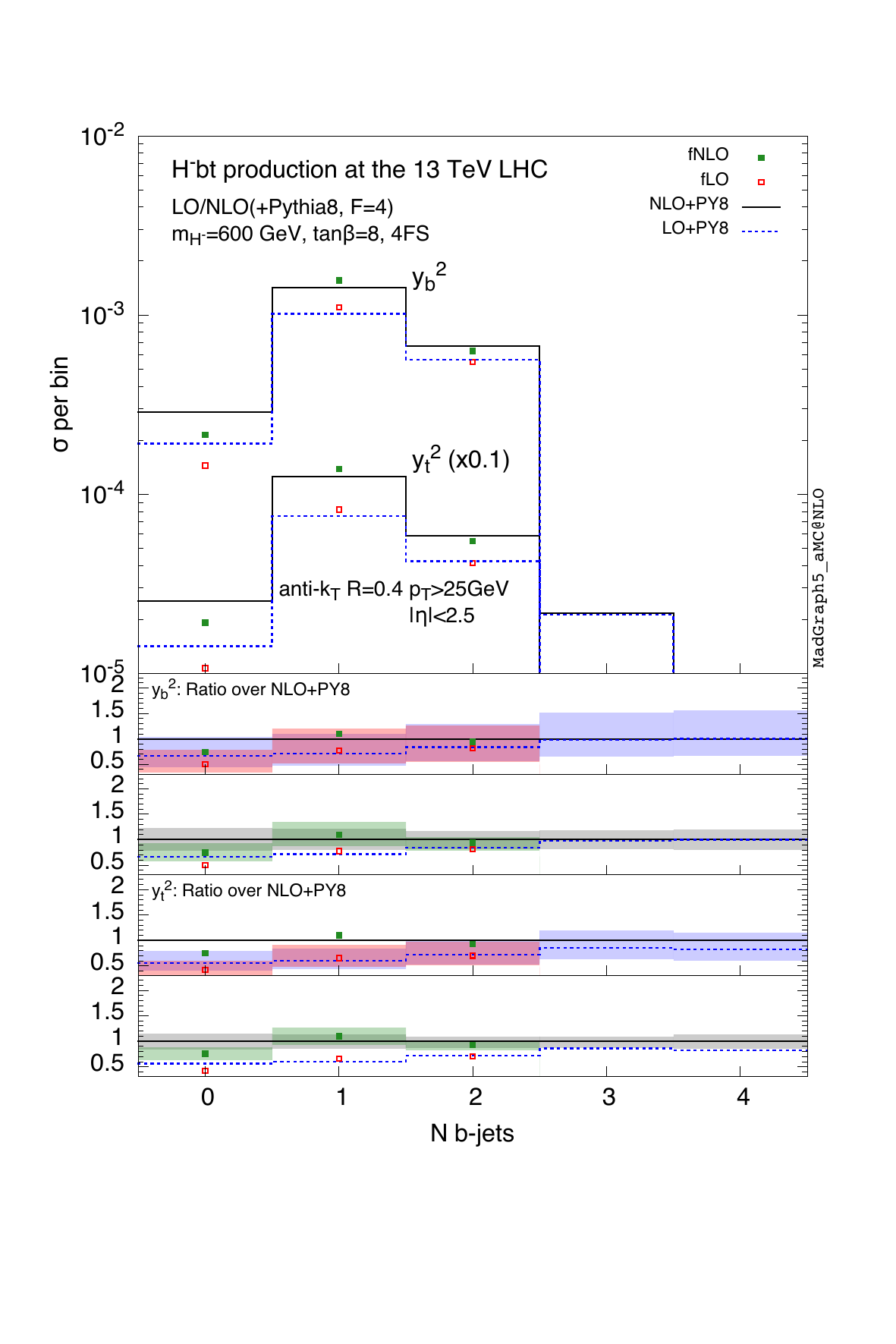}
\caption{\label{fig:4fnjet}
Same as Fig.~\ref{fig:4fpth}, but for the jet rates (left panels) and $b$-jet rates (right panels) with $m_{H^-}=200\gev$ (upper panels) 
and $m_{H^-}=600\gev$ (lower panels).}
\end{figure}

Let us now conclude our 4FS study by considering jet rates, displayed in Fig.~\ref{fig:4fnjet}. 
In the left panel, we show jet multiplicities without requiring any $b$ tagging, 
while in the right panel we show $b$-jet multiplicities.
First of all, we recall
that in our setup the top quark decays (leptonically) both in the shower and 
at fixed order. For this reason, up to two/three jets and 
up to two $b$ jets can appear at fLO/fNLO. 
Looking at the $\mH=200\gev$ results first (upper plots), we can appreciate 
the effects of the NLO corrections and of the matching with parton showers. 
NLO effects have the largest impact in the two-jet bin, where the cross section 
is increased by 35\% (65\%) for the $y_b^2$ ($y_t^2$) term. Their effect is minor
for lower multiplicities: almost no
correction occurs for the $y_b^2$ term, while the $y_t^2$ one is increased by 20\%.
If in turn we consider the effects of the parton shower and compare the fNLO histograms to 
the NLO+PS ones, we infer that the main effect is to populate higher multiplicities. Therefore, 
the zero- and one-jet rates are reduced 
as a consequence of unitarity. 
Such a reduction
is quite important, as the NLO+PY8 prediction falls outside the fNLO uncertainty band,
particularly in the one-jet bin.
The two-jet bin is left almost unchanged by the shower. 
We also find that shower effects 
have a much larger impact at LO, which reflects the large uncertainties associated with the LO computations, particularly at fixed order. 

The distribution of events with respect to the number of $b$ jets is 
displayed in the upper right panel of Fig.~\ref{fig:4fnjet}.
In this case NLO corrections have the largest effect on the zero- and one-$b$ jet bins. 
Since the majority of events lies in the one-$b$ jet bin, 
the shower moves events from this bin into the higher 
and lower multiplicities, although the effect is in general moderate 
and within the fNLO uncertainties. 
Only in the zero-$b$ jet bin, whose rate is however suppressed, the differences 
between fNLO and NLO+PS results are larger than in the 
flavour-unspecific case and the uncertainties barely overlap.
Overall, the NLO predictions are reasonably close to each other, since higher multiplicities (beyond two 
$b$ jets) are phase-space suppressed in the NLO+PS simulations. 

For both the jet and $b$-jet rates, the fraction of events with jet multiplicities beyond the ones already present 
at the hard-matrix element level (more than three jets and more than two $b$ jets) is in good agreement between the LO+PS and NLO+PS predictions, the LO ones being slightly enhanced though. 
At this point, we remark that these multiplicities are 
utterly Monte Carlo dependent and a substantial disagreement between 
{\sc Pythia8} and {\sc Herwig++} is found for these bins. For the lower multiplicities, however, 
their agreement is excellent, as will be discussed in detail in Sect.~\ref{sec:4vs5FS}.

The results of Fig.~\ref{fig:4fnjet} analysed so far are for $\mH=200\gev$. 
Since these observables are particularly relevant for experimental analyses based on jet and $b$-jet categories, 
let us discuss explicitly the results for a charged Higgs 
mass of $\mH=600\gev$, which are displayed in the lower plots of Fig.~\ref{fig:4fnjet}.
While for jet multiplicities the general features are not so different, some
specific features are.
Apart from a larger Higgs mass changing the LO normalisation, the 
distribution of events at LO appears 
shifted towards low multiplicities as compared to LO+PS, without 
any overlap of the corresponding uncertainties in the first two bins. However, given our findings for the lower Higgs 
mass case, this is expected: the shower shifts events from lower towards
higher jet multiplicities; this is enhanced for $\mH=600\gev$ due to a 
generally increased hardness of the process. Indeed, the two-jet bin
has the largest rate, and the three- and four-jet bins are 
less suppressed than in the lighter Higgs case.
NLO corrections slightly improve the agreement of showered and fixed-order results, 
albeit fNLO and NLO+PS still fall outside the respective uncertainties 
in the zero- and one-jet bins.

In the case of $b$ jets, on the other hand, the features of the 
relative curves reflect those discussed for the lighter Higgs and no further comments are needed.

%%%%%%%%%%%%%%%%%%%%%%%%%%%%%%%%%%%%%%%%%%%%%%%%%%%%%%%%%%%%%%%%%%%%%5
\subsubsection{The $y_by_t$ contribution}
\label{sec:ybyt}
%%%%%%% PLOTS FOR THE YBYT TERM AT LO
 
\begin{figure}[t]
\centering
\includegraphics[width=0.46\textwidth, clip=true, trim=0.cm 8.0cm 0.7cm 1cm]{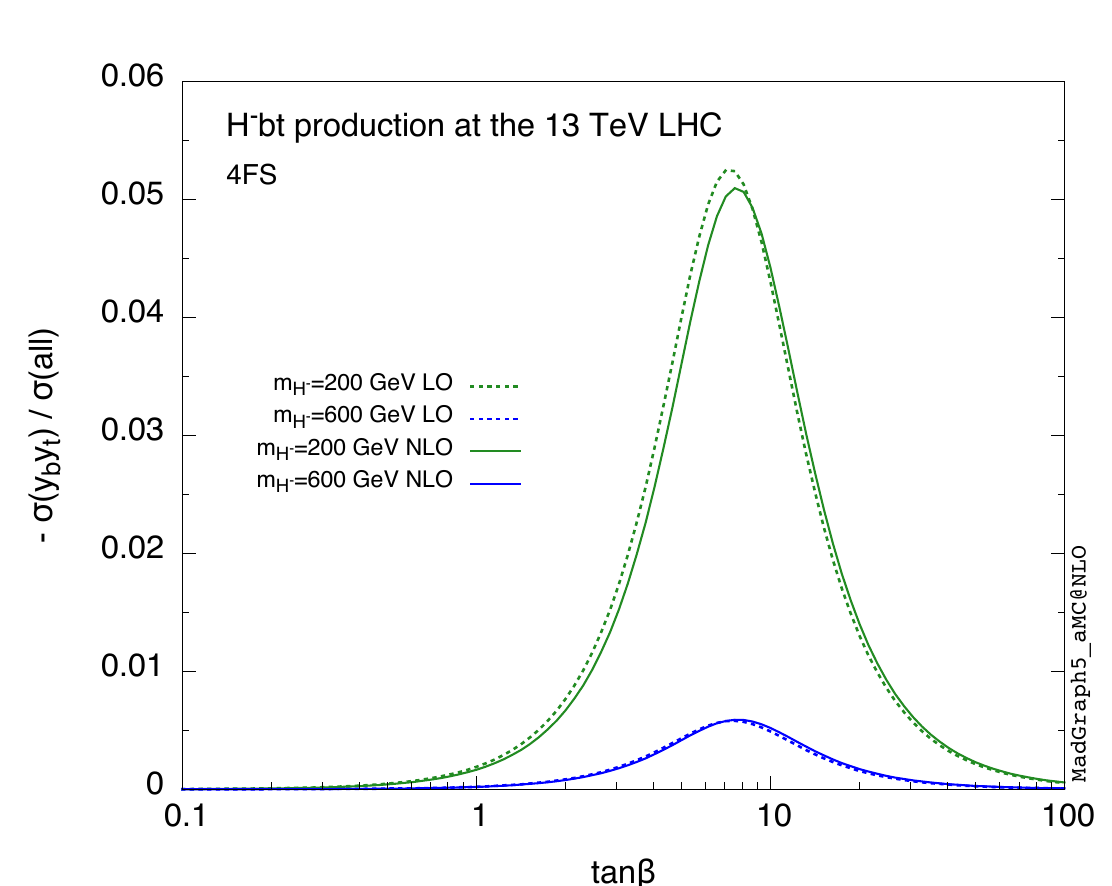}\\[-8pt]
\caption{\label{fig:ybytsigma}Relative contribution of the $y_by_t$ term at LO (dashed) and NLO (solid) in the 4FS, with respect to the total 
(N)LO cross section, as a function of $\tan\beta$. The two cases 
$m_{H^-}=200\gev$ (green) and $m_{H^-}=600\gev$ (blue) are shown.}
\end{figure}

\begin{figure}[p]
\centering
\includegraphics[width=0.46\textwidth, clip=true, trim=0.cm 6.0cm 0.7cm 1cm]{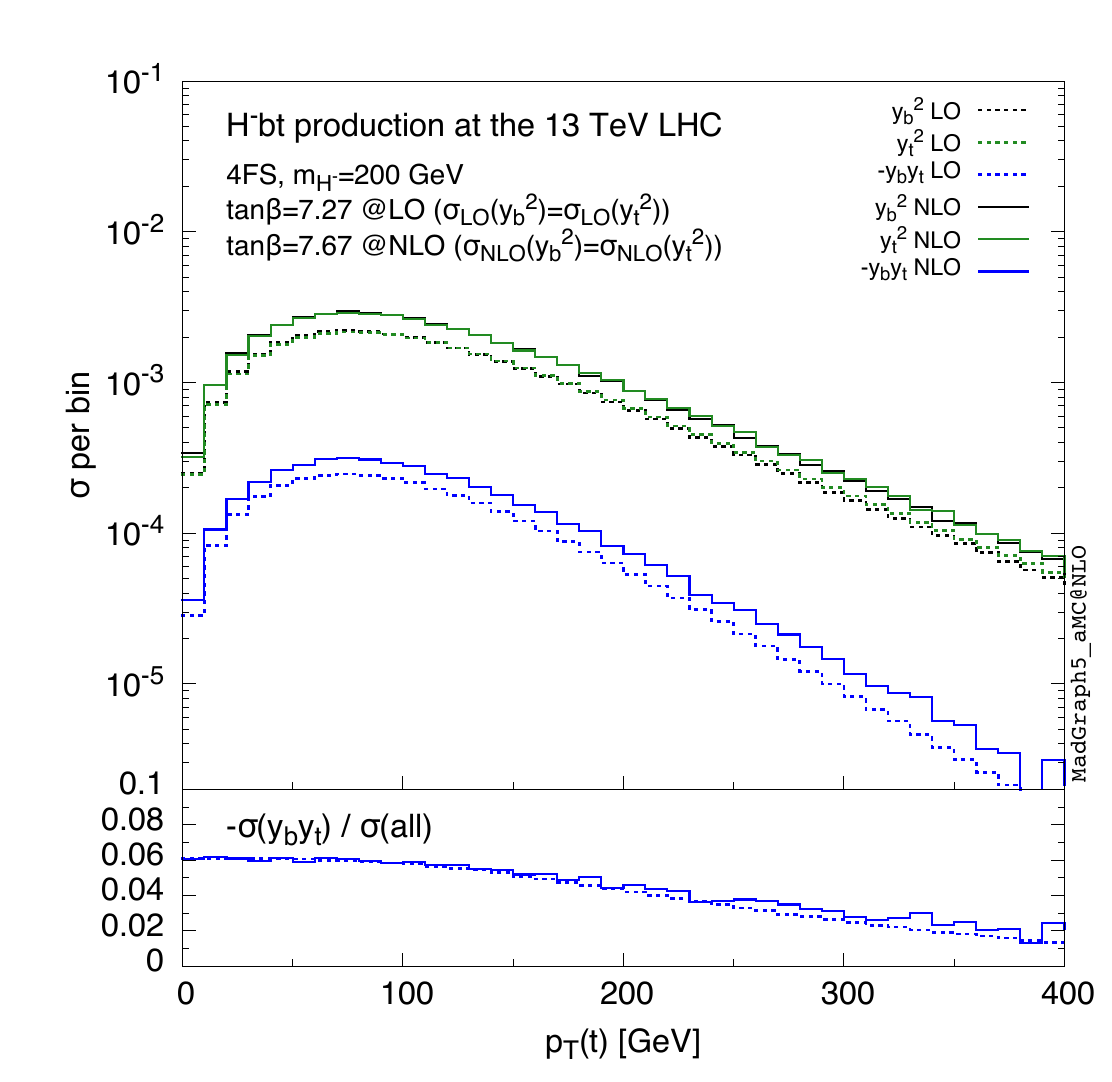}
\includegraphics[width=0.46\textwidth, clip=true, trim=0.cm 6.0cm 0.7cm 1cm]{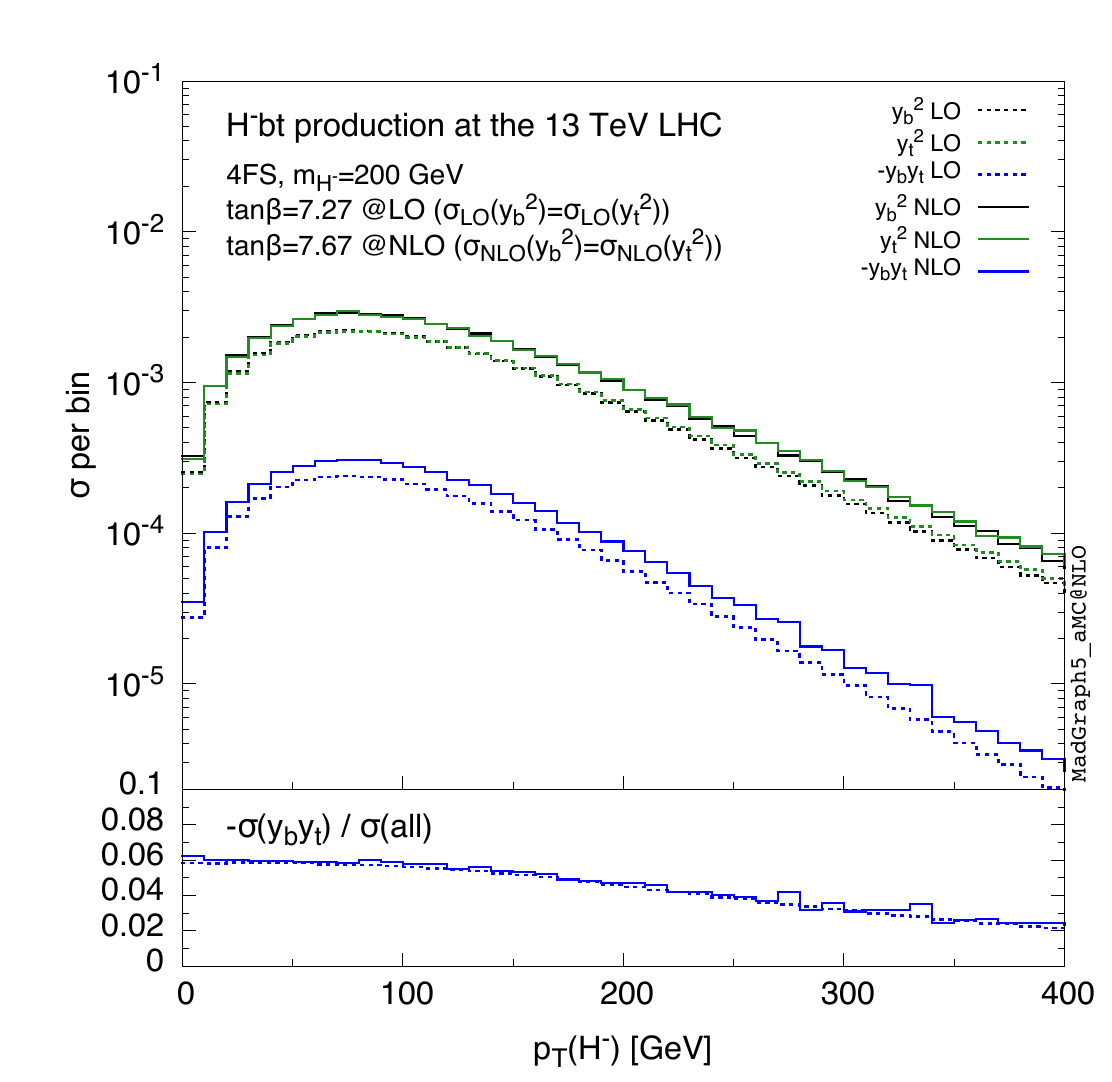}\\[-8pt]
\includegraphics[width=0.46\textwidth, clip=true, trim=0.cm 6.0cm 0.7cm 1cm]{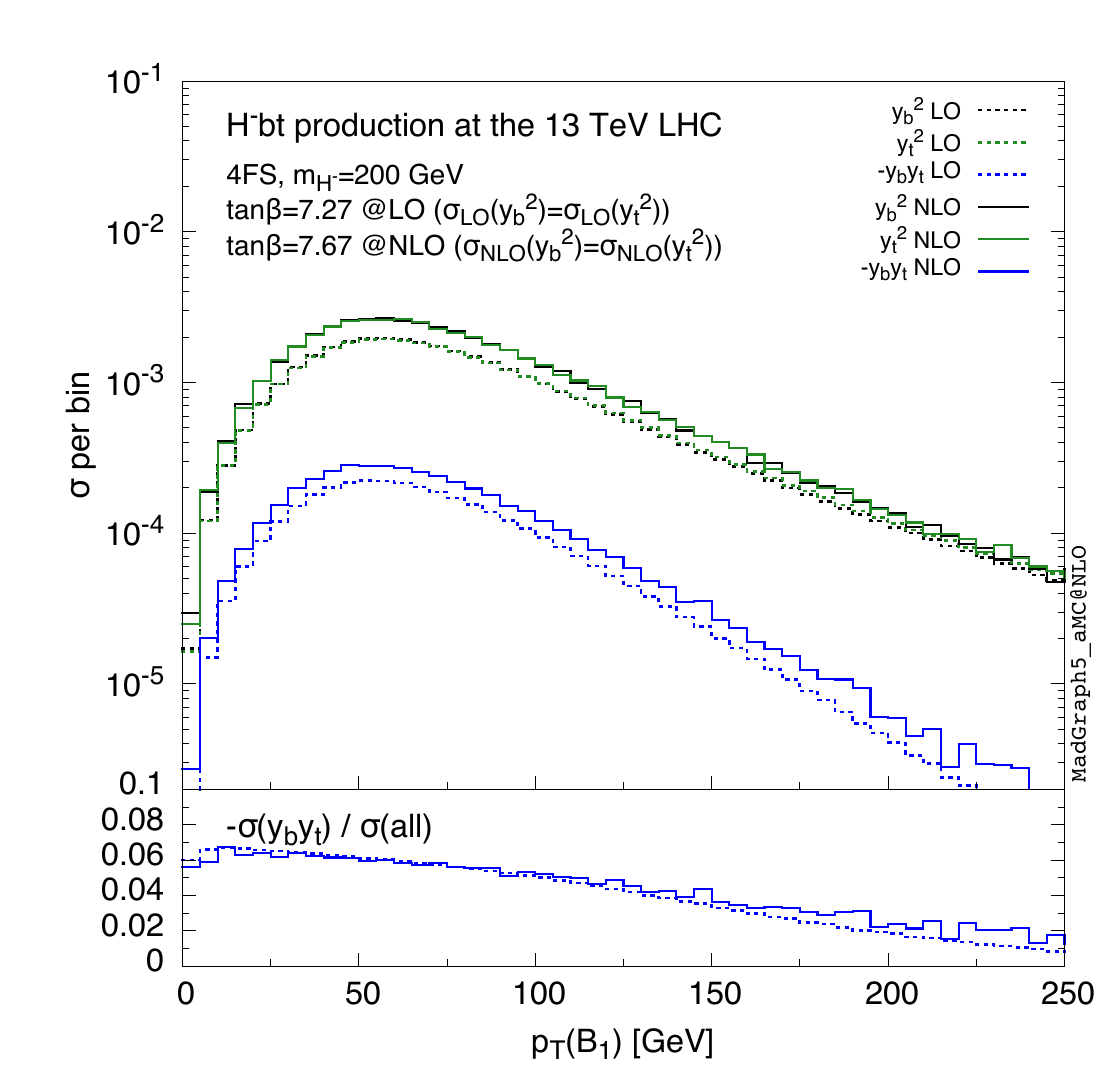}
\includegraphics[width=0.46\textwidth, clip=true, trim=0.cm 6.0cm 0.7cm 1cm]{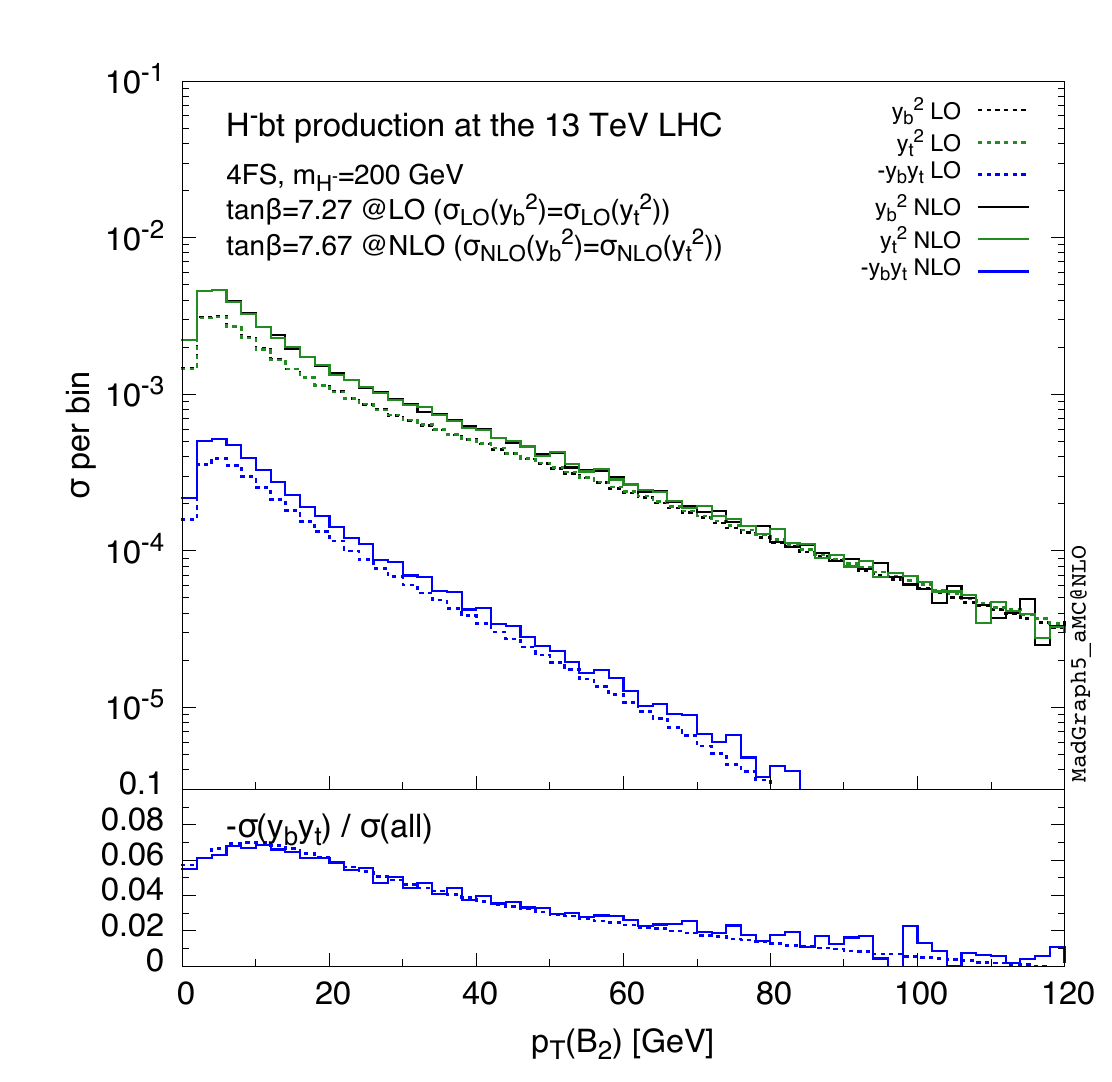}\\[-8pt]

\caption{\label{fig:ybytdiff}Differential comparisons at the between the $y_by_t$ term and the $y_b^2$, $y_t^2$ ones, for $m_{H^-}=200\gev$. The transverse momentum of the top quark (top left),
of the Higgs boson (top right) and of the hardest (bottom left) and second hardest (bottom right) $B$ hadron are considered. In the main frame the 
$y_b^2$ (black), $y_b^2$ (green) and $y_by_t$ (blue) distribution are plotted at LO (dashed) and NLO (solid), while the inset shows the ratio $\sigma(y_by_t)/\left(\sigma(y_b^2)+\sigma(y_by_t)+\sigma(y_b^2)\right)$ at LO and NLO.}
\end{figure}

As we mentioned before, in the 5FS the NLO cross section receives contributions either proportional to $y_b^2$ or to $y_t^2$. 
No $y_b y_t$ term appears, given that it would come from the interference of left-handed with right-handed massless bottom quarks. 
If in turn $b$ quarks are massive, as in the 4FS, the $y_b y_t$ term does not vanish any longer, and it is proportional to 
$m_b^2/Q^2$, where $Q$ is some hard scale of the process.
So far, we have limited our 4FS analysis to the $y_b^2$ and $y_t^2$ contributions, assuming the $y_b y_t$ one to be suppressed. 
In this section, we show that this is indeed the case.

To this purpose, we consider the total cross section for charged Higgs production in the 4FS at LO and NLO, 
and plot the relative contribution $-\sigma_{y_b y_t}/\sigma_{\rm all}$ as a function of 
$\tan \beta$ in Fig.~\ref{fig:ybytsigma}, with $\sigma_{\rm all}$ being the sum 
of all terms. The results are shown for $m_{H^-}= 200, 600 \gev$. 
The minus sign takes into account the fact that the $y_by_t$ term is negative. 
We stress that the $y_by_t$ contribution is independent of $\tan \beta$. As can be inferred from the plots, the relative size
of the $y_by_t$ term is below 5\% for $m_{H^-}= 200\gev$, and  0.5\% for $m_{H^-}= 600\gev$. 
The relative contribution to the cross section proportional to $y_by_t$ is maximal when the $y_b^2$ and $y_t^2$ terms are equal, i.\,e. when
\begin{equation}
y_b^2 \tan\beta^2 = {y_t^2}/{\tan\beta^2} \quad \Rightarrow \quad \tan\beta =7.27(7.67),
\end{equation}
at LO (NLO), for $m_{H^-}=200\gev$.\footnote{The difference between the LO and NLO values is due to the different perturbative order in the running of $y_b$.}

Let us further investigate the potential impact of the inclusion of the  $y_by_t$
term on some differential observables, for such a value of $\tan\beta$. 
In particular, we look at the transverse momentum of the Higgs, the top and the two hardest 
$B$ hadrons for $m_{H^-}= 200\gev$, displayed in Fig.~\ref{fig:ybytdiff}. 
From these plots we notice that the
effect of the $y_b y_t$ term is peaked at low scales, by reaching at most $6-7\%$ of the full cross section, 
and is almost the same at LO and NLO. 
We stress again that these numbers have been computed for the value of $\tan\beta$ for which the relative $y_by_t$ contribution is 
maximal: for larger (smaller) values of $\tan \beta$, this contribution is suppressed by a factor $1/\tan^2\beta$ ($\tan^2\beta$) 
with respect to the $y_t^2$ ($y_b^2$) contribution and further reduced for 
heavier charged Higgs bosons. The typical scale uncertainties
at NLO ($\sim10-15\%$) justify  our choice to neglect the $y_by_t$ contribution in the current analysis.
A viable alternative would be to include the relative contribution of the $y_by_t$ term only at LO, which was shown to be very similar to the NLO one.

%%%%%%%%%%%%%%%%%%%%%%%%%%%%%%%%%%%%%%%%%%%%%%%%%%%%%%%%%%%%%%%%%%%%%5
\subsection{Four- and Five-flavour scheme comparison}
\label{sec:4vs5FS}
%%%%%%% FIGURES SHOWING NF=4 VS NF=5 AND LO+PS VS NLO+PS
\begin{figure}[t]
\centering
\includegraphics[width=0.48\textwidth, clip=true, trim=0.5cm 3cm 0.7cm 1cm]{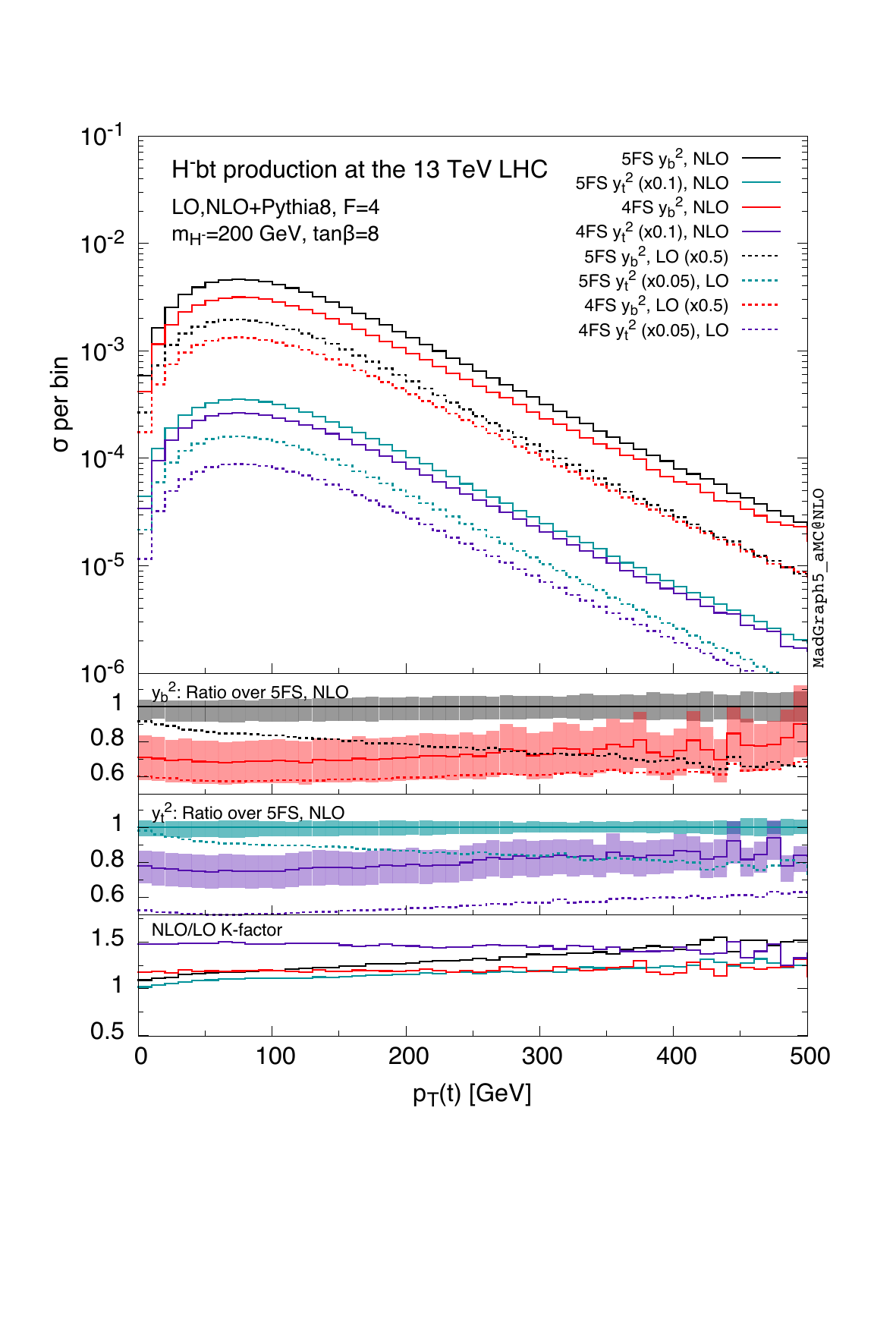}
\includegraphics[width=0.48\textwidth, clip=true, trim=0.5cm 3cm 0.7cm 1cm]{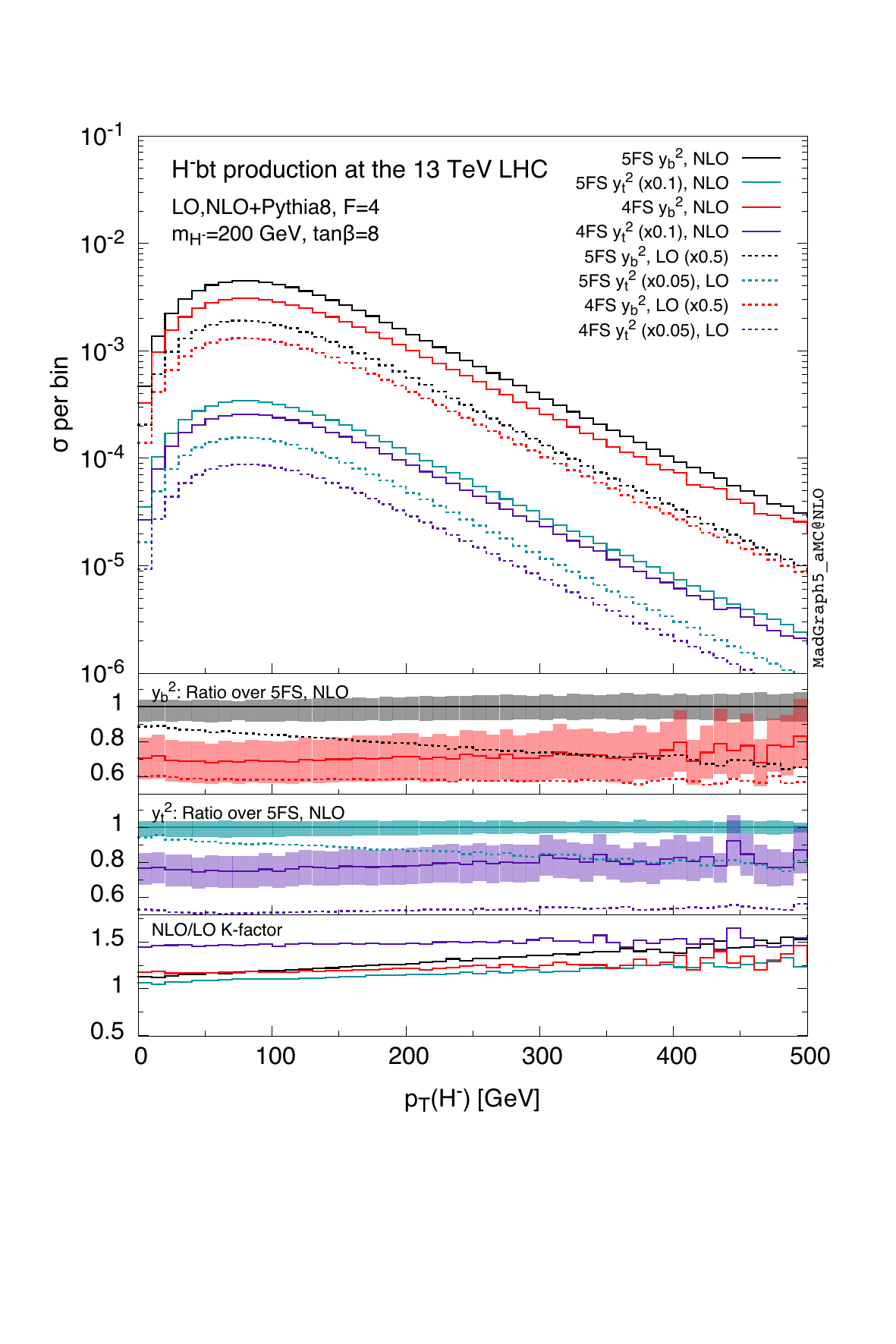}
\caption{\label{fig:LOvsNLOpth}LO (dashed) and NLO (solid) predictions matched with {\sc Pythia8} in the 4FS 
(red for $y_b^2$, violet for $y_t^2$) and 5FS (black for $y_b^2$, light blue for $y_t^2$), for the transverse momentum of the top quark (left) and
of the charged Higgs boson (right). Rescaling factors are introduced in the main frame for better readability. The first and second insets show the 
ratio over the NLO prediction in the 5FS for the $y_b^2$ and $y_t^2$ term respectively, and the scale uncertainty band for the NLO curves. 
The third inset show the differential $K$-factor (NLO/LO) for the four predictions. A charged Higgs boson mass $m_{H^-}=200\gev$ is considered.}
\end{figure}
\begin{figure}[t]
\centering
\includegraphics[width=0.47\textwidth, clip=true, trim=0.5cm 3cm 0.7cm 1cm]{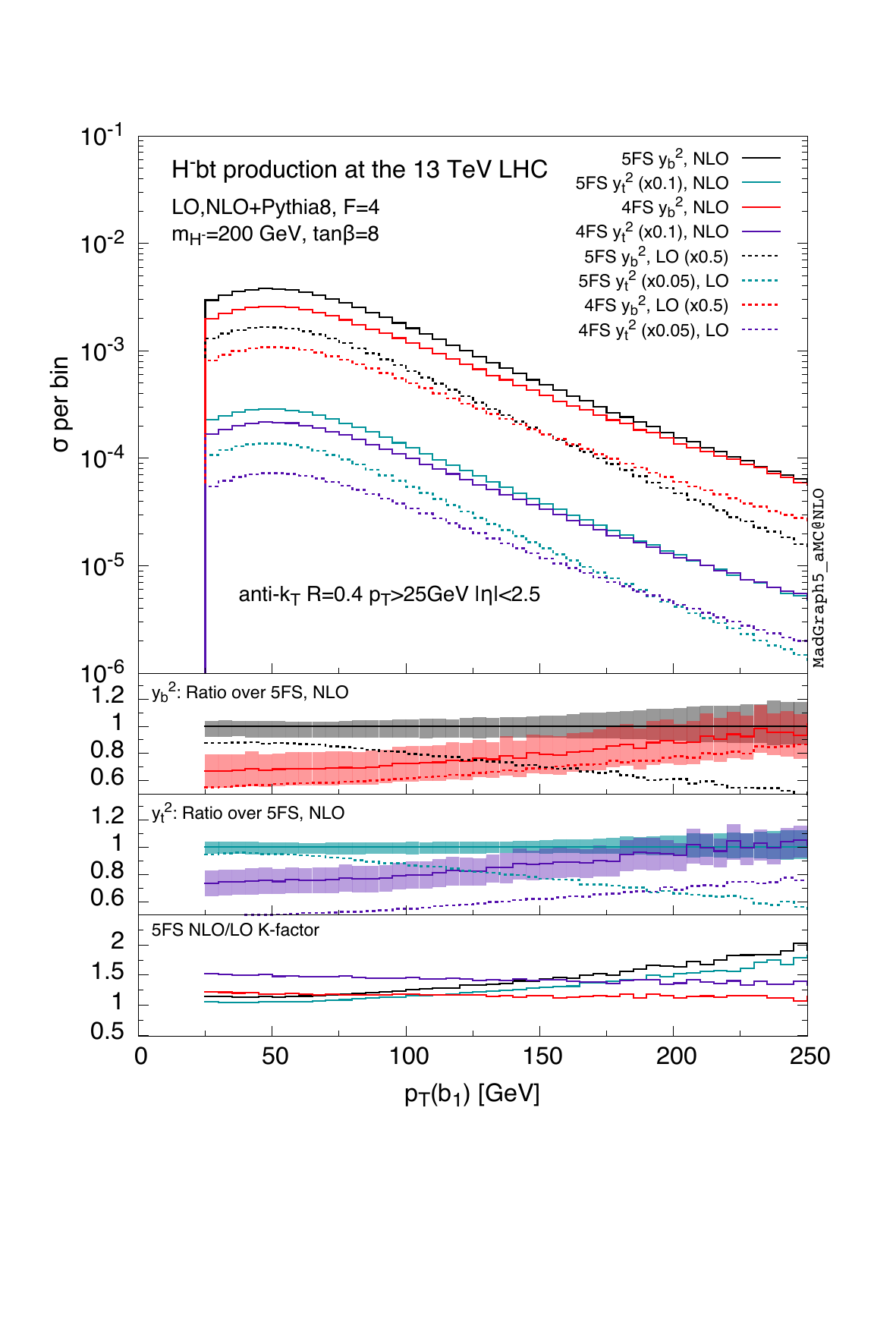}
\includegraphics[width=0.47\textwidth, clip=true, trim=0.5cm 3cm 0.7cm 1cm]{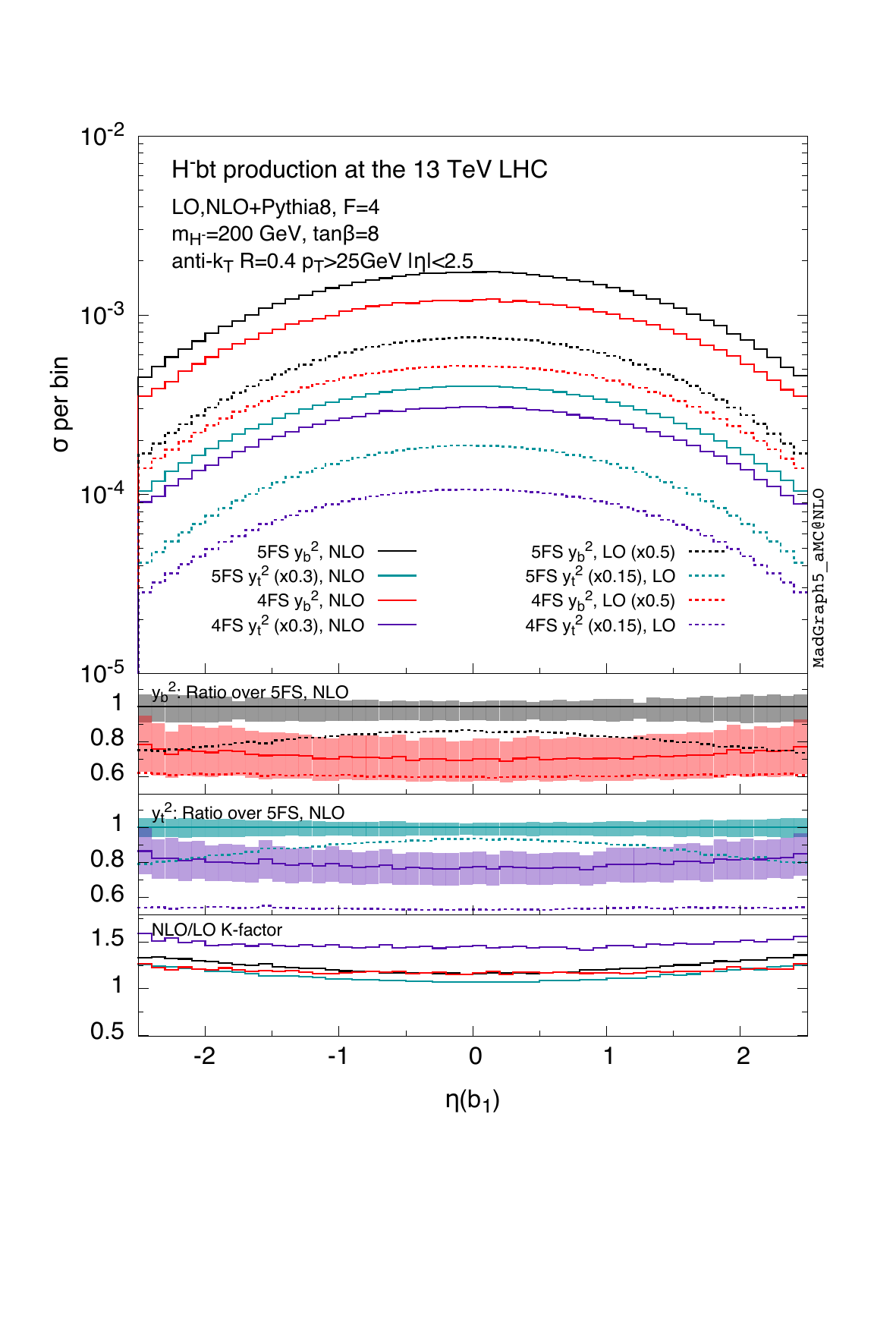}
\caption{\label{fig:LOvsNLOptbj12}Same as Fig.~\ref{fig:LOvsNLOpth}, but for the transverse momentum (left) and pseudo-rapidity (right) of the hardest $b$ jet.}
\end{figure}

\begin{figure}[t]
\centering
\includegraphics[width=0.47\textwidth, clip=true, trim=0.5cm 3cm 0.7cm 1cm]{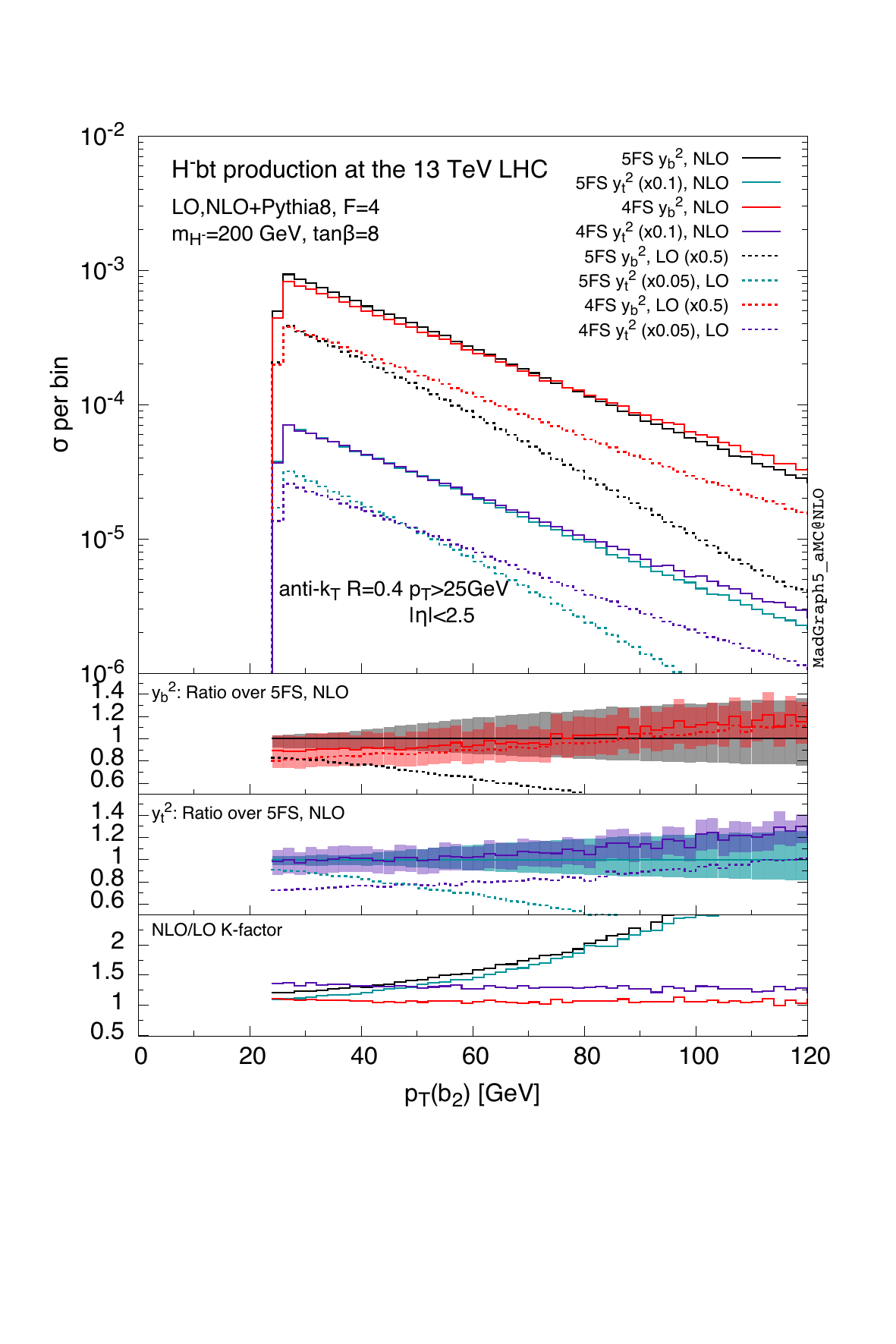}
\includegraphics[width=0.47\textwidth, clip=true, trim=0.5cm 3cm 0.7cm 1cm]{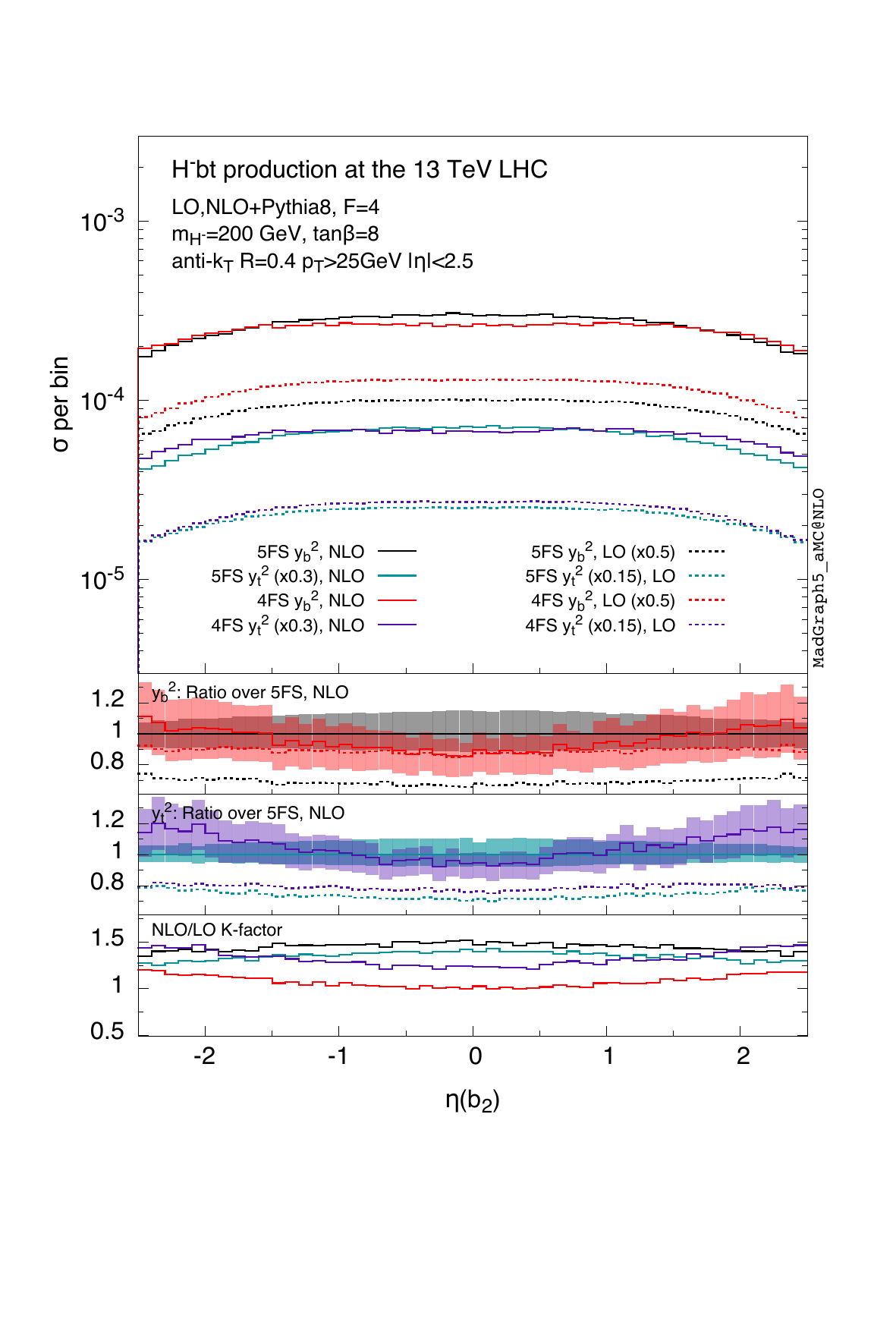}
\caption{\label{fig:LOvsNLOetabj12}Same as Fig.~\ref{fig:LOvsNLOpth}, but for the transverse momentum (left) and pseudo-rapidity (right) of the second-hardest $b$ jet.}
\end{figure}

\begin{figure}[t]
\includegraphics[width=0.47\textwidth, clip=true, trim=0.5cm 3cm 0.7cm 1cm]{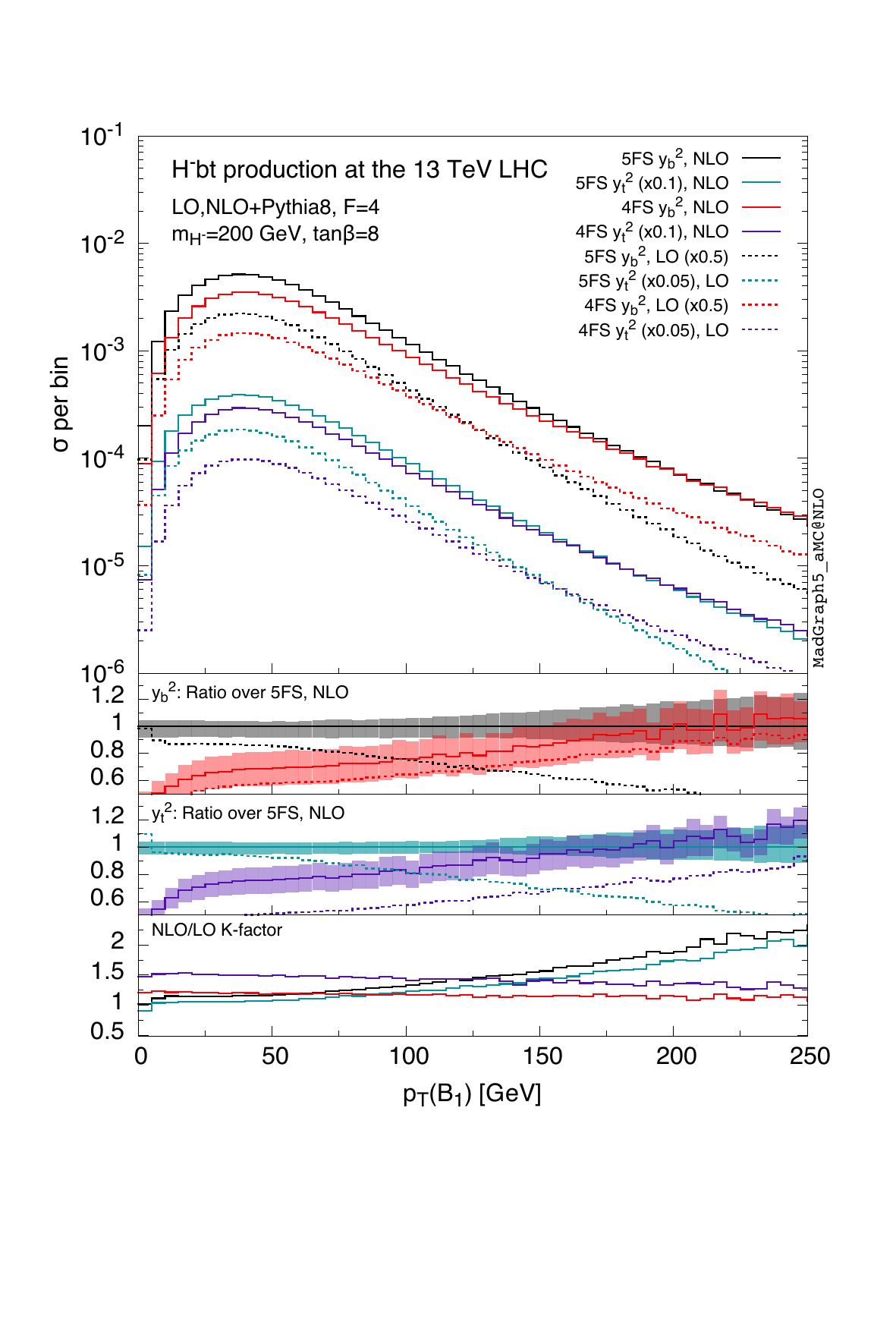}
\includegraphics[width=0.47\textwidth, clip=true, trim=0.5cm 3cm 0.7cm 1cm]{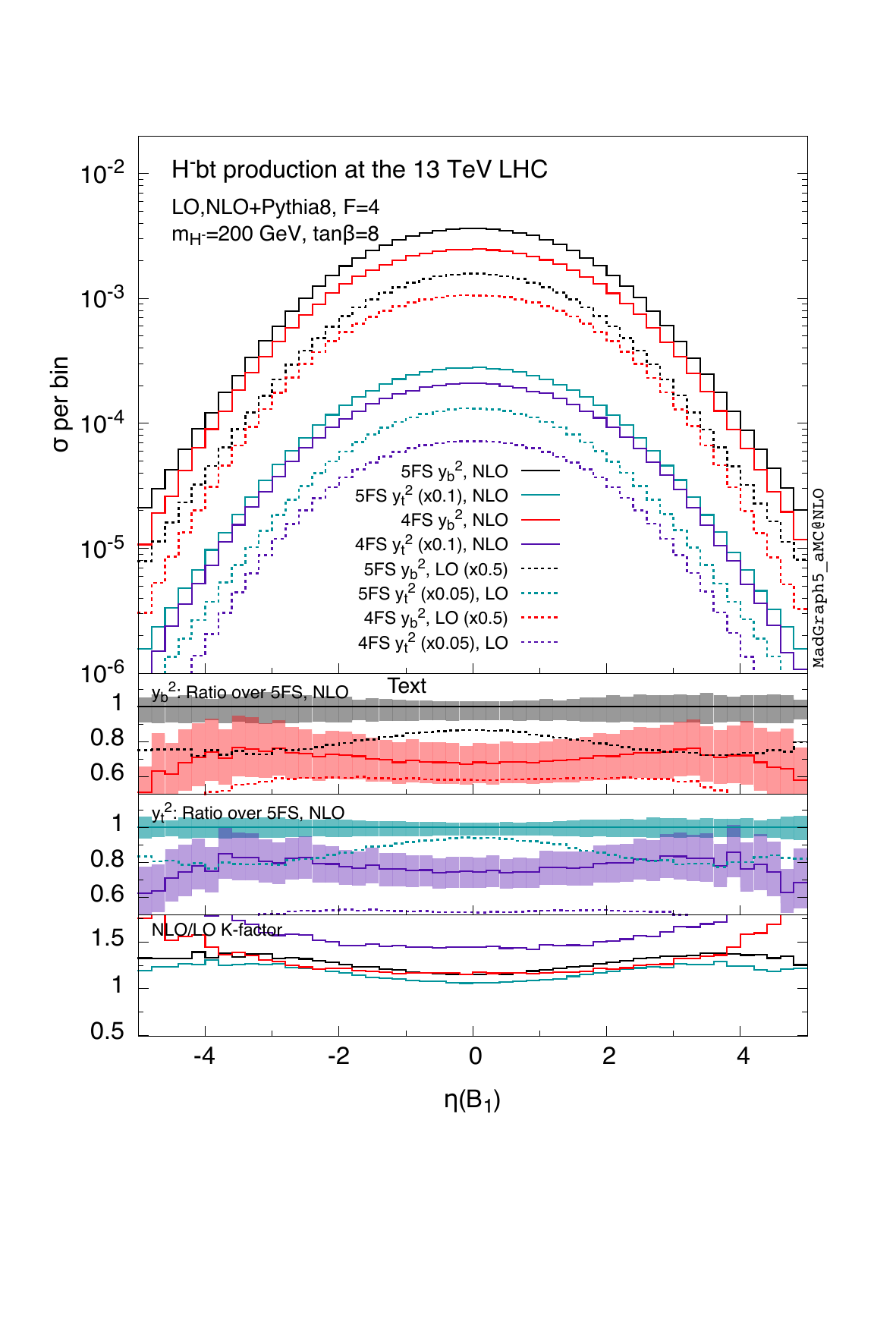}
\caption{\label{fig:LOvsNLOptetabh1}Same as Fig.~\ref{fig:LOvsNLOpth}, but for the transverse momentum (left) and pseudo-rapidity (right) of the hardest $B$ hadron.}
\end{figure}
We turn now to investigate how predictions obtained in the four- and five-flavours schemes compare. 
The two schemes are actually identical up to $b$-mass power suppressed terms when computed to all orders in 
perturbation theory, but the way of ordering the perturbative series 
is different. As a consequence, the results in the two schemes 
may be different at any finite order, while the inclusion of higher 
orders necessarily brings the predictions in the two schemes 
closer to each other.  
We start by quantifying how 
the inclusion of NLO corrections improves their mutual agreement. 
In Figs.~\ref{fig:LOvsNLOpth}-\ref{fig:LOvsNLOetabj12} we show, for
some relevant observables, the LO and NLO predictions (matched with {\sc Pythia8}) in the two schemes. 
All figures have the same pattern: a main frame with the absolute
predictions in the 5FS (black for $y_b^2$ and light blue for $y_t^2$) and the 4FS 
(red for $y_b^2$ and violet for $y_t^2$) at LO (dashed) and NLO (solid). 
In the first and second insets we show the ratio of the curves in the main frame
over the 5FS NLO prediction, for the $y_b^2$ and $y_t^2$ contributions respectively. 
For the NLO predictions, a band
indicating the scale uncertainty\footnote{We recall that we vary both renormalisation
and factorisation scales by a factor of 2 independently about their central values.} 
is attached to the curves. In the third inset, 
the four differential NLO/LO $K$-factors ($y_b^2$ and $y_t^2$ for 4FS and 5FS) are displayed. 

A general observation is that, as expected, the NLO predictions in the
two schemes are much closer to each other than the LO ones, in particular as far as shapes are concerned. 
Differences in the overall normalisation reflect the differences in the total cross section, which
were already discussed in Ref.~\cite{Flechl:2014wfa}, while in this comparison we are mostly interested in the shapes.
In Fig.~\ref{fig:LOvsNLOpth} we observe that for the transverse momentum  
of the top quark and the Higgs boson the difference between the two schemes can be compensated 
by a simple overall rescaling of the total rates 
($\sigma_{\rm tot}^{\rm 4FS}/\sigma_{\rm tot}^{\rm 5FS}\simeq 0.7$)
at NLO, while LO predictions in the two schemes have considerably different shapes. 
The same level of agreement should be found also for observables related to the (leptonic) decay 
products of the top quark and the Higgs. 
Let us recall that in our simulation we do not decay the Higgs boson, but we decay leptonically the top quark. 
The $b$ quark from the top decay mostly ends up in the hardest $b$ jet. This 
explains why the $p_T$ spectrum of the hardest $b$ jet (left plot in Fig.~\ref{fig:LOvsNLOptbj12}) 
displays a flat ratio between the 4FS and 5FS at NLO, up to
$\sim 120\gev$. Above that value, secondary $g\to b\bar b$ splittings from hard gluons become more relevant, 
which is also reflected in the growth of the 5FS uncertainty band and $K$-factor. 
A similar behaviour has been observed in the case of $tH$ production 
in the SM~\cite{Demartin:2015uha}. The 
pseudo-rapidity
of the hardest $b$ jet (right plot in Fig.~\ref{fig:LOvsNLOptbj12}) is mostly dominated 
by the low-$p_T$ region, and it therefore also displays a good agreement between 4FS and 5FS shapes at NLO.

Larger differences between the two schemes appear for the second-hardest $b$ jet, 
which is expected to be poorly described in the 5FS. In particular, its kinematics in the 5FS 
at LO is determined by the shower, while at NLO it is driven by a 
tree-level matrix element (therefore being formally only LO accurate). 
Predictions for the transverse momentum of the second $b$ jet and its pseudo-rapidity 
are shown in 
the left and right panels of Fig.\,\ref{fig:LOvsNLOetabj12}. 
The 5FS develops large $K$-factors and larger uncertainties, 
since its LO prediction stems only from the shower 
evolution. Therefore, the 4FS description has to be preferred for these observables, 
both because of its better perturbative behaviour and the proper modelling of the 
final-state $b$ jets.

The effects of the different treatment of the bottom quark in the two schemes is even more visible for the 
differential observables related to the hardest $B$ hadron
(see Fig.\,\ref{fig:LOvsNLOptetabh1}). At medium and large $p_T(B_1)$ and at central $\eta(B_1)$ 
similar effects as for the hardest $b$ jet are observed. 
At variance, the 4FS prediction is suppressed with respect to the 5FS one at 
low $p_T(B_1)$ and at large $\eta(B_1)$. This is most likely due to mass effects:  
these kinematical regions correspond to one $b$ quark being collinear to the beam. 
In the 5FS these configurations are enhanced because of the collinear singularities, while 
in the 4FS such a singularities are screened by the $b$-quark mass. Therefore, even after the PS, the 5FS
is reminiscent of the collinear enhancement. 
In the case of the second-hardest $B$ hadron (not shown) these effects are further enhanced.

Let us make a final remark on the inclusion of the NLO corrections. 
The NLO/LO $K$-factor is quite different in the two schemes: in the 4FS the $K-$factor appears much more
pronounced for the $y_t^2$ than for the $y_b^2$ term, while in the 5FS it is similar for 
both contributions. 
Despite that, a remarkable compensation in shape between the LO differential cross sections 
and the NLO corrections takes place, 
such that the 4FS/5FS ratio at NLO is quite similar for the $y_b^2$ and $y_t^2$ terms.

All the plots discussed so far are relevant to the lighter Higgs under consideration ($m_{H^-}=200\gev$). 
For a heavier charged Higgs boson ($m_{H^-}=600\gev$), the picture does not change significantly.
The only thing that may be worth mentioning is the fact that, for the second $b$ jet, the $K$-factor 
in the 5FS lies much closer to unity than for the lighter Higgs. 
Such a behaviour may be due to the increased weight of the initial-state 
collinear logarithms resummed by the bottom PDFs in the 5FS, which 
are enhanced at larger masses of the produced particle.
Besides, as already pointed out in Ref.~\cite{Maltoni:2012pa}, 
collinear logarithms become increasingly relevant the larger the fraction 
of the momentum carried by the initial partons is.\\

%%%%%%% FIGURES SHOWING NF=4 VS NF=5 AND FRAC=1 VS FRAC=4
\begin{figure}[p]
\centering
\includegraphics[width=0.48\textwidth, clip=true, trim=0.5cm 3cm 0.7cm 1cm]{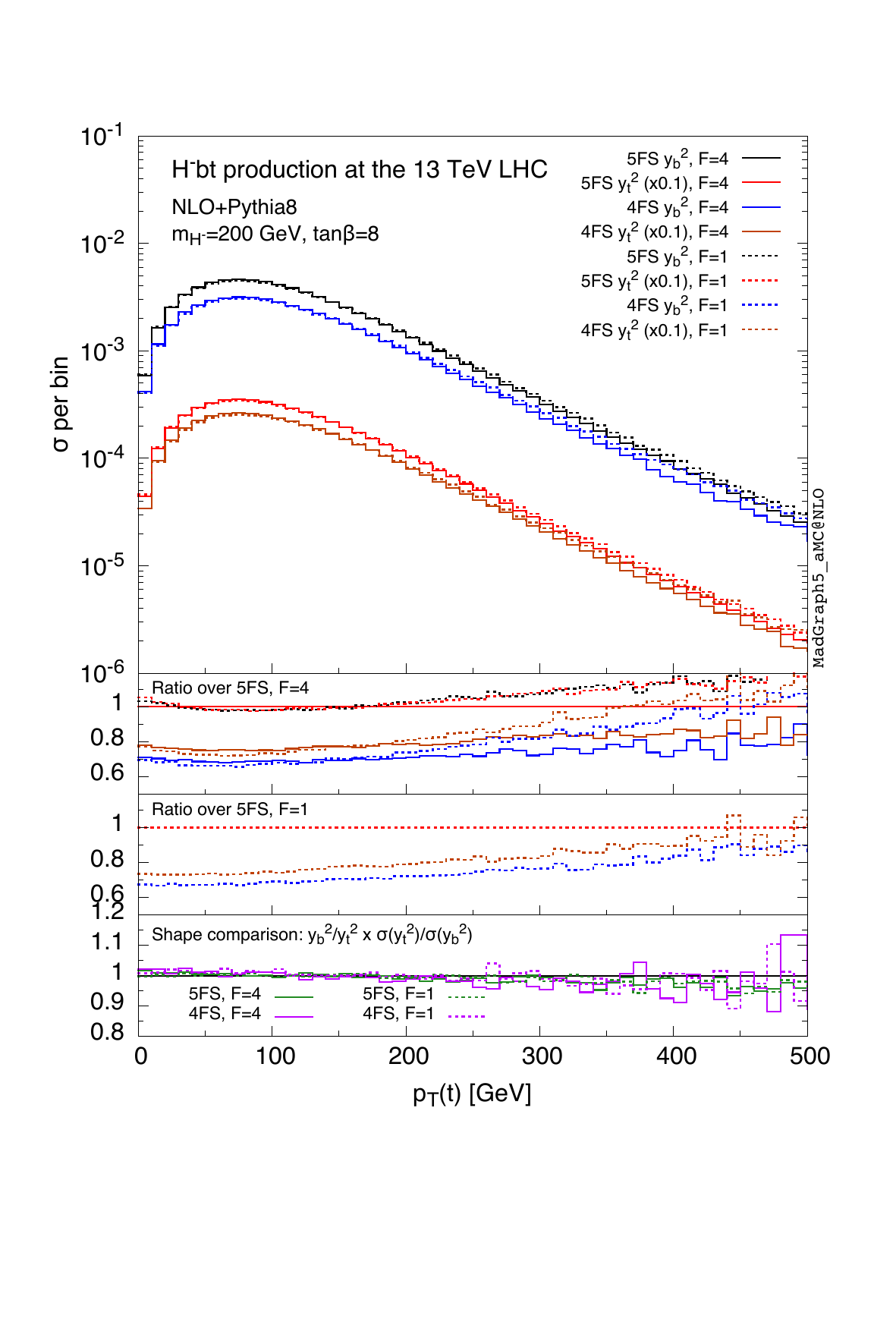}
\includegraphics[width=0.48\textwidth, clip=true, trim=0.5cm 3cm 0.7cm 1cm]{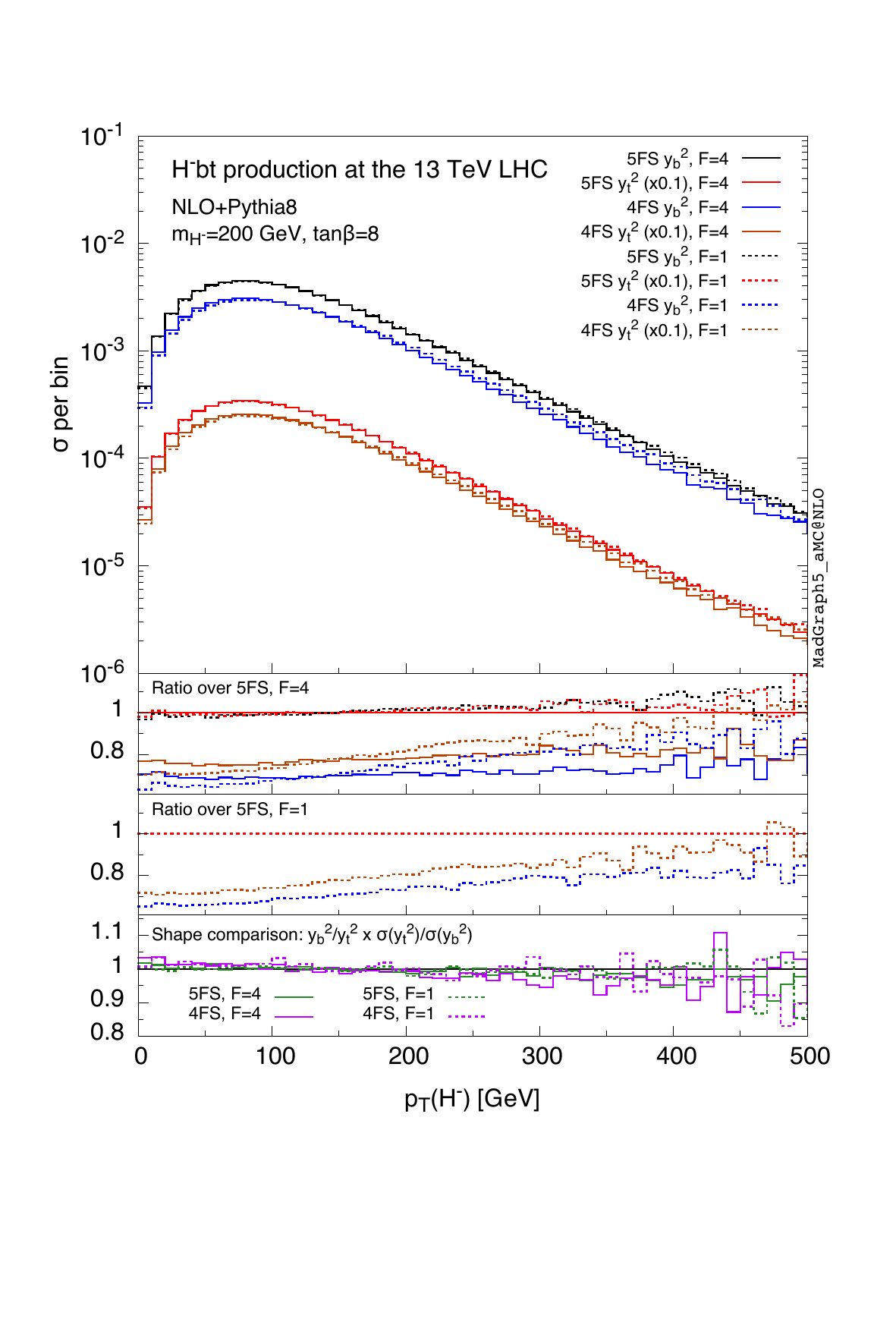}
\caption{\label{fig:frac1vs4pth} 4FS (blue for the $y_b^2$ term, orange for the $y_t^2$ one) and 5FS (black for the $y_b^2$ term, red for the $y_t^2$ one)
distributions for the transverse momentum of the top (left) and charged Higgs (right), 
for $m_{H^-}=200\gev$; the $F=1$ (dashed) and $F=4$ (solid) predictions are shown. The first inset shows the ratios of the histogram in the main frame
over the 5FS, $F=4$ prediction. The second inset shows the ratios of the $F=1$ histograms over the corresponding 5FS ones. The third inset
shows the ratio of the normalised $y_b^2$ histograms in the main frame over the corresponding $y_t^2$ ones.}

\centering
\includegraphics[width=0.47\textwidth, clip=true, trim=0.5cm 3cm 0.7cm 1cm]{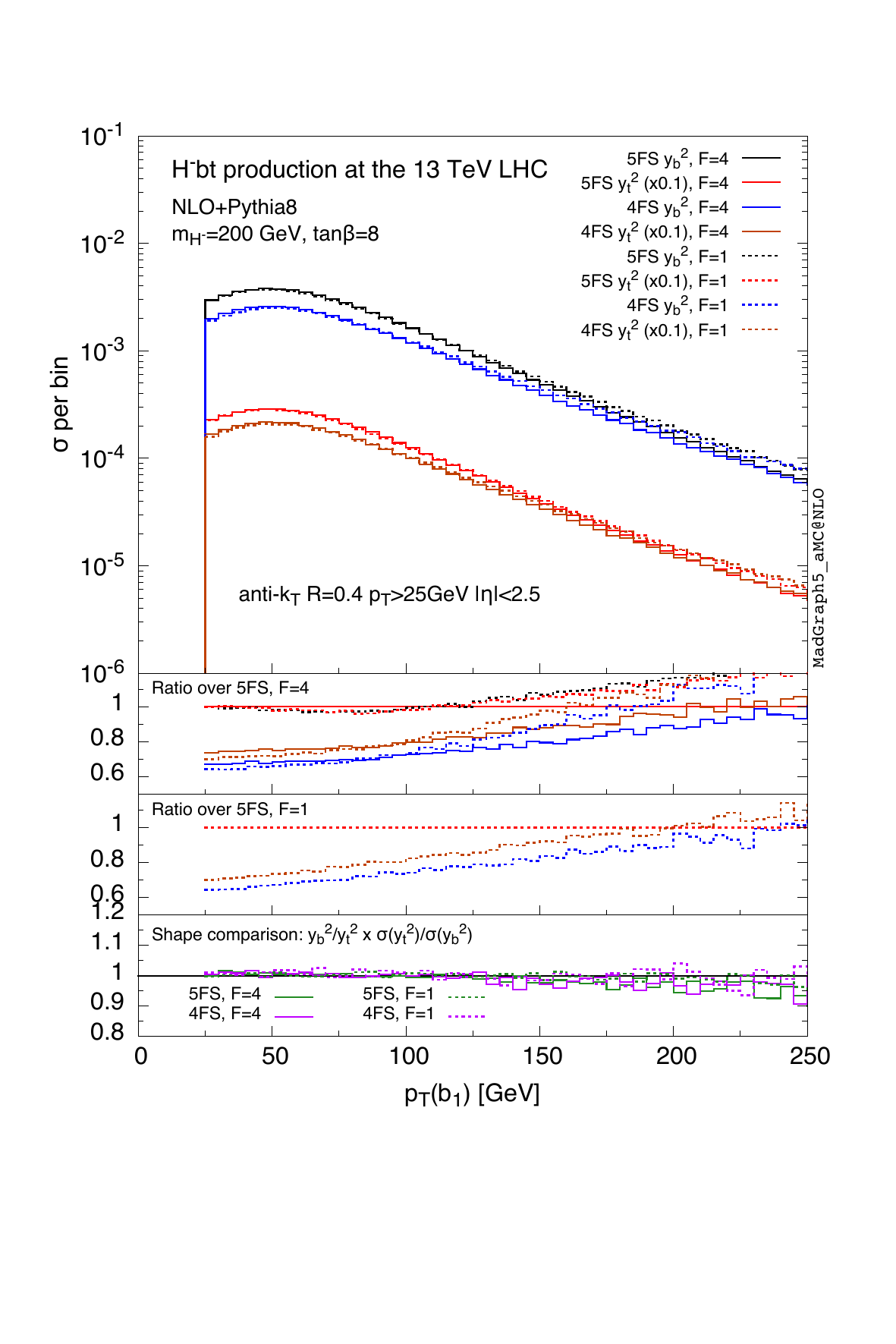}
\includegraphics[width=0.47\textwidth, clip=true, trim=0.5cm 3cm 0.7cm 1cm]{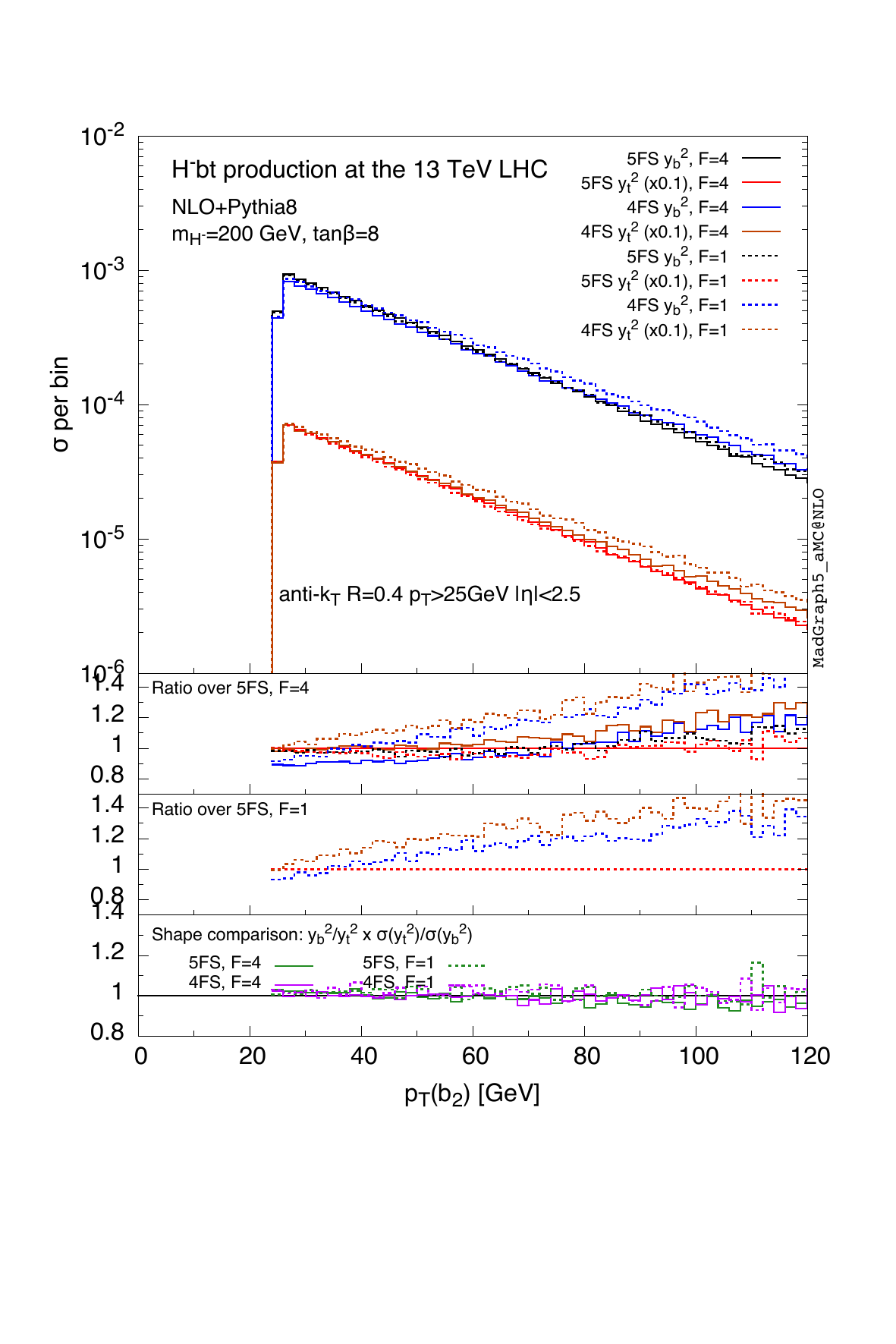}
\caption{\label{fig:frac1vs4ptbjh1} Same as Fig.~\ref{fig:frac1vs4pth}, but for the transverse momentum of the hardest (left) and second-hardest (right) $b$ jet .}
\end{figure}
\begin{figure}[t]
\centering
\includegraphics[width=0.48\textwidth, clip=true, trim=0.5cm 3cm 0.7cm 1cm]{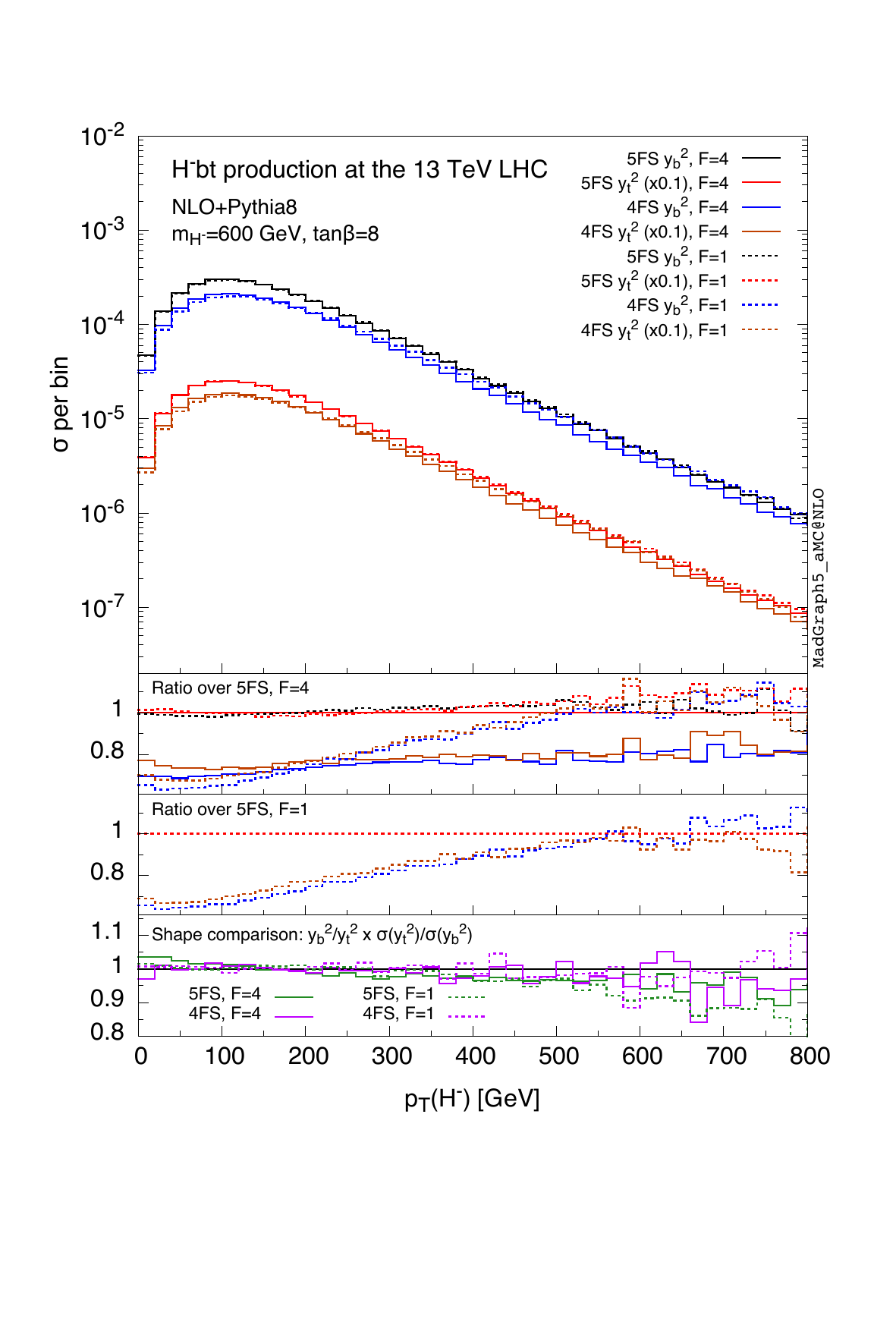}
\includegraphics[width=0.47\textwidth, clip=true, trim=0.5cm 3cm 0.7cm 1cm]{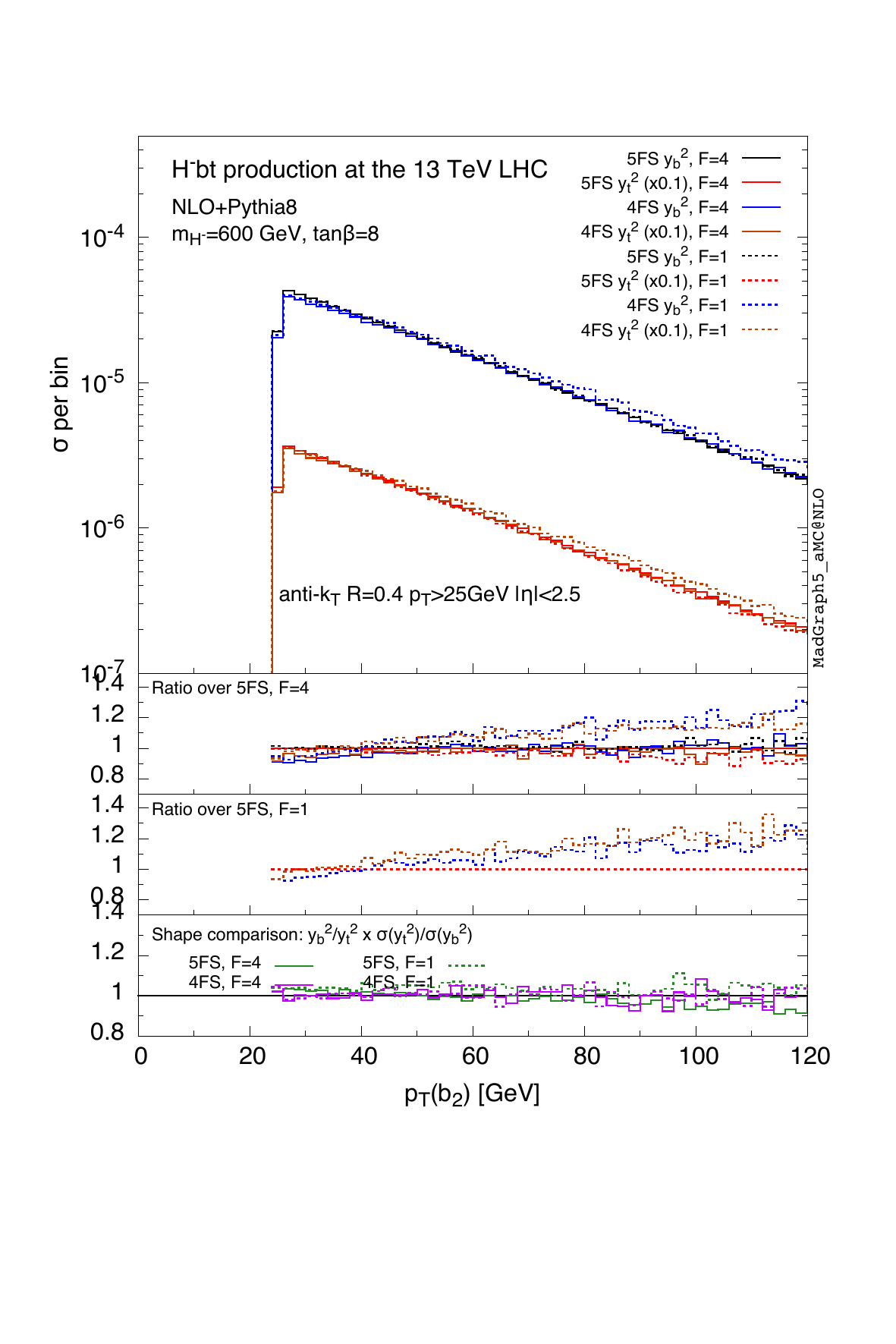}
\caption{\label{fig:600-frac1vs4pth} Same as Fig.~\ref{fig:frac1vs4pth}, but for 
transverse momentum distribution of the top (left) and the second-hardest $b$ jet (right) with $m_{H^-}=600\gev$.}
\end{figure}
\begin{figure}
\centering
\includegraphics[width=0.48\textwidth, clip=true, trim=0.5cm 3cm 0.7cm 1cm]{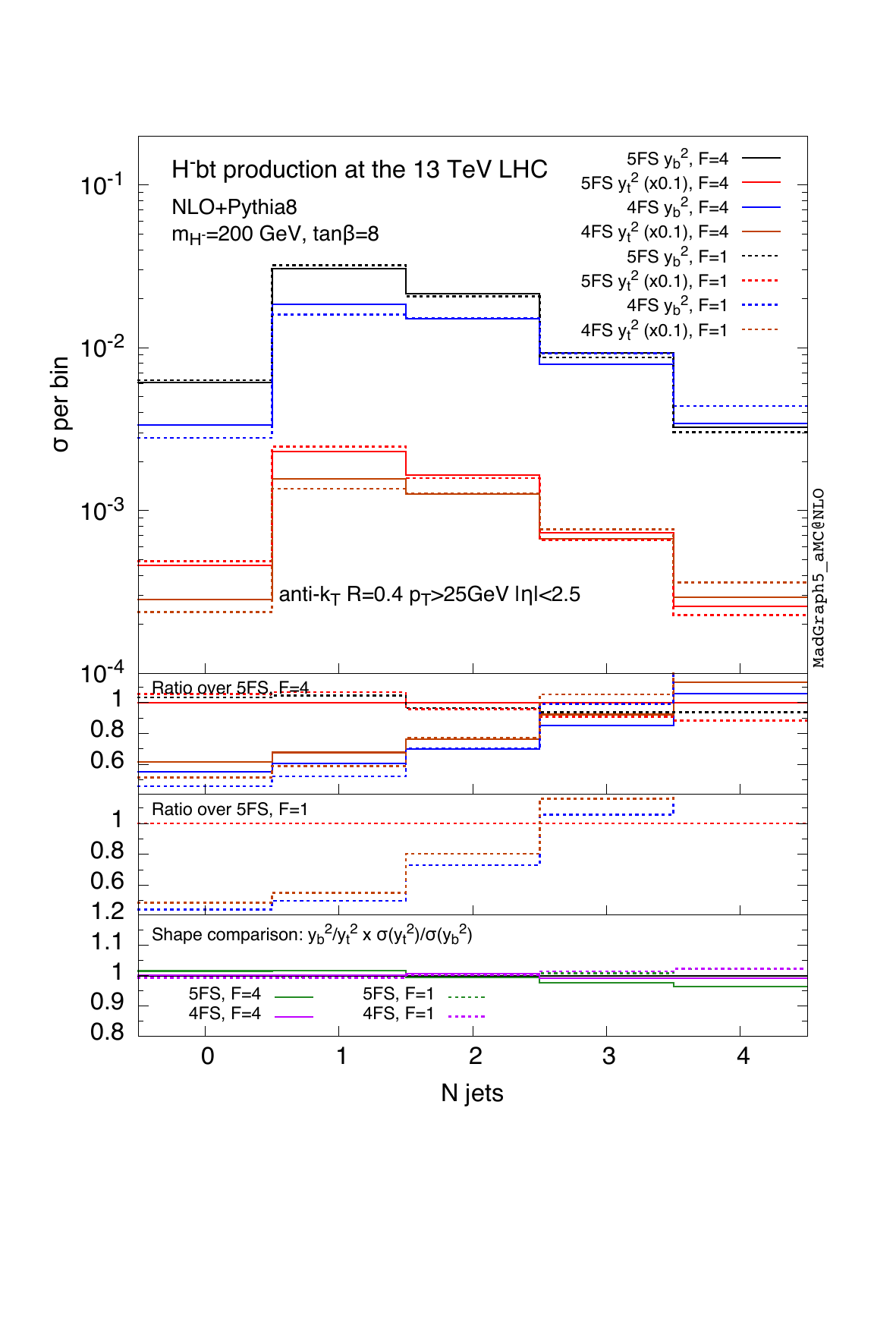}
\includegraphics[width=0.48\textwidth, clip=true, trim=0.5cm 3cm 0.7cm 1cm]{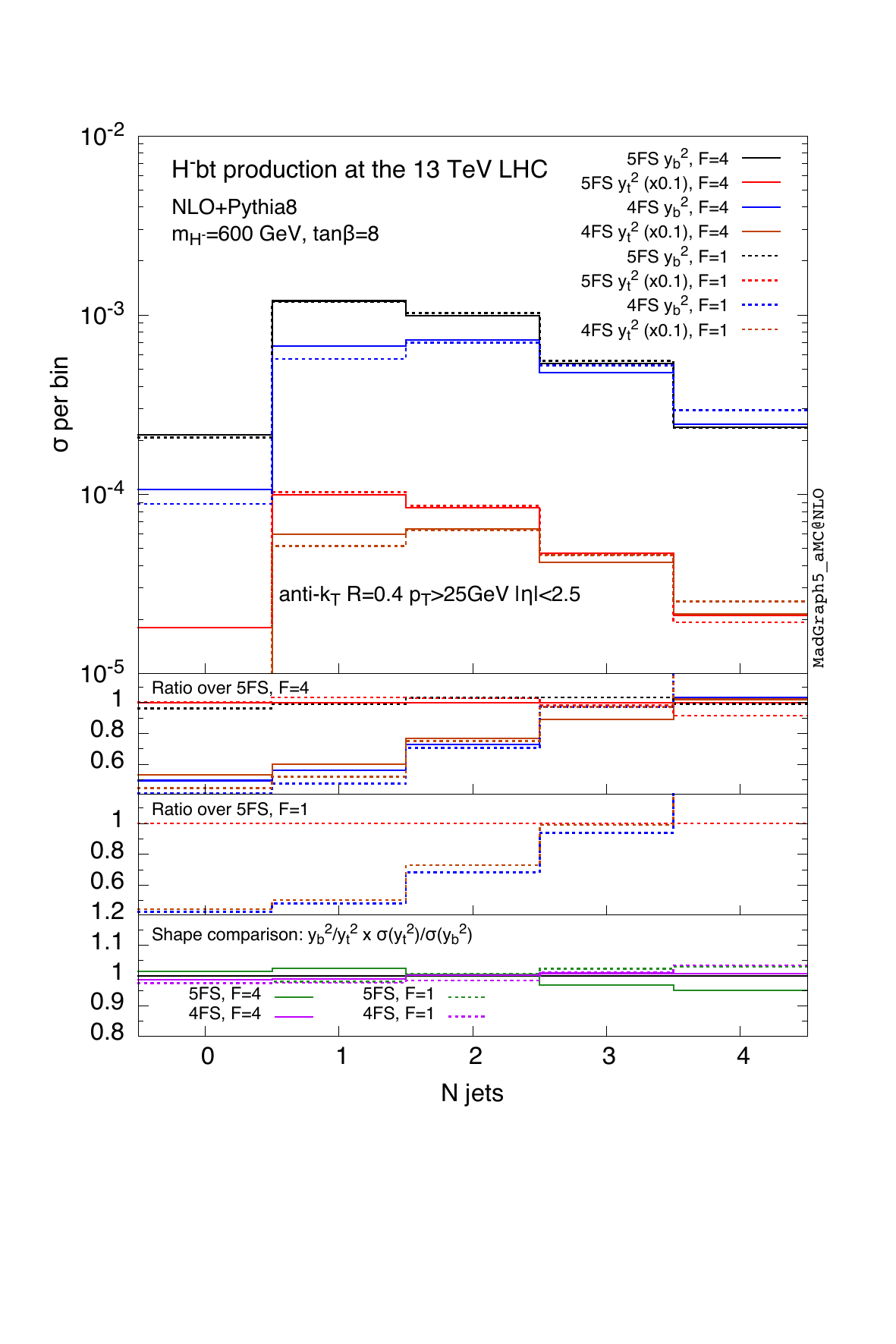}
\caption{\label{fig:frac1vs4njet} Same as Fig.~\ref{fig:frac1vs4pth}, but for the jet multiplicity with $m_{H^-}=200\gev$ (left) 
and $m_{H^-}=600\gev$ (right).}
\end{figure}
\noindent
We continue our analysis by investigating how the choice of the shower scale 
affects the results in the two schemes. 
We stress once more that our default choice ($F=4$) is physically well-motivated by the arguments given in Sect.~\ref{sec:settings}.
Below we show that this choice also improves the mutual agreement between the NLO predictions in the
two schemes. A number of differential distributions in the 4FS and 5FS for $F=1$ and $F=4$ are shown in Figs.~\ref{fig:frac1vs4pth}-\ref{fig:frac1vs4njet}, for both $m_{H^-} = 200\gev$ and $600 \gev$. The main frame displays
predictions in the 5FS for the $y_b^2$ term (black) and $y_t^2$ term (red) as well as in the 4FS
(in blue and orange respectively). Solid curves are used
for our reference predictions with $F=4$, while dashed curves refer to the default choice in 
\amc\ ($F=1$). The first inset displays the ratio of the curves in the main
frame over the corresponding ones ($y_b^2$ or $y_t^2$) in the 5FS for $F=4$. 
The second inset shows the ratio of the curves with $F=1$ in the 4FS over the corresponding ones in
the 5FS. In these two insets we can analyse whether $F=1$ or $F=4$ 
yields a flatter 4FS/5FS ratio.
In the last inset, the ratio of the normalised $y_b^2$ and $y_t^2$ distributions is plotted, 
for the 4FS and 5FS and for $F=1,4$. The purpose of this inset is to study whether  the $y_t^2$ and $y_b^2$ contributions develop similar shapes.

Overall, we observe a smaller dependence on $F$ in the 5FS than in the 4FS distributions. 
A similar behaviour was found in the context of Higgs production in association 
with bottom quarks~\cite{Wiesemann:2014ioa}. In any case, the choice of $F$ is almost irrelevant
when considering observables that are not sensitive to the $b$-quark 
degrees of freedom. 
In Fig.~\ref{fig:frac1vs4pth}, where $m_{H^-}=200\gev$,
we notice that charged Higgs and top $p_T$ distributions 
are slightly harder in the 4FS 
(10-15\% at large $p_T$) for $F=1$. Moreover, 
the $F=4$ choice yields a remarkably flat 4FS/5FS ratio, much flatter than for $F=1$. 
Consequently, $F=4$  improves the agreement between the two schemes. 

This improvement becomes even more visible for observables sensitive to the $b$ kinematics. For example, Fig.~\ref{fig:frac1vs4ptbjh1} shows the transverse momentum of the hardest (left panel) and second-hardest (right panel) $b$ jet,
where $F=1$ provides much harder spectra in the 4FS than $F=4$. A similar pattern 
is visible in the case of $B$ hadrons. In this case, however, the general agreement 
between 4FS and 5FS significantly deteriorates, as it was already observed.
 
In general, similar conclusions can be drawn for $m_{H^-}=600\gev$. 
The only two distributions that exhibit a different behaviour as compared to the $m_{H^-}=200\gev$ case 
are shown in Fig.~\ref{fig:600-frac1vs4pth}. The transverse momentum distribution of the heavy
charged Higgs boson (left panel) in the 4FS is much more affected by the choice of $F$, with
effects that reach up to 40\% in the tail of the distribution. In this case
the 4FS and 5FS mutual agreement is significantly improved by the choice $F=4$. 
The transverse momentum distribution of the second-hardest $b$ jet (right panel)
is less sensitive to the choice of $F$ than in the lighter-Higgs scenario.

\begin{figure}[t]
\centering
\includegraphics[width=0.47\textwidth, clip=true, trim=0.5cm 4cm 0.7cm 1cm]{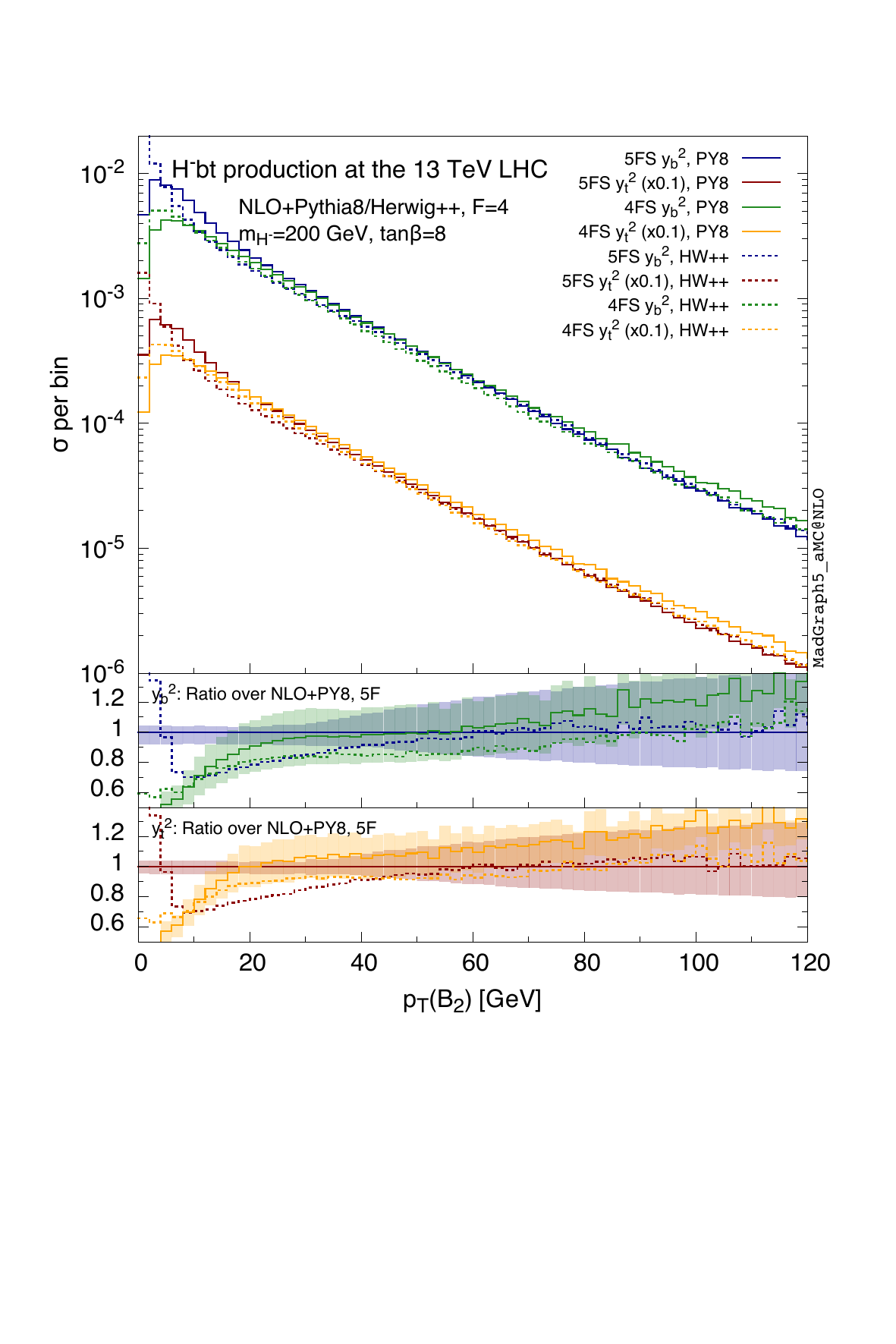}
\includegraphics[width=0.47\textwidth, clip=true, trim=0.5cm 4cm 0.7cm 1cm]{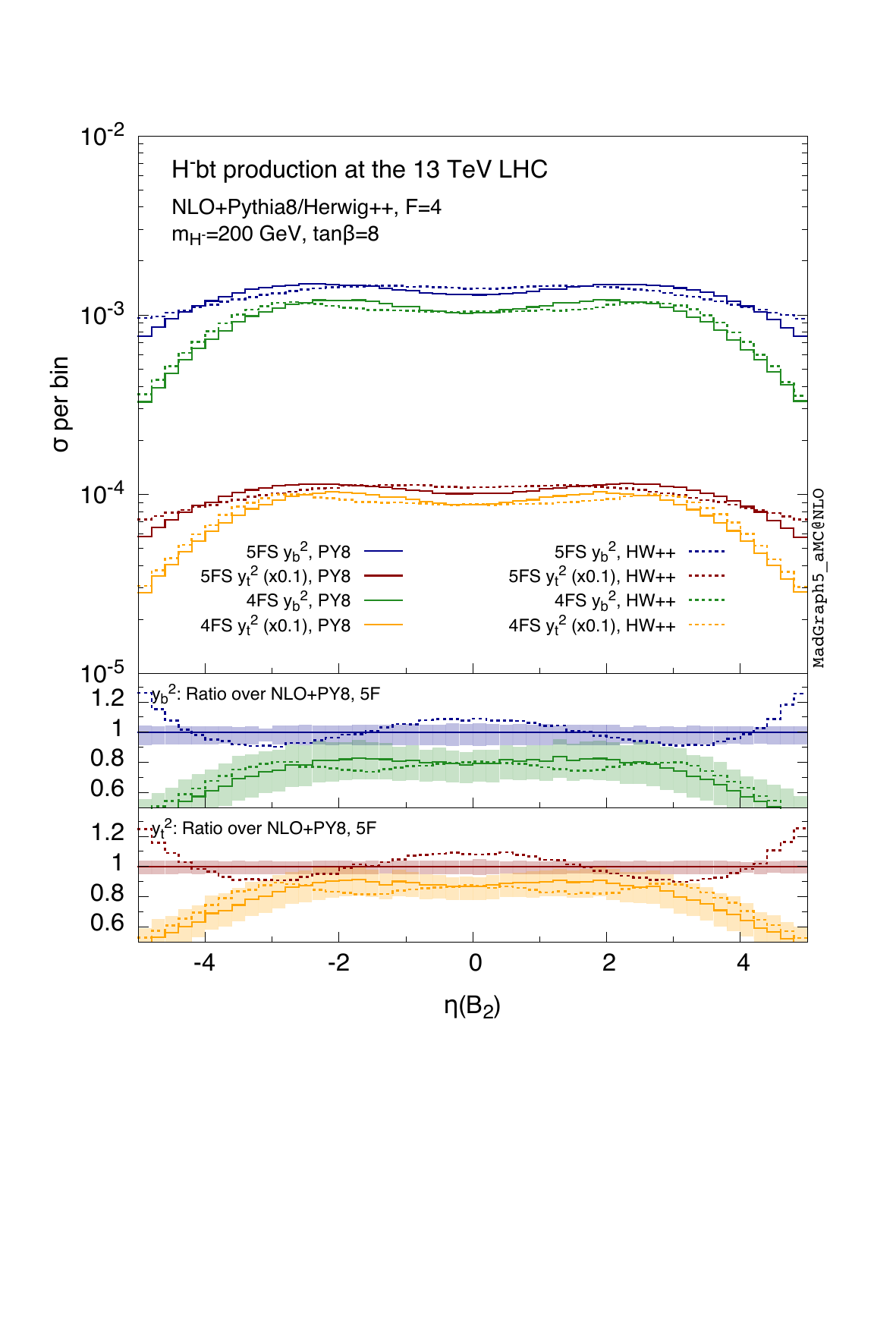}
\caption{\label{fig:hwvspyptetabh2}Transverse momentum (left) and pseudo-rapidity (right) distribution of the second-hardest $B$ hadron. NLO
curves matched with {\sc Pythia8}\ (solid) and {\sc Herwig++}\ (dashed) are shown, in the 5FS (blue for the $y_b^2$ term, dark-red for the $y_t^2$ one) 
and 4FS (green for the $y_b^2$ term, light-orange for the $y_t^2$ one). The two insets show the ratios of the histogram in the main frame over the 
5FS prediction matched with {\sc Pythia8}, for the $y_b^2$ and $y_t^2$ terms separately. Scale uncertainty bands are shown around the {\sc Pythia8}
predictions.}
\end{figure}

Jet multiplicities are a class of observables that are particularly sensitive to the excess 
of radiation generated by using $F=1$.  
From Fig.~\ref{fig:frac1vs4njet} we conclude that the 4FS generally 
prefers higher jet multiplicities than the 5FS. A similar behaviour 
was observed also in Ref.~\cite{Demartin:2015uha} and it was explained 
by considering the different colour structure of the initial state in the two schemes, besides 
the fact that the process is generally harder in the 4FS.
This tendency is slightly reduced---and as a consequence the agreement slightly improves---by the
choice of a smaller shower scale ($F=4$).
We remark that the dependence on the shower scale is even less apparent 
in the heavy charged Higgs case, and for $b$-jet multiplicities (not shown).

Finally, the ratio between the $y_b^2$ and $y_t^2$ terms (last insets of Figs.~\ref{fig:frac1vs4pth}-\ref{fig:frac1vs4njet}) 
illustrates that the two 
contributions give remarkably similar shapes for all observables under consideration. 
Any difference emerges from the dynamical scale choice at which
 the bottom Yukawa is computed ($\bar{m}_b(\mu_R)$ with $\mu_R=H_T/3$, 
see Sect.~\ref{sec:settings}). However, such difference is never larger 
than 10\%, with the $y_b^2$ distributions being slightly softer 
than the $y_t^2$ ones. \\

\noindent
Finally, we analyse the sensitivity of various observables to the parton shower in the four- and five-flavour
schemes. To this purpose, we compare results at NLO matched with the {\sc Pythia8} and {Herwig++} Monte Carlos. 
Having verified that the relative behaviour of the two Monte Carlos hardly depends on 
the specific choice of the charged Higgs mass under consideration, 
we limit the discussion to the $m_{H^-}=200\gev$ results.

As already pointed out in the introduction, 
the parton-shower matching for bottom-quark initial states 
in the 5FS involves some approximations: the initial-state backward evolution is based 
on leading-log accurate gluon splittings and requires the reshuffling of massless into 
massive bottom quarks. 
For these reasons, the 5FS predictions are extremely sensitive to the specific 
treatment of bottom quarks in a given Monte Carlo. This is most 
remarkable for $b$-jet/$B$-hadron related observables, such as the transverse momentum distribution of 
the second-hardest $B$ hadron (displayed in the left panel of Fig.~\ref{fig:hwvspyptetabh2}). 
In the 5FS, the {\sc Herwig++} prediction displays a significant shape distortion, both at small
and large values of the transverse momentum, falling outside the {\sc Pythia8} uncertainty bands at small $p_T$. 
A similar discrepancy is evident in the forward region 
of the rapidity spectrum, shown in the right panel of Fig.~\ref{fig:hwvspyptetabh2}. In this plot, the
5FS prediction is larger for {\sc Herwig++} than for {\sc Pythia8}. For both observables, the 4FS 
results display a significantly smaller Monte Carlo dependence.

\begin{figure}[p]
\centering
\includegraphics[width=0.47\textwidth, clip=true, trim=0.5cm 4cm 0.7cm 1cm]{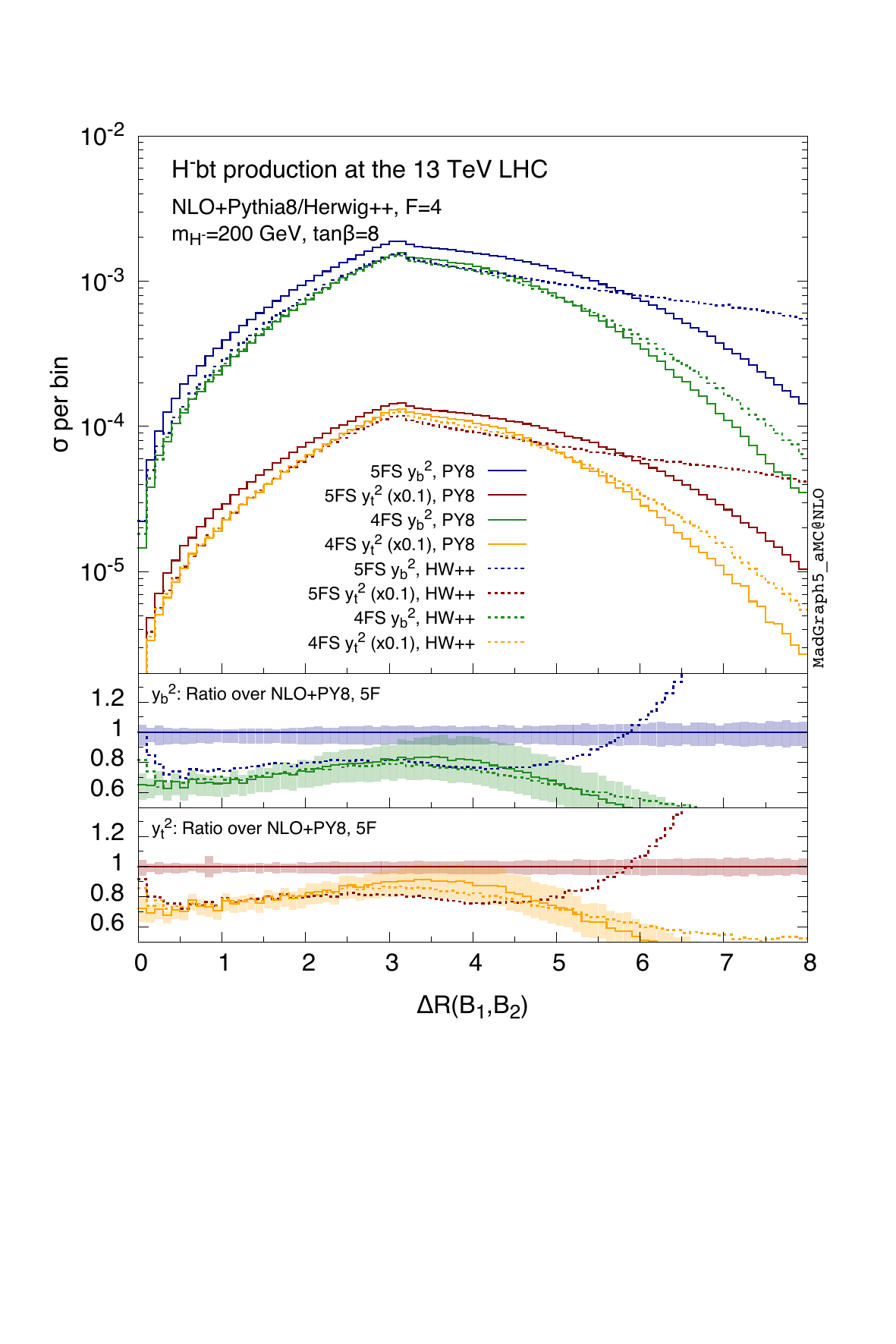}
\includegraphics[width=0.47\textwidth, clip=true, trim=0.5cm 4cm 0.7cm 1cm]{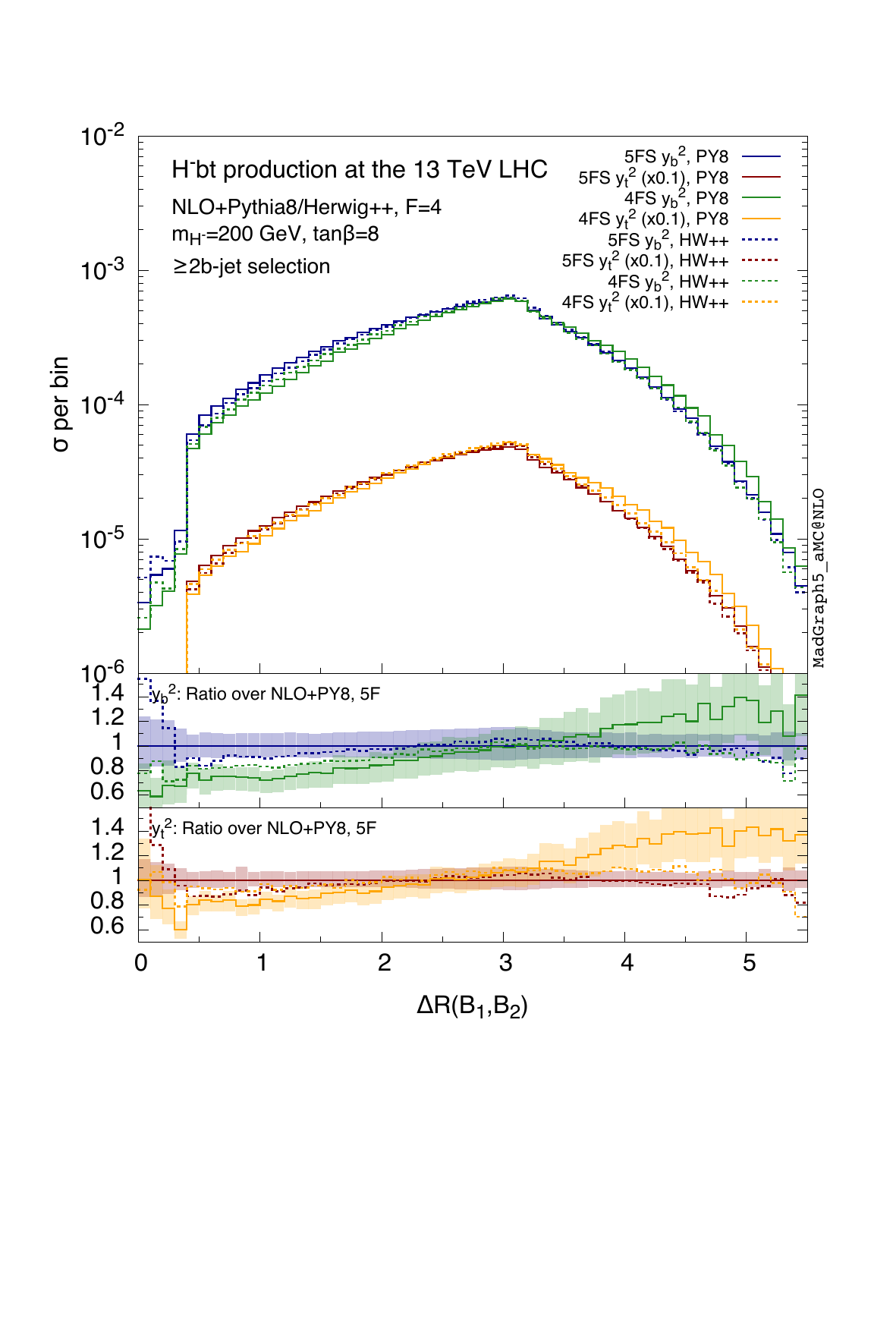}
\caption{\label{fig:hwvspydrbh}Same as Fig.~\ref{fig:hwvspyptetabh2}, but for the $\eta-\phi$ distance of the two hardest $B$ hadrons with no cuts (left) 
and requiring at least two $b$ jets (right).}
\includegraphics[width=0.47\textwidth, clip=true, trim=0.5cm 4cm 0.7cm 1cm]{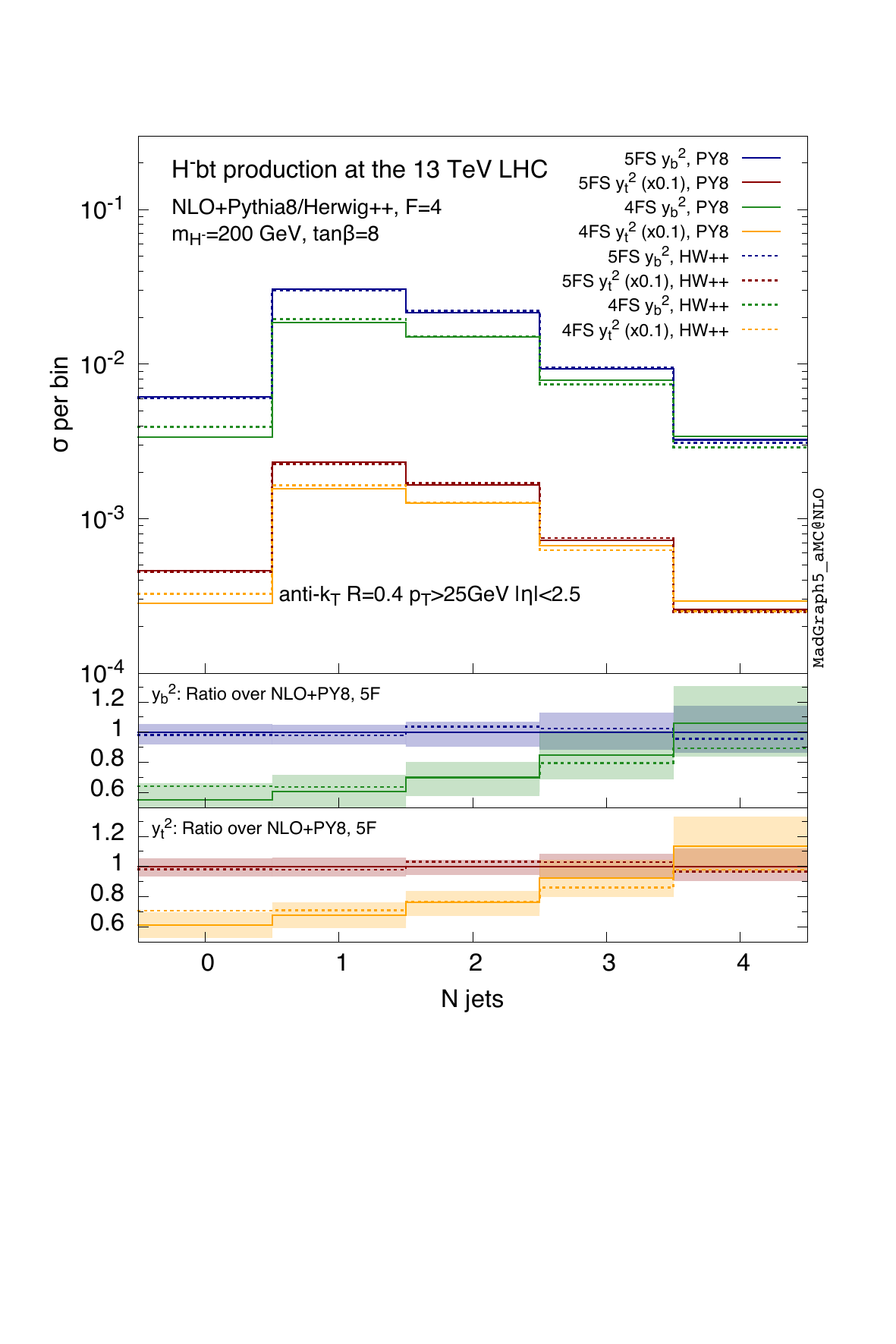}
\includegraphics[width=0.47\textwidth, clip=true, trim=0.5cm 4cm 0.7cm 1cm]{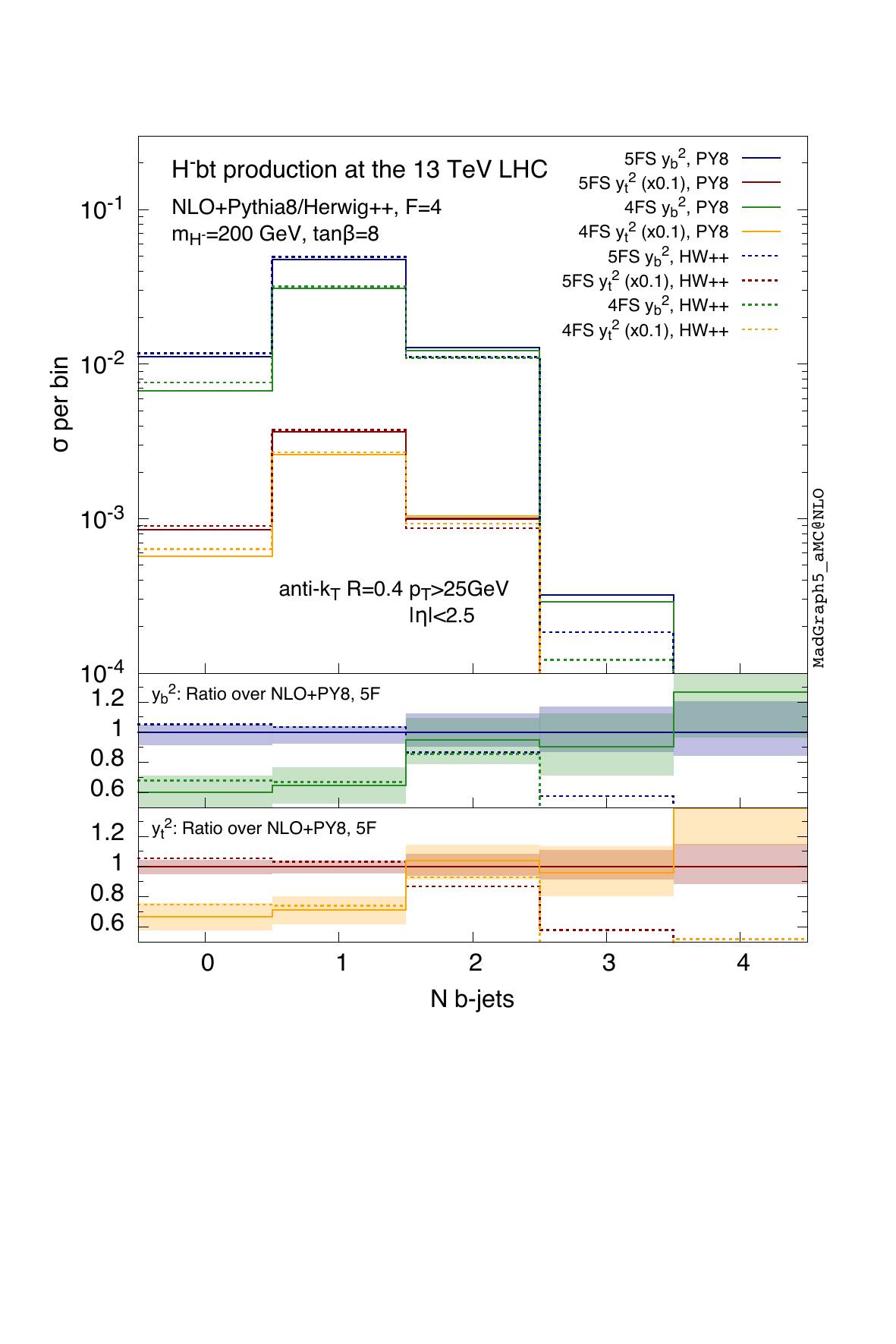}
\caption{\label{fig:hwvspynjet}Same as Fig.~\ref{fig:hwvspyptetabh2}, but for the jet (left) and $b$-jet (right) multiplicity.}
\end{figure}

An even more spectacular example is provided by the $\eta-\phi$ distance between the two hardest 
$B$ hadrons. The results are shown in the left panel of Fig.~\ref{fig:hwvspydrbh}. 
The {\sc Herwig++} prediction in the 5FS features a much higher tail at large separations.
This can be traced back to the fact that {\sc Herwig++} tends to produce $B$ hadrons 
much closer to the beam line when simulations are performed in the 5FS. The same behaviour was observed 
in Ref.~\cite{Wiesemann:2014ioa}, in which it was pointed out that such effects 
are however not relevant when realistic cuts on the $B$ hadrons
are imposed, for example, when $b$ jets are required. As a matter of fact,
we observe a neatly improved agreement among all curves 
in Fig.~\ref{fig:hwvspydrbh} when requiring at least two $b$ jets (right panel).

Looking further at jet (left panel) and $b$-jet  (right panel) multiplicities in Fig.~\ref{fig:hwvspynjet}, 
we observe instead that the two flavour-schemes display a similar Monte Carlo dependence. 
In the flavour--unspecific case 
this dependence is quite small for all jet-multiplicity bins. On the other hand, for $b$ jets 
{\sc Herwig++} and {\sc Pythia8} are in good agreement up to one $b$ jet in the 5FS and two $b$ jets 
in the 4FS, that is up to the multiplicities described by the hard matrix element at NLO. The two $b$-jet
bin in the 5FS---described only at LO by the hard matrix element---is affected by larger discrepancies between
the two Monte Carlos, which exceed the uncertainty bands. Larger $b$-jet multiplicities, generated only
at the shower level via gluon splittings, are affected by discrepancies as large as 100\%: {\sc Herwig++} 
predicts less $b$ jets than {\sc Pythia8}.

Generally speaking, the difference between the Monte Carlos is smaller for the 4FS than for 5FS predictions. 
Indeed, the 4FS has more differential information 
at the matrix-element level, which reduces the effects of the shower. 
The only case worth mentioning, in which the Monte Carlo dependence is larger in the 4FS than in the 5FS
is the distance between two hardest $b$ jets at large separations. This observable 
is closely related to the distance of the two hardest $B$ hadrons when two $b$ jets are required 
(see right panel in Fig.~\ref{fig:hwvspydrbh}). For both observables, at large separations,
the 4FS predictions matched with {\sc Herwig++} lie very close to the 5FS NLO+PS predictions, while the 
matching with {\sc Pythia8} yields a higher tail. 
However, we point out that such discrepancy between {\sc Herwig++} and {\sc Pythia8} in the 4FS barely exceeds the scale uncertainty bands. 

Finally, we remark that similar conclusions can be drawn for the heavier charged Higgs 
case for most of the studied observables. However, a heavier charged Higgs reduces the
differences between the two Monte Carlos (in particular in the 5FS) in the transverse 
momentum distribution of the two hardest $B$ hadrons/$b$ jets.

%%%%%%%%%%%%%%%%%%%%%%%%%%%%%%%%%%%%%%%%%%%%%%%%%%%%%%%%%%%%%%%
\section{Conclusions}
\label{sec:conclusion}
%%%%%%%%%%%%%%%%%%%%%%%%%%%%%%%%%%%%%%%%%%%%%%%%%%%%%%%%%%%%%%%
We have presented predictions for the production of a heavy charged Higgs boson in a type-II 2HDM, by
explicitly considering a lighter Higgs scenario ($m_{H^-}$ = 200 GeV) and a heavier one 
($m_{H^-}$ = 600 GeV). Our predictions have been presented for $\tan\beta =8$, but they are applicable 
to any $\tan\beta$ value through a simple rescaling. Furthermore, these results can be straightforwardly
extended to a type-I 2HDM by rescaling the Yukawa couplings. Details are given in Sect.~6 of Ref.~\cite{Flechl:2014wfa}.

For the first time, a fully differential computation has been performed in the 4FS at fNLO and NLO+PS accuracy. 
We have exploited the automatised \amc\ framework, by obtaining the NLO version of the 2HDM via 
the \nlo\ package. The model has been supplemented by the computation of the bottom Yukawa coupling
in the $\overline{\text{MS}}$ scheme, which has the advantage with respect to the default on-shell scheme 
of resumming large logarithms of $m_b/\mu_R$.

Our results indicate that a reduced shower scale with respect 
to the default one in \amc\ improves the matching between parton shower and
fixed-order results at large transverse momenta. 
For this scale choice, we discussed the effects
of incorporating NLO corrections and the matching to the parton shower 
in the 4FS simulations by considering a number of differential 
observables. We 
found NLO corrections to be generally flat
with our choice of the renormalisation and factorisation scales. 
Effects due to the parton shower are important in the 
Sudakov-dominated regions (jets and $B$ hadrons at low $p_T$), 
or when observables sensitive to the $b$-quark fragmentation are 
considered.
These observations have been made separately for the $y_b^2$ and $y_t^2$ terms. 
On the other hand, we argued that the $y_by_t$ contribution, appearing only in the 4FS, 
can be safely neglected, since its size is smaller than $\sim 5$\% of the total
cross section (for $\tan\beta=8$ and $m_{H^-}=200$ GeV)
and is well within the scale uncertainty of the computation. For larger Higgs masses or different values of
$\tan\beta$ it is further suppressed.

Besides discussing the new results in the 4FS, we provided a 
comprehensive comparison to the ones
in the 5FS, consistently generated within \amc. The inclusion of NLO(+PS) corrections
in the two schemes improves their mutual agreement at the level of shapes. 
This agreement follows, although to a minor extent, from the 
reduced shower scale choice. Differences remain, however, and they are particularly
sizeable for observables related to $b$ jets and $B$ hadrons. Given
these differences, it was vital to carry out an unbiased analysis of our results 
in order to acquire the most reliable predictions for this class of observables.
The proper simulation of the signal will be crucial 
for the experiments to fully exploit the potential of the data collected in
charged Higgs searches at the LHC. 

Our final recommendation is to use 4FS predictions for any realistic
signal simulation in experimental searches. 
This recommendation is backed by two sets of evidences: first we have proven that, 
for a large number of observables, the 4FS prediction provides 
a better description of the final state kinematics; second, it reduces the systematic 
error related to the 
usage of a given parton shower. 
Moreover, when matching the NLO calculation to the shower, we recommend to use a lower shower 
scale (by setting $F=4$ in our case). 
This choice provides a better matching to the fixed-order computation at large 
transverse momenta, slightly reduces the parton shower dependence 
and improves the agreement of four- and five-flavour scheme computations.

Any user interested in the simulation of charged Higgs production with \amc\ is strongly 
encouraged to contact the authors.

\section*{Acknowledgements}
We are indebted to Fabio Maltoni for his constant encouragement and his valuable suggestions. 
We thank Michael Spira, Paolo Torrielli and Rikkert Frederix for useful discussions
and Michael Kr\"amer for having provided the 4FS code that we used for comparison.
Thanks to the SUSY working group in Cambridge for comments and discussions.
Finally, we are grateful to Liron Barack and her experimental colleagues for their suggestions and requests.
MW and MZ thank the Cavendish Laboratory and the HEP group for their hospitality in Cambridge
during the course of this work. CD is a Durham International Junior Research
Fellow. The work of MU is supported by the UK Science and Technology Facilities Council. 
MW is supported by the Swiss National Science
Foundation (SNF) under contract 200020-141360. The work of MZ is supported by the ERC
grant ``Higgs@LHC'', in part by the Research Executive Agency (REA) of the European
Union under the Grant Agreement number PITN-GA-2010-264564 (LHCPhenoNet), and
in part by the ILP LABEX (ANR-10-LABX-63), in turn supported by French state funds
managed by the ANR within the ``Investissements d'Avenir'' programme under reference
ANR-11-IDEX-0004-02. This work was supported in part by the European Union as part of 
the FP7 Marie Curie Initial Training Network MCnetITN (PITN-GA-2012-315877).

\renewcommand{\em}{}
\bibliographystyle{UTPstyle}
\bibliography{cHdiff}

\end{document}